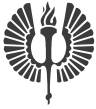

# MEASUREMENT SIMULABILITY AND INCOMPATIBILITY IN QUANTUM THEORY AND OTHER OPERATIONAL THEORIES

Leevi Leppäjärvi

# Acknowledgements

Even before my undergraduate studies I have been fascinated by the inner workings of Nature and the laws of physics that dictate it. Especially hearing about some of the peculiarities of quantum mechanics led me down the road I am currently on of trying to understand those peculiarities and the theory behind them. This Thesis is a result of that journey so far, and it is my first attempt at trying to answer some of the questions surrounding the foundations of quantum theory. However, the work and research presented in this Thesis could not have been possible without several other people.

First and foremost I want to thank my supervisors Teiko and Sergey. Teiko, you have been teaching, guiding and helping me through every level of my studies ever since I came to your office with the vague idea of investigating the proof of Gleason's theorem for my bachelor's thesis (a proof which was, to my huge disappointment, skipped in your lectures). You also introduced me to Sergey who evolved from a co-author to also become my mentor and eventually my official co-supervisor. Thank you both for your guidance and support in my journey.

Next I want to thank my other collaborators Martin, Miso, Oskari and Stan without whom many of the results presented in this thesis would not have been published. I also thank Marcin Pawłowski and Cécilia Lancien for pre-examining this thesis, and Tom for proofreading it and helping me in making it readable.

Of course I cannot forget the environmental impact of the whole laboratory of theoretical physics and its personnel. The supportive working atmosphere and sense of community in the laboratory and our research group is something I will truly miss. Thank you Sabrina for creating and sustaining this atmosphere for as long as it lasted. Everyone in the laboratory has had a positive impact on me and for that I am grateful. I thank all of the former and current fellow doctoral students for sharing the journey with me. Especially Henri, Juho, Oskari and Laura, thank you for keeping the mood light during the coffee breaks and for being my friends outside of the work as well. And our great postdoc Tom, just so know, you are also included in this group of people (mainly because of all the rounds of Sopio that we played) but I wanted to separate you from the others for the sole purpose of making you stand out which I know makes you uncomfortable. So thank you, Tom!

As interesting and fulfilling as doing research is, solely thinking in equations can get tiresome for ones mind and body in the long run. To balance out the sitting and thinking, the solution that I found that works for me for this is the sport of powerlifting. I want to thank all the people that have shared this part of my life as well. Most notably I am grateful for: Henri, for making this discovery with me; I will never forget our training in Faraday and Roddis. My brother Joona, for helping me with all things related as well as setting an example in the world of iron. And Oskari, for prepping and competing with me in my first meet. Thank you for keeping me sane.

Last but not least I want to thank my family and my relatives as well as my old friends from Lapland; you have always supported me, offered me an escape from my work during holidays and helped me not to forget where I come from.

# Contents

# Abstract

Quantum theory is a particularly important instance of an operational theory. By looking at quantum theory from the perspective of more abstract operational framework one is able to study its properties in a wider context. This allows us to identify some of the physical features characteristic of quantum theory and it helps us to understand what makes quantum theory special among other theories. From the information-theoretic point of view this might give us insight into the foundations behind the advantages of quantum information processing over its classical counterpart.

In this thesis, based on Publications **I** − **VI**, we consider the properties of measurements in quantum theory and other operational theories. After having introduced the framework of operational theories, we consider a communication scheme based on an experimental prepare-and-measure scenario and demonstrate this with different communication tasks. This gives us context for how the different communication tasks can be implemented in different theories and how operational theories can be compared to each other, in doing so establishing quantum theory intuitively as an operational theory among other theories.

The main property of measurements we focus on in this work is the simulation of measurements, which consists of manipulating the inputs and outputs of the measurement devices. We study how using this process on existing measurement devices can be used to operationally imitate new devices, and what kind of structure the simulation process induces on measurements. We look at the components of simulability, analysing and demonstrating them in quantum theory as well as various toy theories. This gives us structural information that differentiates quantum theory from other theories.

We also consider applications of simulability. Firstly, we consider operational restrictions imposed upon measurements. We argue that the restricted set of physical measurements must be closed with respect to the simulation process since the simulation of physical devices must lead to other physically feasible devices. We demonstrate different types of restrictions by classifying them and analysing their structure.

As a second application we see how the simulation of measurements relates to joint measurability, i.e. compatibility of measurements, and how it can be viewed as a generalisation of it. This allows us to present an operational principle previously known to quantum theory, the no-free-information principle, according to which any measurement that is compatible with all other measurement must not provide any useful, and therefore free, information about the system. Whilst this principle holds in quantum theory, there are non-classical theories for which it is violated, and so enforcing this principle may be considered a way to exclude some unphysical theories.

# Tiivistelmä

Kvanttiteoria on operationaalisten teorioiden tärkeä erikoistapaus. Tarkastelemalla kvanttiteoriaa abstraktien operationaalisten teorioiden näkökulmasta voidaan kvanttiteorian tärkeimpiä piirteitä tutkia laajemmassa mittakaavassa. Tämän avulla voimme yrittää ymmärtää kvanttiteoriaa karakterisoivia fysikaalisia ominaisuuksia, ja samalla pyrimme näkemään mikä tekee kvanttiteoriasta niin erikoisen ja tärkeän. Informaatioteoreettisesta näkökulmasta katsottuna tämä voi auttaa oivaltamaan mikä antaa kvantti-informaatiolle ja sen prosessoinnille edun klassiseen informaatioteoriaan nähden.

Tässä väitöskirjassa, perustuen julkaisuihin **I** − **VI**, tutkin mittauksia ja niiden ominaisuuksia kvanttiteoriassa ja muissa operationaalisissa teorioissa. Operationaalisten teorioiden määrittelyn jälkeen jatkan tutkimalla fysikaaliseen koejärjestelyyn perustuvaa kommunikaatiota eri teorioissa. Esittelen tähän viitekehykseen sopivia erilaisia kommunikointitehtäviä, selvitän miten hyvin niitä voidaan toteuttaa eri teorioissa ja tämän perusteella vertaan teorioita toisiinsa. Tämä auttaa ymmärtämään kvanttiteoriaa luonnollisena osana operationaalisten teorioiden joukkoa.

Väitöstyössäni keskityn erityisesti mittausten simulointiin, joka operationaalisesti tapahtuu manipuloimalla eri mittalaitteistojen syöte- ja tulosteportteja. Tällä tavalla saatavilla olevia mittalaitteita manipuloimalla voidaan simuloidan jonkin toisen, mahdollisesti uuden mittalaitten toimintaa. Simulonti synnyttää mittalaitteiden teoreettiseen kuvaukseen matemaattista rakennetta, jota tässä työssä erityisesti tutkin. Analysoimalla simulointia kvanttiteoriassa ja useissa leluteorioissa pystyn osittain erottamaan kvanttiteorian luonteenomaista rakennetta.

Simuloinnin eräänä sovelluksena tutkin kuinka mittauksille voidaan asettaa mahdollisia operationaalisia rajoitteita, joiden kuitenkin simuloinnin operationaalisen luonteen takia pitää olla suljettuja simuloinnin suhteen. Näin ollen simulointiprosessilla ei pystytä kiertämään asetettuja rajoitteita. Erityisesti annan esimerkkejä erilaisista operationaalisista rajoitteista,

**9**

jaan rajoitteet eri luokkiin ja tukin näiden luokkien eroja.

Toisena sovelluksena linkitän simuloinnin eri fysikaalisten suureiden yhteismittauksen (tai yhteensopivuuden) käsitteeseen, jonka mukaan yhteensopivat suureet voidaan mitatata samanaikaisella yhteismittalaitteella. Tällöin voidaan nähdä kuinka yhteismittaus on vain simuloinnin erikoistapaus. Simuloinnin ja yhteensopivuuden yhteyden perusteella yleistän kvanttiteoriasta tutun operationaalisen periaatteen, jonka mukaan kaikkien mittauksien kanssa yhteismitattavat suureet eivät voi antaa olennaista tietoa mitattavasta fysikaalisesta systeemistä. Muotoilemalla tämän periaatteen operationaalisissa teorioissa pystyn osoittamaan, että periaate ei päde kaikissa teorioissa osoittaen, että sen täytyy vangita jotain kvanttiteorialle ominaisia piirteitä.

# List of papers

This thesis consists of a review of the subject and the following original research articles:

    **I**    **A necessary condition for incompatibility of observables in general probabilistic theories**,
S. N. Filippov, T. Heinosaari, *L. Leppäjärvi*, Phys. Rev. A **95**, 032127 (2017).

    **II**    **Simulability of observables in general probabilistic theories**,
S. N. Filippov, T. Heinosaari, *L. Leppäjärvi*, Phys. Rev. A **97**, 062102 (2018).

    **III**    **No-free-information principle in general probabilistic theories**,
T. Heinosaari, *L. Leppäjärvi*, M. Plávala, Quantum **3**, 157 (2019).

    **IV**    **Operational restrictions in general probabilistic theories**,
S. N. Filippov, S. Gudder, T. Heinosaari, *L. Leppäjärvi*, Found. Phys. **50**, 850–876 (2020).

    **V**    **Communication tasks in operational theories**,
T. Heinosaari, O. Kerppo, *L. Leppäjärvi*, J. Phys. A: Math. Theor. **53**, 435302 (2020).

    **VI**    **Postprocessing of quantum instruments**,
*L. Leppäjärvi*, M. Sedlák, Phys. Rev. A **103**, 022615 (2021).

# Other published material

This is a list of the publications produced which have not been chosen as a part of this doctoral thesis

- **Entanglement protection via periodic environment resetting in continuous-time quantum-dynamical processes**,
  T. Bullock, F. Cosco, M. Haddara, S. Hamedani Raja, O. Kerppo, *L. Leppäjärvi*, O. Siltanen, N. W. Talarico, A. De Pasquale, V. Giovannetti, S. Maniscalco, Phys. Rev. A **98**, 042301 (2018).

- **Phase covariant qubit dynamics and divisibility**,
  S. N. Filippov, A. N. Glinov, *L. Leppäjärvi*, Lobachevskii J. Math. **41**, 617–630 (2020).

# Introduction

Historically quantum theory emerged from the pursuit of describing the fundamental physical properties of Nature at the scale of atoms and subatomic particles, where the laws of classical mechanics were shown to be insufficient. Since its birth over 100 years ago, quantum theory has evolved from just a theoretical curiosity into an experimentally applicable theory of information and computation. In particular, it has been demonstrated that quantum information theory holds the potential to make communication and computation more secure and more efficient. It is because of the success of quantum information theory that the most important and most studied quantum system is the qubit, a quantum version of the classical unit of information, the bit. Similar to the bit, the qubit is a two-level system, making it also the simplest non-trivial quantum system. By taking into account the importance and the simplicity of the qubit, it is not hard to see why the qubit has become the primary quantum system that we are taught in modern-day introductory courses on quantum theory.

One of the first things we learn about qubits is that they can be parameterized with three real numbers and that the space of possible parameters form a regular 3-dimensional ball, called the *Bloch ball*. This is contrary to the much more intuitive classical bit which can be represented as a line segment whose end points correspond to the two possible values of the bit. As both of these systems are two-level systems, i.e., at any given time it is only possible to distinguish between two different states, what is it that distinguishes the qubit from its classical counterpart? And why is it specifically a ball and not a disc, a regular polygon or a Platonic solid that Nature chose to represent the possible qubit states? Which known and used quantum properties are specifically because of the spherical shape, and what would happen if the shape were to change? Why are not all higher-dimensional quantum systems represented by higher dimensional balls? These are some of the questions one might have after hearing about qubits and the Bloch ball for the first time.

Whilst in basic courses these types of questions can be, to some degree, answered by exploring the mathematical structure of the theory (such as the superposition principle) given by the mathematical axioms of quantum theory as laid out by Dirac [1] and von Neumann [2], the arguments involving the physical and operational properties of quantum theory seems to be lost and forgotten in the translation. To bridge this gap between the clear and well-understood mathematical structure of the theory and the possible physical principles either leading to or emerging from that structure, one has to consider quantum theory in a wider class of *operational theories*. Based on the primitive concepts of *physical systems*, *states* and *measurements*, an operational theory specifies what kind of states the physical system can be prepared in, and it determines the rules on how the outcome statistics of an experiment involving said primitives can be calculated. In its simplest terms, operational theories aim to capture the minimal requirements that one wishes an empirical physical theory to satisfy.

By considering theories with different rules for the preparation and measurement procedures one is able to study how these affect the resulting physical properties of these theories. In particular, this allows us to consider what is uniquely 'quantum' about quantum theory and what gives it an advantage over the classical theory. Furthermore, we can formulate and study the operational non-classical properties of quantum theory and see how they behave in other theories to see how quantum theory is set apart and how these properties can be used to characterize different theories.

The central ideas and motivation for operational theories were previously given in the works of Mackey [3], Ludwig [4–7], Dähn [8], Stolz [9, 10], Davies & Lewis [11], Edwards [12, 13], Mielnik [14, 15], and Gudder [16] with the aim of finding an axiomatization for quantum theory. More recent topics that have been studied in the framework of operational theories, some of which are based more on information-theoretical concepts after the success of quantum information theory, include (information-theoretic) axiomatization of quantum theory [17–21], no-cloning and no-broadcasting [22, 23], non-locality [24–26], joint measurability [27–35], steering [25, 31, 36], entropies [37–39], thermodynamics [40, 41], entanglement [40, 42, 43] and contextuality [44, 45]. For more extensive historical and topical reviews, see [46, 47].

As one of the primitives, measurements and their properties play a central role in the research of different operational theories. As one of the most important examples, the concept of *joint measurability*, i.e., *compatibility of measurements* (see, e.g., [48]), captures the idea that given a set of measurement devices one is able to build a joint measurement device capable of implementing all of the given devices simultaneously. It is an elementary observation in quantum theory that there are measurements on quantum systems that cannot be implemented jointly, the paradigmatic example being the impossibility to sharply determine both the position and momentum of a particle at the same time (rooted in the Heisenberg uncertainty principle [49]). As we will see later this impossibility to measure everything jointly is not only a property of quantum theory but in fact it is a property of all theories that are not determined classical.

Another operational property of measurements is the *simulation of measurements* (see, e.g., [30]). Defined as a combination of the operational notions of mixing (using different measurement devices with different probabilities in each round of the experiment) and post-processing (manipulating the outputs of the measurement devices), simulation captures the idea of how one can operationally obtain new measurement devices from known ones. Thus, given a set of measurement devices, the simulation process can be used to obtain new devices that might not be directly at hand or that otherwise might be hard to implement or build. Furthermore, in this context, jointly measurable devices can be simulated from a single device so that simulability involving multiple devices can also be seen as a generalization of joint measurability.

The main purpose of this thesis is to study the properties arising from and the mathematical structures given by the simulation of measurements in quantum theory and other operational theories, consisting of original research conducted in Publications **I** − **VI**. In Chapter 1 we start by considering the basics of operational theories and present the convex formulation of operational theories; namely the framework of *general probabilistic theories (GPTs)*. We discuss the basic components (states, effects, observables, channels, instruments and composite systems) of an operational theory and demonstrate these in the context of quantum theory. We also present some toy theories that we will use later on to demonstrate various concepts.

In Chapter 2 we study a physical communication scheme arising from a prepare-and-measure scenario describing a standard physical experiment and the structure it induces. Following Publication **V**, we use the communication scheme to demonstrate various concepts and communication tasks that one may consider in operational theories, introduce a way to compare the difficulty of implementing such tasks and use this to compare different theories to each other based upon which tasks can be implemented. This helps us establish quantum theory as an operational theory among other theories.

In Chapter 3 we start our journey towards the simulation of measurements by considering one of the main components of the simulation scheme, namely the post-processing of measurements. We start by considering the classical manipulation of the measurement outcome statistics and then, following Publication **VI**, continue to generalize the post-processing relation to measurements which output not only classical measurement outcomes but also the post-measurement state, thereby accommodating sequential measurements and their manipulation.

In Chapter 4 we present the full simulation scheme for measurements with classical measurement outcomes, as described in Publication **II**. We show that every measurement can be reduced into a simulation of *simulation irreducible measurements* and characterize their structure and properties. Following Publication **IV**, we demonstrate measurement simulability by considering possible operational restrictions on measurements that are closed with respect to the process of simulation, characterize these restrictions in different classes and demonstrate their differences.

In the final Chapter of this thesis, Chapter 5, we explore the connection between simulability and compatibility. We show how joint measurability can be seen as a particular instance of simulability, and derive a simulation-based condition for joint measurability originally shown in Publication **I**. Furthermore, we show how the structure given by the simulation scheme, the simulation irreducible measurements in particular, can be used to characterize the set of measurements that are jointly measurable with any other measurement in a given theory. In accordance with Publication **III**, we show that although such measurements provide no information about the measured system in quantum theory, i.e., such measurements are trivial, in general we might have theories where the *no-free-information principle* is violated so that such non-trivial measurements exist and they can be thus implemented freely with any other measurement.

In addition to Publications **I − VI**, this thesis consists of some new unpublished or generalized results that are notated with a * in front of the respective Proposition/Corollary.

# Chapter 1

# Operational theories

For an operational theory, the primitives are physical devices: *state preparators*, *measurement devices* and *channels*, which together can be used to conduct experiments, giving information about the systems described by the theory. A physical experiment is typically separated into two parts: the *preparation* of the physical system followed by a *measurement* of the system including the registration of the measurement outcome. In order for the experiment to provide information about the physical system the measurement outcome should depend (at least probabilistically) on the chosen preparation and measurement. This is the premise of any statistical operational theory [50–52]. In this Chapter we present the most important operational notions and principles, and argue how they lead to the convex formulation of the *general probabilistic theories (GPTs)* which we will use as the framework for studying the operational properties of measurements that form the content of this Thesis.

## 1.1 Basic operational notions and principles

Let us denote the set of possible preparations by $\mathcal{P}$ and the set of measurements by $\mathcal{M}$. Given a preparation $P \in \mathcal{P}$ and a measurement $M \in \mathcal{M}$ with a set of possible outcomes $\Omega$ (which is taken to be a finite set for simplicity) there must be a probability distribution $p_M^P$ over $\Omega$ such that $p_M^P(x)$ is the probability that outcome $x \in \Omega$ is registered when the system goes through the preparation $P$ and is measured with $M$. When repeating the same experiment with fixed $P \in \mathcal{P}$ and $M \in \mathcal{M}$, the relative frequency of registering each outcome $x \in \Omega$ in each round should better approximate $p_M^P(x)$ as the number of repetitions increases. A schematic picture of a physical experiment is depicted in Fig. 1.1.

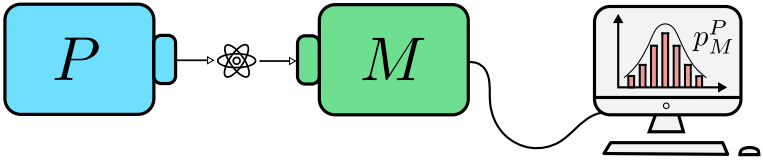

Figure 1.1: In a physical experiment a preparation device $P$ prepares a physical system in a state that is measured with a measurement device $M$ resulting in a measurement outcome probability distribution $p_M^P$.

In an experiment, two preparations $P$ and $\tilde{P}$ are considered equivalent if they produce the same measurement statistics for every measurement, i.e., $p_M^P = p_M^{\tilde{P}}$ for all $M \in \mathcal{M}$. Similarly two measurements $M$ and $\tilde{M}$ are equivalent if $p_M^P = p_{\tilde{M}}^P$ for all preparations $P \in \mathcal{P}$. The equivalence classes of preparations and measurements are called *states* and *observables* respectively. The set of states that a given physical system can be prepared in is denoted by $\mathcal{S}$ and the set of all observables that can be used to measure the system is denoted by $\mathcal{O}$.

One of the most basic operational properties is the probabilistic *mixing* of states and observables. If we are measuring an observable $O \in \mathcal{O}$ and have two different preparators preparing states $s_1 \in \mathcal{S}$ and $s_2 \in \mathcal{S}$, we can assign a probability $\lambda \in [0,1]$ and decide that with probability $\lambda$ we will use the preparator with state $s_1$ and with probability $1 - \lambda$ we will use the preparator with state $s_2$ in each round of the experiment. The result of the experiment thus leads to a probability distribution $\lambda p_O^{s_1} + (1-\lambda) p_O^{s_2}$. The mixing procedure described above can also be considered as a recipe to prepare a state, and this resulting state $s(\{\lambda, s_1, s_2\})$ must therefore satisfy $p_O^{s(\{\lambda, s_1, s_2\})} = \lambda p_O^{s_1} + (1-\lambda) p_O^{s_2}$. We call $s(\{\lambda, s_1, s_2\})$ the *mixture* of $s_1$ and $s_2$. Similarly to state preparators, one can choose to mix the measurement devices. If a system is prepared in a state $s \in \mathcal{S}$ and we have two observables $O_1$ and $O_2$ that we mix with weights $\lambda \in [0,1]$ and $1-\lambda$ respectively, the resulting probability distribution reads as $\lambda p_{O_1}^s + (1-\lambda) p_{O_2}^s$. This can be interpreted as a new mixed observable $O(\{\lambda, O_1, O_2\})$ that is dependent on $\lambda$, $O_1$ and $O_2$ and satisfies $p_{O(\{\lambda, O_1, O_2\})}^s = \lambda p_{O_1}^s + (1-\lambda) p_{O_2}^s$.

By accepting the mixing procedure as a new way to prepare states we impose a *convex structure* on the set of states $\mathcal{S}$. These kinds of structures can be formalized in various ways (see [16, 53–56]) but under some reasonable assumptions they can be reduced to the typical notion of convexity in real vector spaces [53, 56, 57]. This leads to one of the most common formulation of operational theories, namely the convex formulation of general probabilistic theories (GPTs).

## 1.2 The framework of general probabilistic theories

General probabilistic theories constitute an operational framework for considering quantum theory in a more abstract setting. In addition to quantum theory, GPTs include classical theory as well as countless toy theories. Once formally defined, one can start to consider various non-classical features of quantum theory in a more general setting, see how these features manifest themselves in different theories, quantify them and use them to compare theories to each other with the goal of understanding more about these features and what they tell us about quantum theory. We start by introducing the basic components of GPTs; for a more detailed presentation of the present convex formulation of GPTs, see [46, 47, 58]. For the mathematical details of the notions and results in convex analysis we refer to [59].

### 1.2.1 States

As was described in the previous section, the set of states $\mathcal{S}$ consists of equivalent preparation procedures of physical systems. The mixing of state preparators imposes a convex structure on $\mathcal{S}$ and this structure can be expressed as a convex set within a real vector space[1] (under some natural assumptions). Thus, we make the following formal definition for $\mathcal{S}$.

**Definition 1.** A state space $\mathcal{S}$ is a compact convex subset of a real finite-dimensional vector space $\mathcal{V}$.

We assume that the underlying vector space $\mathcal{V}$ is finite-dimensional in order to simplify the treatment of the theory. The purpose of this work is to examine various features of quantum theory in a broader operational context, and we focus on achieving this in the finite-dimensional setting without getting too lost in the technicalities that would arise in the infinite dimensional case. It is worth pointing out that the most recent application of GPTs focuses heavily on information theory and processing information where finite-dimensional classical and quantum systems have been traditionally considered. Works on infinite dimensional GPTs include [33, 34, 46, 60, 61].

---

[1]A subset $\mathcal{C}$ of a real vector space $\mathcal{V}$ is *convex* if $tx + (1-t)y \in \mathcal{C}$ for all $x, y \in \mathcal{C}$ and $t \in [0, 1]$.

We also make some technical assumptions about the topological properties of $\mathcal{S}$. Firstly, we assume that the underlying vector space is Hausdorff: if we have a limit of converging states we want the limit to be a unique state. Another technical assumption that we make is the compactness of the state space. In our finite-dimensional setting this is equivalent with closedness and boundedness of $\mathcal{S}$ [62]. By the Krein-Milmann theorem [59], the compactness guarantees that the convex state space $\mathcal{S}$ can be completely characterized by its extreme points[2], or *pure states*, so that every state has a convex decomposition into pure states. However, this convex decomposition is not unique unless $\mathcal{S}$ is a *simplex*, i.e., a convex hull of its finitely many affinely independent extreme points [59]. A state that is not pure is called a *mixed state*.

Given a $d$-dimensional state space $\mathcal{S}$, i.e., $\dim(\text{aff}(\mathcal{S})) = d$, where $\text{aff}(\mathcal{S})$ denotes the affine span of $\mathcal{S}$, one can choose the vector space $\mathcal{V}$ such that $\dim(\mathcal{V}) = d + 1$ and $\mathcal{S}$ is a compact base for a closed generating proper cone $\mathcal{V}_+$[3]. The proper cone $\mathcal{V}_+$ induces a partial order on $\mathcal{V}$: for $x, y \in \mathcal{V}$ we denote $x \geq y$ (or $x \geq_{\mathcal{V}_+} y$ if we want to be more specific) if and only if $x - y \in \mathcal{V}_+$. Thus, $\mathcal{V}_+$ is the set of positive elements in $\mathcal{V}$ with respect to this order. Then $\mathcal{S}$ can be expressed as

$$\mathcal{S} = \{x \in \mathcal{V} \,|\, x \geq 0, \; u(x) = 1\}, \tag{1.1}$$

where $u$ is a strictly positive functional on $\mathcal{V}$. This is depicted in Fig. 1.2.

---

[2] An element $x \in \mathcal{K}$ of a convex set $\mathcal{K}$ is *extreme* (or *extremal*) if any *convex decomposition* of the form $x = ty + (1-t)z$ for some $y, z \in \mathcal{K}$ and $t \in (0, 1)$ implies that $x = y = z$.

[3] A subset $\mathcal{C} \subset \mathcal{V}$ of a vector space $\mathcal{V}$ is a (convex) cone if $\mathcal{C} + \mathcal{C} \subseteq \mathcal{C}$ and $\alpha \mathcal{C} \subseteq \mathcal{C}$ for every $\alpha \in \mathbb{R}_+$. Furthermore, $\mathcal{C}$ is a proper cone if $\mathcal{C} \cap (-\mathcal{C}) = \{0\}$ and generating if $\mathcal{C} - \mathcal{C} = \mathcal{V}$. A subset $\mathcal{B} \subset \mathcal{C}$ is a base of $\mathcal{C}$ if for every $x \in \mathcal{C} \setminus \{0\}$ there exists unique $\beta > 0$ and $b \in \mathcal{B}$ such that $x = \beta b$.

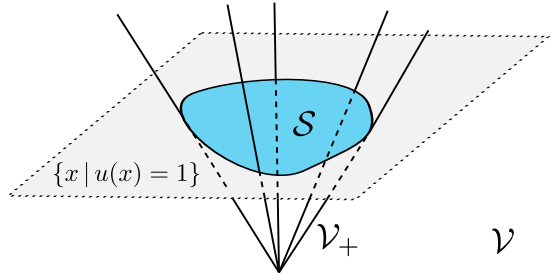

Figure 1.2: State space $\mathcal{S}$ as a base for a cone $\mathcal{V}_+$ in a vector space $\mathcal{V}$.

**Example 1** (Quantum theory)**.** Let $\mathcal{H}$ be a $d$-dimensional Hilbert space. The state space $\mathcal{S}(\mathcal{H})$ of a $d$-dimensional quantum system is defined as [63]

$$\mathcal{S}(\mathcal{H}) = \{\varrho \in \mathcal{L}_s(\mathcal{H}) \,|\, \varrho \geq O, \ \mathrm{tr}\,[\varrho] = 1\},$$

where $\mathcal{L}_s(\mathcal{H})$ denotes the real vector space of self-adjoint operators on $\mathcal{H}$, $O$ is the zero operator on $\mathcal{H}$ and the partial order used is induced by the proper cone of positive semi-definite matrices on $\mathcal{H}$. The elements in $\mathcal{S}(\mathcal{H})$ are often called *density operators* or *density matrices* on $\mathcal{H}$.

### 1.2.2   Effects and observables

The most basic types of measurements are the 'yes-no' type of questions regarding some property of the systems. These are represented by *effects* [64]. Mathematically we describe them by affine functionals giving probabilities on states.

**Definition 2.** The set of effects $\mathcal{E}(\mathcal{S})$ consists of affine functionals $e : \mathcal{S} \to [0, 1]$.

We interpret the positive number $e(s) \in [0, 1]$ as the probability that the event described by the effect $e \in \mathcal{E}(\mathcal{S})$ is detected when the system is in the state $s \in \mathcal{S}$. The affinity of effects follows from the basic statistical interpretation of a physical experiment [50] so that when we measure mixed states, the measurement statistics are constructed from measurements of the states used in the mixture, i.e.,

$$e(\lambda s_1 + (1 - \lambda)s_2) = \lambda e(s_1) + (1 - \lambda)e(s_2)$$

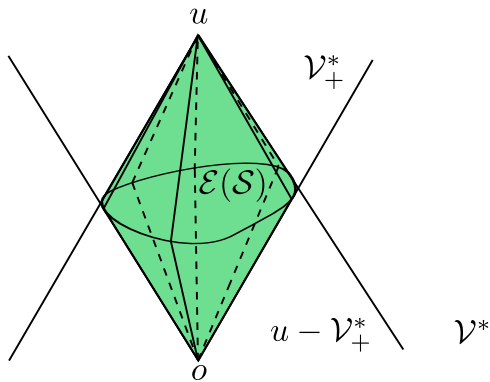

Figure 1.3: Effect space $\mathcal{E}(\mathcal{S})$ as an intersection $\mathcal{E}(\mathcal{S}) = \mathcal{V}_+^* \cap (u - \mathcal{V}_+^*)$.

for all $\lambda \in [0,1]$, $s_1, s_2 \in \mathcal{S}$ and $e \in \mathcal{E}(\mathcal{S})$.

The assumption of treating all mathematically valid affine functionals that give probabilities on states as physical effects is called the *no-restriction hypothesis* [65]. Even though this hypothesis holds in quantum and classical theories, in principle there is no operational justification to require it to hold in every theory. We will take a closer look at theories with operational restrictions beyond the no-restriction hypothesis in Chapter 4, but unless otherwise stated, we take our set of effects to be unrestricted.

Let us consider a $d$-dimensional state space $\mathcal{S}$ as a compact base for a positive cone $\mathcal{V}_+$ in a $(d+1)$-dimensional vector space $\mathcal{V}$ as in Eq. (1.1). Since the state space $\mathcal{S}$ is a base for $\mathcal{V}_+$ and since $\mathcal{V}_+$ is a generating cone for $\mathcal{V}$, we can extend the effects to the whole vector space $\mathcal{V}$. By interpreting the elements of the form $\alpha s \in \mathcal{V}_+$ with $\alpha \in [0,1]$ as *subnormalized states*, where $\alpha$ denotes the probability of success in the preparation of the state with an imperfect preparation device, we can set $e(0) = 0$ as the probability for the empty measurement where an event is trying to be detected on a system that was never prepared. For the affine functionals this means fixing the origin so that, in particular, we can consider the effects as linear functionals, which means that $\mathcal{E}(\mathcal{S}) \subset \mathcal{V}^*$, where $\mathcal{V}^*$ is the dual of $\mathcal{V}$. It follows that we can represent the (extended) set of effects as

$$\mathcal{E}(\mathcal{S}) = \{e \in \mathcal{V}^* \,|\, o \leq e \leq u\} = \mathcal{V}_+^* \cap (u - \mathcal{V}_+^*), \qquad (1.2)$$

where the partial order is induced by the proper dual cone $\mathcal{V}_+^{*\,4}$ of $\mathcal{V}_+$, and where the zero effect $o$ and the unit effect $u$ are defined as $o(s) = 0$ and $u(s) = 1$ for all $s \in \mathcal{S}$ respectively. This is depicted in Fig. 1.3.

An important class of effects are the *indecomposable effects* [37]. We say that a non-zero effect $e \in \mathcal{E}(\mathcal{S})$ is indecomposable if $e$ decomposes into a sum $e = f + g$ of any other two effects $f, g \in \mathcal{E}(\mathcal{S})$ only when $f$ and $g$ are proportional to $e$, i.e., when there are $\alpha, \beta > 0$ such that $e = \alpha f = \beta g$. It is known that every effect can be decomposed into a finite sum of indecomposable effects [37]. Together with extreme effects the indecomposable effects are an essential part of the geometric picture of the effect space as indecomposable effects correspond to the extreme rays[5] of the dual cone $\mathcal{V}_+^*$.

**Example 2** (Quantum theory)**.** In $d$-dimensional quantum theory with a $d$-dimensional Hilbert space $\mathcal{H}$ the set of effects $\mathcal{E}(\mathcal{S}(\mathcal{H}))$ as linear functionals on states can be shown [63] to be isomorphic to the set of positive semidefinite unit-bounded selfadjoint operators on $\mathcal{H}$:

$$\mathcal{E}(\mathcal{S}(\mathcal{H})) \cong \mathcal{E}(\mathcal{H}) := \{E \in \mathcal{L}_s(\mathcal{H}) \,|\, O \leq E \leq \mathbb{1}\},$$

where $O$ and $\mathbb{1}$ denote the zero and identity operators on $\mathcal{H}$. The isomorphism is given my the map $\mathcal{E}(\mathcal{S}(\mathcal{H})) \ni e \mapsto E \in \mathcal{E}(\mathcal{H})$ defined by $e(\varrho) = \mathrm{tr}\,[E\varrho]$ for all $\varrho \in \mathcal{S}(\mathcal{H})$. The previous equation for obtaining probabilities in quantum theory is often called the *Born rule*. The indecomposable effects are exactly those that have rank equal to one [37].

**Example 3** (Qubit states and effects)**.** An example of a quantum system worth considering is the qubit system. In this case the Hilbert space that we consider is $\mathbb{C}^2$. The elements of $\mathcal{L}_s(\mathbb{C}^2)$ are given by Hermitian $2 \times 2$ matrices; such matrices can be parametrised in terms of the so-called *Bloch representation* (see [63] for details): An operator $A \in \mathcal{L}_s(\mathbb{C}^2)$ can be expressed in the form

$$A = \frac{1}{2}(\alpha \mathbb{1}_2 + \vec{a} \cdot \vec{\sigma}), \tag{1.3}$$

---

[4]The *dual cone* $\mathcal{C}^* \subset \mathcal{V}^*$ of a cone $\mathcal{C} \subset \mathcal{V}$ on a vector space $\mathcal{V}$ is defined as $\mathcal{C}^* = \{f \in \mathcal{V}^* \,|\, f(x) \geq 0 \;\forall x \in \mathcal{C}\}$.

[5]A *face* $\mathcal{F}$ of a convex set $\mathcal{K}$ is a convex subset $\mathcal{F} \subset \mathcal{K}$ such that if $tx + (1-t)y \in \mathcal{F}$ for some $x, y \in \mathcal{K}$ and $t \in (0, 1)$, then also $x, y \in \mathcal{F}$. An *extreme ray* of a convex cone $\mathcal{C}$ is a face that is a half-line emanating from the origin.

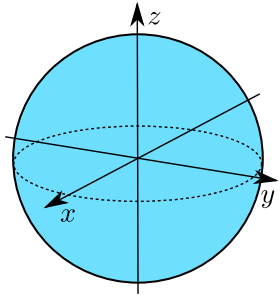

Figure 1.4: The Bloch ball.

where $\mathbb{1}_2$ is the $2 \times 2$ identity matrix, $\vec{a}$ is referred to as the *Bloch vector* of the operator, and $\vec{\sigma} = (\sigma_x, \sigma_y, \sigma_y)$ is composed of the traceless Pauli matrices. The eigenvalues of such an operator are $\frac{1}{2}(\alpha \pm \|\vec{a}\|_2)$, where $\|\cdot\|_2$ refers to the standard Euclidean 2-norm $\|\vec{x}\|_2 = (\sum_i x_i^2)^{1/2}$. From this we see that an operator $A$ is positive semidefinite if $\|\vec{a}\|_2 \leq \alpha$.

In the case of states we require that they are both positive semidefinite and of unit trace. The second of these conditions implies that $\alpha = 1$ and so, by the positivity condition, we require that its Bloch vector satisfies $\|\vec{a}\|_2 \leq 1$. This collection of vectors define the so-called *Bloch ball* (see Fig. 1.4). Normalised Bloch vectors correspond to pure states, as it can be readily calculated that such vectors lead to idempotent operators, i.e., projections, whilst subnormalised Bloch vectors describe mixed states. In the case of $\vec{a} = \vec{0}$ our state is *maximally mixed*.

For an effect $E$, we not only require positivity but also that $E \leq \mathbb{1}$, which is equivalent to the eigenvalues of $E$ being less than or equal to 1. This means that we require $\|\vec{a}\|_2 + \alpha \leq 2$. Combining this with the positivity requirement, we have that for an effect $E \in \mathcal{E}(\mathbb{C}^2)$ given by the Bloch representation (1.3), the parameters $(\vec{a}, \alpha)$ satisfy

$$\|\vec{a}\|_2 \leq \alpha \leq 2 - \|\vec{a}\|_2. \tag{1.4}$$

As in the case with states, if we possess a normalised Bloch vector then $\alpha = 1$ and the effect is a rank-one projection. Furthermore, we call an effect *unbiased* when $\alpha = 1$ so that the probability given by the Born rule with the maximally mixed state is exactly half. If we have an effect $E$ given by parameters $(\vec{a}, \alpha)$, then the complement effect $I - E$ is given by $(-\vec{a}, 2 - \alpha)$.

As we mentioned earlier, the effects correspond to the simplest 'yes-no'-type dichotomic measurements. For more general types of measurements, namely *observables*, we make the following definition.

**Definition 3.** Let $\mathcal{S}$ be a state space. An observable $\mathsf{A}$ with a finite number of outcomes is a mapping $\mathsf{A} : x \mapsto \mathsf{A}_x$ from a finite outcome set $\Omega$ to the set of effects $\mathcal{E}(\mathcal{S})$ such that $\sum_{x \in \Omega} \mathsf{A}_x(s) = 1$ for all $s \in \mathcal{S}$.

The set of observables on $\mathcal{S}$ with an outcome set $\Omega$ is denoted by $\mathcal{O}(\Omega, \mathcal{S})$, and the set of all observables on $\mathcal{S}$ by $\mathcal{O}(\mathcal{S})$. In terms of the unit effect $u$ the normalization criteria can also be expressed as $\sum_{x \in \Omega} \mathsf{A}_x = u$. The definition of an observable captures the idea that each effect $\mathsf{A}_x$ of an observable $\mathsf{A}$ corresponds to some possible measurement outcome $x$ in the measurement of that observable. The normalization criteria then guarantees that some measurement outcome is always registered. Thus, we interpret $\mathsf{A}_x(s)$ as the probability that outcome $x \in \Omega$ was observed when the system in state $s \in \mathcal{S}$ was measured with an observable $\mathsf{A}$.

We note that observables with countably infinite or continuous outcome sets can be considered by defining the observables on some $\sigma$-algebra of $\Omega$ (see, e.g., [34]). However, in this work we only consider observables with a finite number of outcomes for simplicity. As a special class of obervables we consider the set of *trivial observables* $\mathcal{T}(\mathcal{S})$ that consists of observables of the form $\mathsf{T}_x = p_x u$ for all outcomes $x \in \Omega$ for some probability distribution $(p_x)_{x \in \Omega}$ over $\Omega$. As $\mathsf{T}_x(s) = p_x$ for all states $s \in \mathcal{S}$ they do not provide any information about the state when measured. Another important class of observables are the *indecomposable observables* which consists of observables whose every nozero effect is indecomposable. We will consider indecomposable observables more closely in Chapter 3.

**Example 4** (Quantum theory)**.** In a $d$-dimensional quantum theory with a Hilbert space $\mathcal{H}$ an observable $\mathsf{A}$ with a finite outcome set $\Omega$ corresponds to a *positive operator-valued measure (POVM)* [63] $A$ defined as a mapping $A : x \mapsto A(x)$ from $\Omega$ to $\mathcal{E}(\mathcal{H})$ such that $\sum_{x \in \Omega} A(x) = \mathbb{1}$. Thus, we see that the definition of an observable is a direct generalization of that of a POVM. The trivial POVMs then are of the form $T(x) = p_x \mathbb{1}$ for some probability distribution $(p_x)_{x \in \Omega}$. Indecomposable observables are exactly the rank-1 observables, i.e., observables with rank-1 effects.

### 1.2.3   Operations, channels and instruments

In addition to preparations and measurements, the third type of device operational theories are built on are the *channels* that can transform the system from one state to another. Mathematically this means that the channel must preserve the positivity as well as the normalization of the states. Whilst positivity is always required, we can relax the normalization when considering probabilistic transformations, known as *operations*, where we only require that the normalization does not increase:

**Definition 4.** Let $\mathcal{S} \subset \mathcal{V}_+ \subset \mathcal{V}$ and $\mathcal{S}' \subset \mathcal{V}'_+ \subset \mathcal{V}'$ be two state spaces. An operation is a linear mapping $\Phi : \mathcal{V} \to \mathcal{V}'$ such that $\Phi(x) \in \mathcal{V}'_+$ and $u'(\Phi(x)) \leq u(x)$ for all $x \in \mathcal{V}_+$.

For an operation $\Phi$, we interpret $u(\Phi(s))$ as the probability that the transformation succeeds when the system is in a state $s \in \mathcal{S}$. We call $\Phi$ a *channel* if the equality $u'(\Phi(x)) = u(x)$ holds for all $x \in \mathcal{V}_+$ so that the transformation is deterministic instead of probabilistic. The set of channels from $\mathcal{S}$ to $\mathcal{S}'$ is denoted by $\mathrm{Ch}(\mathcal{S}, \mathcal{S}')$, or simply by $\mathrm{Ch}(\mathcal{S})$ if $\mathcal{S}' = \mathcal{S}$.

We can also use state transformations to define measurement devices that allow for sequential measurements. In the measurement of an observable we were only interested in the (classical) measurement outcomes, but if we wish to make further measurements on the already measured system, we must be able to describe how the measurement device has interacted with the system and how that transformed the system. To this end, we consider this type of measurement device, *instruments*, to consist of a collection of probabilistic state transformations (operations) such that the realization of one of these transformations is considered to be the measurement outcome.

**Definition 5.** Let $\mathcal{S}$ and $\mathcal{S}'$ be two state spaces. An instrument $\mathcal{I}$ with a finite outcome set $\Omega$ is a mapping $\mathcal{I} : x \mapsto \mathcal{I}_x$ from $\Omega$ to the set of operations from $\mathcal{S}$ to $\mathcal{S}'$ such that $\sum_{x \in \Omega} \mathcal{I}_x \in \mathrm{Ch}(\mathcal{S}, \mathcal{S}')$.

The set of instruments with outcome set $\Omega$ from $\mathcal{S}$ to $\mathcal{S}'$ is denoted by $\mathrm{Ins}(\Omega, \mathcal{S}, \mathcal{S}')$, or simply by $\mathrm{Ins}(\Omega, \mathcal{S})$ if $\mathcal{S}' = \mathcal{S}$. The observable $\mathsf{A}^\mathcal{I}$ defined by $\mathsf{A}^\mathcal{I}_x(s) = u'(\mathcal{I}_x(s))$ for all $x \in \Omega$ and $s \in \mathcal{S}$ is called the *induced observable of* $\mathcal{I}$. The interpretation is that when measured with an instrument $\mathcal{I}$, a system initially in state $s \in \mathcal{S}$ is transformed into the *conditional output state* $\tilde{s}^\mathcal{I}_x := \mathcal{I}_x(s)/u'(\mathcal{I}_x(s))$ with probability $u'(\mathcal{I}_x(s))$ (in the case that $u'(\mathcal{I}_x(s)) > 0$) while an outcome $x \in \Omega$ is obtained for the observable $\mathsf{A}^\mathcal{I}$. Thus, instruments can also be considered as conditional state preparators.

An important class of instruments are the *measure-and-prepare instruments* which perform a (demolishing) measurement of some observable and prepare a new state based on the measurement outcome. Formally, an instrument $\mathcal{I} \in \text{Ins}(\Omega, \mathcal{S}, \mathcal{S}')$ is a measure-and-prepare instrument if it is of the form $\mathcal{I}_x(s) = \mathsf{A}_x(s)s'_x$ for all $x \in \Omega$ for some observable $\mathsf{A} \in \mathcal{O}(\Omega, \mathcal{S})$ and some set of states $\{s'_x\}_{x \in \Omega} \subset \mathcal{S}'$. We note that in this case the conditional output state $\tilde{s}^\mathcal{I}_x = \mathcal{I}_x(s)/u'(\mathcal{I}_x(s)) = s'_x$ whenever $\mathsf{A}_x(s) \neq 0$ is independent of the input state and only depends on the measurement outcome. Furthermore, as a special case we have the *trash-and-prepare instruments* when the measured observable is trivial, i.e., $\mathsf{A}_x = p_x u$ for all $x \in \Omega$ for some probability distribution $(p_x)_{x \in \Omega}$, so that $\mathcal{I}_x(s) = p_x s'_x$ for all $x \in \Omega$. We will consider measure-and-prepare and trash-and-prepare instruments more closely in Chapter 3.

**Example 5** (Quantum theory). Let $\mathcal{H}$ and $\mathcal{K}$ be two finite-dimensional Hilbert spaces. In quantum theory an operation from $\mathcal{S}(\mathcal{H})$ to $\mathcal{S}(\mathcal{K})$ is described by a *completely positive (CP)* and *trace-nonincreasing* map $\Phi : \mathcal{L}(\mathcal{H}) \to \mathcal{L}(\mathcal{K})$ meaning that $id \otimes \Phi$ is positive on $\mathcal{L}(\mathcal{H}' \otimes \mathcal{H})$ where $id$ is the identity channel on $\mathcal{H}'$ for all finite-dimensional $\mathcal{H}'$, and $\text{tr}\,[\Phi(\varrho)] \leq \text{tr}\,[\varrho]$ for all $\varrho \geq O$ [63]. While being trace-nonincreasing corresponds to not increasing the normalization in Def. 4, the notion of complete positivity is strictly stronger than positivity: not only are completely positive maps positive but in addition they are positive on a larger system when considered as transformations on a subsystem of the larger system where rest of the system is left unchanged. A quantum operation $\Phi$ is a quantum channel if it is *trace-preserving (TP)*, i.e., $\text{tr}\,[\Phi(\varrho)] = \text{tr}\,[\varrho]$ for all $\varrho \in \mathcal{L}(\mathcal{H})$. The definition of quantum instruments mirrors Def. 5 as mappings from an outcome set to the set of quantum operations. In particular, a measure-and-prepare quantum instrument $\mathcal{I} \in \text{Ins}(\Omega, \mathcal{H}, \mathcal{K})$ is of the form $\mathcal{I}_x(\varrho) = \text{tr}\,[A(x)\varrho]\,\sigma_x$ for all $x \in \Omega$ for some POVM $A \in \mathcal{O}(\Omega, \mathcal{H})$ and some set of states $\{\sigma_x\}_{x \in \Omega} \subset \mathcal{S}(\mathcal{K})$.

Quantum channels and operations have a well-known representation in an operator-sum form [63]: a linear map $\Phi : \mathcal{L}(\mathcal{H}) \to \mathcal{L}(\mathcal{K})$ is a quantum operation if and only if there are linear operators $K_i : \mathcal{H} \to \mathcal{K}$ for all $i \in \{1, \ldots, n\}$ for some $n \in \mathbb{N}$ such that $\Phi(\varrho) = \sum_{i=1}^n K_i \varrho K_i^*$ for all $\varrho \in \mathcal{L}(\mathcal{H})$ and $\sum_{i=1}^n K_i^* K_i \leq \mathbb{1}_\mathcal{H}$, where $K_i^*$ is the adjoint of $K_i$. Further-more, $\Phi$ is a channel if $\sum_{i=1}^{n} K_i^* K_i = \mathbb{1}_{\mathcal{H}}$. The operator-sum form is often called the *Kraus representation* of $\Phi$ and the operators $K_i$ are called the *Kraus operators* of $\Phi$. While the Kraus representation is not unique, it can be shown that it is possible to choose $\dim(\mathcal{H})\dim(\mathcal{K})$ or fewer Kraus operators [63]. The minimal number of Kraus operators for a given operation is called the *Kraus rank* of the operation. For the operations $\mathcal{I}_x$ of an instrument $\mathcal{I} \in \text{Ins}(\Omega, \mathcal{H}, \mathcal{K})$ with Kraus representations $\mathcal{I}_x(\varrho) = \sum_{i=1}^{n_x} K_{xi} \varrho K_{xi}^*$ it follows that the induced POVM $A^{\mathcal{I}} \in \mathcal{O}(\Omega, \mathcal{H})$ can be expressed as $A^{\mathcal{I}}(x) = \sum_{i=1}^{n_x} K_{xi}^* K_{xi}$ for all $x \in \Omega$.

### 1.2.4 Composite systems

So far we have considered only the single-system theory and, although the focus of this work is to consider measurements on single systems, we give a brief outlook on how composite systems are formed and treated in the GPT framework. After all, composite systems are an essential part of the theory as in many cases, instead of an indivisible system, we may have multiple subsystems interacting with each other or its environment. We refer to [46] for more detailed analysis on composite systems.

Let $\mathcal{S}^A \subset \mathcal{V}_+^A \subset \mathcal{V}^A$ and $\mathcal{S}^B \subset \mathcal{V}_+^B \subset \mathcal{V}^B$ be the state spaces of systems $A$ and $B$ respectively. Considered as a composite system $A + B$, one must be able to define the joint state space $\mathcal{S}^{AB} \subset \mathcal{V}_+^{AB} \subset \mathcal{V}^{AB}$. Under the *non-signalling principle* and the *local tomography principle*[6] one can show (see [66, 67], also [68, 69] for details) that that the joint state space $\mathcal{S}^{AB}$ can be considered as a subset of the tensor product $\mathcal{V}^A \otimes \mathcal{V}^B$. We note that there are also ways of forming composites without assuming the local tomography principle (see, e.g., [18, 65, 70]) but we will not focus on them here. As an example, in real quantum theory $\mathcal{S}(\mathbb{R}^d)$, where the field of complex numbers is replaced with real numbers, the local tomography principle is not satisfied.

Although we now have that $\mathcal{S}^{AB} \subset \mathcal{V}^{AB} = \mathcal{V}^A \otimes \mathcal{V}^B$, we still have to specify the positive cone $\mathcal{V}_+^{AB}$ which $\mathcal{S}^{AB}$ is the base of. It turns out that this choice is not unique. Namely, the only operationally motivated requirements are that one should be able to prepare states and perform mea-surements on each subsystem independently. This means that the products $s^A \otimes s^B$ of local states $s^A \in \mathcal{S}^A$ and $s^B \in \mathcal{S}^B$ should be valid states in $\mathcal{S}^{AB}$ and the products $e^A \otimes e^B$ of local effects $e^A \in \mathcal{E}(\mathcal{S}^A)$ and $e^B \in \mathcal{E}(\mathcal{S}^B)$ should be valid effects in $\mathcal{E}(\mathcal{S}^{AB})$.

---

[6]The non-signalling principle states that the marginal probability distribution for the outcomes of a measurement on one of the systems is not affected by the measurement performed on the other system. According to the local tomography principle any joint state of the whole system is uniquely determined by local measurements on the subsystems.

Motivated by only considering the mixtures of product states, we first consider the *minimal tensor product* cone, denoted by $(\mathcal{V}^A \otimes_{min} \mathcal{V}^B)_+$, consisting of all positive linear combinations of products of elements that are positive in their respective subsystem. This leads to the *minimal state space* $\mathcal{S}^A \otimes_{min} \mathcal{S}^B$ that must be included in any joint state space of the composite system $A + B$:

$$\mathcal{S}^A \otimes_{min} \mathcal{S}^B := \left\{ \sum_i \lambda_i s_i^A \otimes s_i^B \,\bigg|\, \forall i : s_i^A \in \mathcal{S}^A, s_i^B \in \mathcal{S}^B, \lambda_i \geq 0, \sum_i \lambda_i = 1 \right\},$$

which now forms a compact base for the cone $(\mathcal{V}^A \otimes_{min} \mathcal{V}^B)_+$.

On the other hand, we can maximally require that our states should just consist of normalized elements that are positive on all product effects. In this case we take the *maximal tensor product* cone $(\mathcal{V}^A \otimes_{max} \mathcal{V}^B)_+ := ((\mathcal{V}^A)^* \otimes_{min} (\mathcal{V}^B)^*)_+^*$ which leads to the *maximal state space* $\mathcal{S}^A \otimes_{max} \mathcal{S}^B$ defined as

$$\mathcal{S}^A \otimes_{max} \mathcal{S}^B := \left\{ s \in \mathcal{V}^A \otimes \mathcal{V}^B \,\bigg|\, (u^A \otimes u^B)(s) = 1, \ (e^A \otimes e^B)(s) \geq 0 \right.$$
$$\left. \forall e^A \in \mathcal{E}(\mathcal{S}^A), \ e^B \in \mathcal{E}(\mathcal{S}^B) \right\}.$$

It is clear that $\mathcal{S}^A \otimes_{min} \mathcal{S}^B \subseteq \mathcal{S}^A \otimes_{max} \mathcal{S}^B$, and it has been shown recently in [42] (see also [71]) that equality holds if and only if either one of the state spaces $\mathcal{S}^A$ or $\mathcal{S}^B$ is a simplex. We call the elements of $\mathcal{S}^A \otimes_{min} \mathcal{S}^B$ *separable* and the elements of $\mathcal{S}^A \otimes_{max} \mathcal{S}^B \setminus \mathcal{S}^A \otimes_{min} \mathcal{S}^B$ *entangled*. To conclude the discussion on composite systems, we make the following formal definition:

**Definition 6.** Let $\mathcal{S}^A \subset \mathcal{V}_+^A \subset \mathcal{V}^A$ and $\mathcal{S}^B \subset \mathcal{V}_+^B \subset \mathcal{V}^B$ be state spaces of systems $A$ and $B$ respectively. Any state space $\mathcal{S}^{AB}$ of the form

$$\mathcal{S}^{AB} = \{x \in \mathcal{V}^A \otimes \mathcal{V}^B \,|\, x \geq_{\mathcal{V}_+^{AB}} 0, \ (u^A \otimes u^B)(x) = 1\}$$

for some proper positive cone $\mathcal{V}_+^{AB} \subset \mathcal{V}^A \otimes \mathcal{V}^B$ satisfying $(\mathcal{V}^A \otimes_{min} \mathcal{V}^B)_+ \subseteq \mathcal{V}_+^{AB} \subseteq (\mathcal{V}^A \otimes_{max} \mathcal{V}^B)_+$ is a joint state space of the system $A + B$.

**Example 6** (Quantum theory)**.** Let $\mathcal{S}(\mathcal{H}_A)$ and $\mathcal{S}(\mathcal{H}_B)$ be the set of density operators of two quantum systems $A$ and $B$ respectively. The joint state space of the composite system $A + B$ is built around the Hilbert space $\mathcal{H}_A \otimes \mathcal{H}_B$, where the inner product is defined on the product elements as $\langle \psi_A \otimes \psi_B | \varphi_A \otimes \varphi_B \rangle = \langle \psi_A | \varphi_A \rangle \langle \psi_B | \varphi_B \rangle$ for all $\psi_A, \varphi_A \in \mathcal{H}_A$ and $\psi_B, \varphi_B \in \mathcal{H}_B$ and then extended to other elements via linearity [63]. Thus, consequently the joint state space of the system $A+B$ is taken to be $\mathcal{S}(\mathcal{H}_A \otimes \mathcal{H}_B)$. It is a basic result of entanglement theory that in this case we have that $\mathcal{S}(\mathcal{H}_A) \otimes_{min} \mathcal{S}(\mathcal{H}_B) \subsetneq \mathcal{S}(\mathcal{H}_A \otimes \mathcal{H}_B) \subsetneq \mathcal{S}(\mathcal{H}_A) \otimes_{max} \mathcal{S}(\mathcal{H}_B)$ as all separable states and some entangled states are included in $\mathcal{S}(\mathcal{H}_A \otimes \mathcal{H}_B)$ but nevertheless not all elements in $\mathcal{S}(\mathcal{H}_A) \otimes_{max} \mathcal{S}(\mathcal{H}_B)$, such as entanglement witnesses, are included in $\mathcal{S}(\mathcal{H}_A \otimes \mathcal{H}_B)$ [63].

**Remark 1.** For quantum operations, in addition to being positive, we required them to be *completely positive*. The reason why complete positivity is not required or even often considered in the GPT framework is twofold: first, as we saw above, the composite state space is not uniquely determined so that positivity on extended systems is not uniquely determined, and second, while the requirement of a map being positive for larger extended systems of all sizes makes sense in quantum theory, where the size of the extension is dictated by the dimension of the Hilbert space, in GPTs there is no clear way to define a unique extension of a state space, let alone an extension of a given size, or even uniquely define the size of the system. Namely, while so far we have only considered the affine dimension of a state space, there are other more physical indicators of the size of the system, and some of these other notions of 'dimensions' are explored in Chapter 2. Thus, in order to define complete positivity in GPTs one has to address which ancillary state space and which tensor product, i.e., which composite state space, one is using. One sees that in this case complete positivity is no longer just a property of the channel alone. This type of consideration has been recently studied in [47, 58].

## 1.3 Non-quantum theories

In the previous Section we defined the key concepts and elements of the GPT framework and demonstrated these in the context of quantum theory. Next we will introduce and consider some of the other important theories that we will use later.

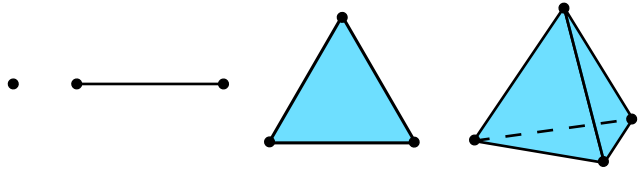

Figure 1.5: The first few simplices in the lowest dimensions.

### 1.3.1 Classical theory

The only other actually physical operational theory besides quantum theory is classical theory. Let us consider the traditional phase space representation of a classical system [50] where each dimension of the phase space corresponds to a degree of freedom of the system so that the points of the phase space determine the state of the system uniquely. By considering this representation as a statistical (operational) theory, we see that the notion of states must be extended to include all probability distributions on the phase space. This is exactly what the notion of a *simplex* captures: if $\Omega$ is a (finite) phase space with $d+1$ points, then the set of probability distributions on $\Omega$ is a $d$-simplex, i.e., it is the convex hull of its $d+1$ affinely independent extreme points. Thus, a theory is said to be classical if and only if the state space is a simplex.

For example, we have that the 0-simplex is a point, a 1-simplex is a line segment (a *bit*), a 2-simplex is a triangle (a *trit*) and a 3-simplex is a tetrahedron. The characteristic feature of simplices is that every point of the simplex has a unique convex decomposition into the extreme points, a feature not shared by any other compact convex set [59]. Some of the first few simplices are depicted in Fig. 1.5.

In the usual representation of the state space of a GPT we consider a $d$-simplex embedded as a compact base for a positive cone in a $d+1$-dimensional real vector space. We denote the state space that is a $d-1$-simplex by $\mathcal{S}_d^{cl}$ and call it the state space of a $d$-dimensional classical system. The set of effects and observables are defined as in the previous section. The set of channels $\text{Ch}(\mathcal{S}_n^{cl}, \mathcal{S}_m^{cl})$ can be shown to coincide with the set of $n \times m$ row-stochastic matrices $\mathcal{M}_{n,m}^{row}$ (see for example [47]). As was pointed out earlier, a composite system of two classical systems (or even a composite of a classical system and any other system) results in a state space with only separable states.

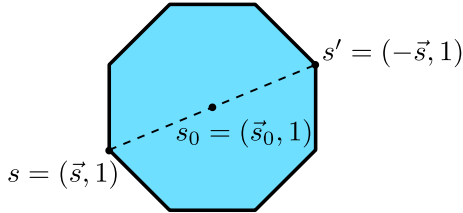

Figure 1.6: A regular octagon is point-symmetric.

### 1.3.2 Point-symmetric theories

Let us generalize the Bloch representation of qubit states and effects considered in Example 3 to a larger class of theories. We start with the following definition.

**Definition 7.** A state space $\mathcal{S}$ is *point-symmetric* if there exists a state $s_0 \in \mathcal{S}$ such that for all $s \in \mathcal{S}$, we have that $s' := 2s_0 - s \in \mathcal{S}$. This means that for each state $s \in \mathcal{S}$ there exists a state $s' \in \mathcal{S}$ such that $s_0$ is an equal mixture of them, i.e., $s_0 = \frac{1}{2}(s + s')$.

Let $\mathcal{S}$ be a point-symmetric state space with $\dim(\text{aff}(\mathcal{S})) = d$. We can embed $\mathcal{S}$ in $\mathbb{R}^d$ such that we can fix the inversion point $\vec{s}_0 = \vec{0} \in \mathbb{R}^d$. This means that for each $\vec{s} \in \mathcal{S}$, we also have $-\vec{s} \in \mathcal{S}$. From the theory of point-symmetric convex sets (Theorem 15.2 in [59]) we then have that the Minkowski functional $\gamma(\vec{s} \,|\, \mathcal{S}) = \inf\{\lambda \geq 0 \,|\, \vec{s} \in \lambda \mathcal{S}\}$, defines a norm $\|\cdot\|_\mathcal{S} := \gamma(\cdot \,|\, \mathcal{S})$ in $\mathbb{R}^d$ such that $\mathcal{S} = \{\vec{s} \in \mathbb{R}^d \,|\, \|\vec{s}\|_\mathcal{S} \leq 1\}$.

We can further embed $\mathcal{S}$ in $\mathbb{R}^{d+1}$ by identifying $\mathcal{S}$ with

$$\tilde{\mathcal{S}} = \{(\vec{s}, 1) \in \mathbb{R}^{d+1} \,|\, \|\vec{s}\|_\mathcal{S} \leq 1\}.$$

Point-symmetric (or often also called *centrally symmetric*) GPTs form an important class of toy theories whose obvious advantage is the symmetric and simple structure. Point-symmetric theories have been considered in [35, 43, 46, 72, 73]. We will denote the state spaces embedded in $\mathbb{R}^d$ and $\mathbb{R}^{d+1}$ both by $\mathcal{S}$, but we make the distinction by denoting the elements of $\mathcal{S}$ in $\mathbb{R}^d$ by the vector notation $\vec{s} \in \mathcal{S} \subset \mathbb{R}^d$ but omit it when considering elements in $\mathbb{R}^{d+1}$, so that $s \in \mathcal{S} \subset \mathbb{R}^{d+1}$. An example of a point-symmetric state space is depicted in Fig. 1.6.

For the effects we can show the following 'Bloch' representation (see also [35]):

**\*Proposition 1.** *For a point-symmetric state space $\mathcal{S}$ we have that*

$$\mathcal{E}(\mathcal{S}) = \left\{ \frac{1}{2} \left( \vec{a}, \alpha \right) \in \mathbb{R}^{d+1} \,\bigg|\, \|\vec{a}\|_{\mathcal{E}} \leq \alpha \leq 2 - \|\vec{a}\|_{\mathcal{E}} \right\}, \tag{1.5}$$

*where $\|\cdot\|_{\mathcal{E}} : \mathbb{R}^d \to \mathbb{R}$ is defined as $\|\vec{a}\|_{\mathcal{E}} = \sup_{\vec{s} \in \mathcal{S}} \vec{a} \cdot \vec{s}$ is a norm in $\mathbb{R}^d$.*

The result follows directly from the fact that the polar of the norm $\|\cdot\|_{\mathcal{S}}$ is also a norm in $\mathbb{R}^d$ due to the point-symmetry of $\mathcal{S}$ (Theorem 15.2 in [59]). Furthermore, because $\|\cdot\|_{\mathcal{S}}$ is finite everywhere, the polar of $\|\cdot\|_{\mathcal{S}}$, denoted by $\|\cdot\|_{\mathcal{E}}$, can be expressed as above, and the derived expression for the effect space $\mathcal{E}(\mathcal{S})$ is just an application of this to the condition $o \leq \frac{1}{2}(\vec{a}, \alpha) \leq u$. Similarly to the qubit case, we call the state $s_0 = (\vec{0}, 1)$ the maximally mixed state, an effect $\frac{1}{2}(\vec{a}, 1)$ an unbiased effect, and for each effect $\frac{1}{2}(\vec{a}, \alpha) \in \mathcal{E}(\mathcal{S})$ we have the complement effect $\frac{1}{2}(-\vec{a}, 2 - \alpha) \in \mathcal{E}(\mathcal{S})$.

### 1.3.3 Polygon theories

A regular $n$-sided polygon (or $n$-gon) state space $\mathcal{S}_n$ embedded in $\mathbb{R}^3$ is the convex hull of its $n$ extreme points

$$s_k = \begin{pmatrix} r_n \cos\left(\frac{2k\pi}{n}\right) \\ r_n \sin\left(\frac{2k\pi}{n}\right) \\ 1 \end{pmatrix}, \quad k = 1, \ldots, n,$$

where we have defined $r_n = \sqrt{\sec\left(\frac{\pi}{n}\right)}$.

Clearly, we now have the zero effect $o = (0, 0, 0)^T$ and the unit effect $u = (0, 0, 1)^T$. Let us denote $s_0 = (0, 0, 1)^T$. Depending on the parity of $n$, the state space may or may not be point-symmetric around $s_0$. As a result of this the effect space $\mathcal{E}(\mathcal{S}_n)$ has different structures for odd and even $n$. For even $n$ we have the non-trivial extreme points

$$e_k = \frac{1}{2} \begin{pmatrix} r_n \cos\left(\frac{(2k-1)\pi}{n}\right) \\ r_n \sin\left(\frac{(2k-1)\pi}{n}\right) \\ 1 \end{pmatrix}, \quad k = 1, \ldots, n, \tag{1.6}$$

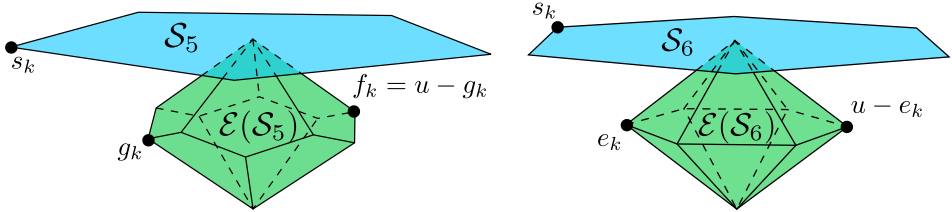

Figure 1.7: The odd (pentagon) and even (hexagon) polygon state spaces and their effects spaces.

so that $\mathcal{E}(\mathcal{S}_n) = \text{conv}(\{o, u, e_1, \ldots, e_n\})$. All of the non-trivial extreme effects lie on a single (hyper)plane determined by the unbiased effects.

In the case of odd $n$, the effect space has $2n$ non-trivial extreme effects

$$g_k = \frac{1}{1+r_n^2} \begin{pmatrix} r_n \cos\left(\frac{2k\pi}{n}\right) \\ r_n \sin\left(\frac{2k\pi}{n}\right) \\ 1 \end{pmatrix}, \qquad f_k = u - g_k \tag{1.7}$$

for $k = 1, \ldots, n$. Now $\mathcal{E}(\mathcal{S}_n) = \text{conv}(\{o, u, g_1, \ldots, g_n, f_1, \ldots, f_n\})$ and the non-trivial extreme effects are scattered on two different planes determined by all those points $g$ and $f$ such that $g(s_0) = \frac{1}{1+r_n^2} =: \sigma_n$ and $f(s_0) = \frac{r_n^2}{1+r_n^2} = 1 - \sigma_n$. We note that the case $n = 3$ corresponds to the classical state space $\mathcal{S}_3^{cl}$ since $\mathcal{S}_3$ is a triangle, i.e., a 2-simplex.

The state and effect spaces for even and odd polygons are depicted in Fig. 1.7. We will use the polygon theories as toy theories to study and demonstrate various properties of measurements in later chapters. We note that one of the useful features of the polygon theories is that in the limit when $n \to \infty$, the state space $\mathcal{S}_n$ becomes a disc, which can be considered as the Bloch disc $\mathcal{S}(\mathbb{R}^2)$ that represents the state space of the real qubit, namely the *rebit*. The rebit shares many of the same properties of the standard (complex) qubit so in some cases one is able to make direct comparisons between polygons, rebit and qubit. The polygon theories were first introduced in [24] and have been used to study various concepts, such as non-locality [24–26], incompatibility [29], self-testing [26], etc., ever since.

## Chapter 2

# Communication tasks

Any kind of communication requires the transmission of a physical system and thus every communication scheme requires a physical theory that it is implemented in. For an operational theory communication between two parties Alice and Bob can be conveniently described as an experiment: Alice wants to communicate some message to Bob and for that Alice has access to some physical systems of a given theory along with a communication line (a channel) that can be used to transmit those systems to Bob. What Alice does is she encodes her message into the systems and sends them to Bob. The task for Bob is then to perform a measurement on the systems that Alice sent him and, based on the measurement outcomes, interpret Alice's intended message. In this Chapter, based on Publication **V**, we consider these kinds of communication schemes in general probabilistic theories and see how various communication tasks can be used to characterize different theories. By considering these types of tasks and comparisons one sees what kinds of features of physical theories it is possible to study in the GPT framework. This also helps us establish quantum theory intuitively as an operational theory among other theories.

## 2.1 Communication matrices

In the communication scheme described above we have Alice preparing systems in certain states and then Bob measuring these states. Such schemes are commonly called as *prepare-and-measure scenarios* and they have been used in several active research areas [74–81]. Let us consider a particular prepare-and-measure scenario in an operational theory determined by the state space $\mathcal{S}$. Alice has access to states $s_1, \ldots, s_n \in \mathcal{S}$ of which she sends a state $s_i$ to Bob. Bob measures the state with some $m$-outcome observable $\mathsf{M}$ and obtains an outcome $j \in [m] := \{1, \ldots, m\}$. After repeating the scenario enough times, they have obtained the full measurement outcome statistics for all states described by the probabilities $\mathsf{M}_j(s_i)$ for all $i \in [n] := \{1, \ldots, n\}$ and $j \in [m] := \{1, \ldots, m\}$. This type of prepare-and-measure scenario is depicted in Fig. 2.1.

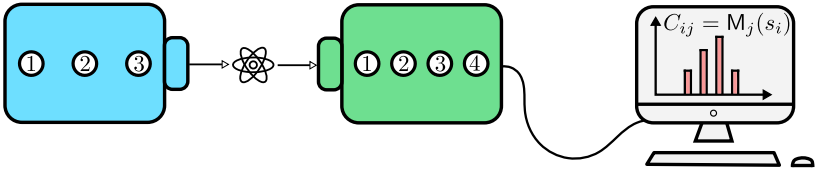

Figure 2.1: A prepare-and-measure scenario and the resulting communication matrix $C$ for some three states and some 4-outcome observable.

Let us denote the $n \times m$ row-stochastic matrix, i.e., a matrix with nonnegative elements such that each row sums to one, that is formed of the probabilities by $C$ so that $C_{ij} = \mathsf{M}_j(s_i)$. We call $C$ a *communication matrix* (also called a *channel matrix* in [82, 83]) and denote the set of all $n \times m$ communication matrices that can be implemented with $n$ states in $\mathcal{S}$ and observables in $\mathcal{O}([m], \mathcal{S})$ by $\mathcal{C}_{n,m}(\mathcal{S})$, and all finite communication matrices in the theory described by $\mathcal{S}$ by $\mathcal{C}(\mathcal{S})$, i.e., $\mathcal{C}(\mathcal{S}) = \cup_{n,m \in \mathbb{N}} \mathcal{C}_{n,m}(\mathcal{S})$.

The given prepare-and-measure scenario can be used to describe various communication tasks and the corresponding communication matrices characterize the specifics of that task. By seeing which tasks can be achieved in a given theory we can start characterizing the theory. Furthermore, by considering different theories and the tasks applicable in each of them, we are able to compare theories to each other. As a concrete example, the task of perfectly distinguishing $n$ states $\{s_1, \ldots, s_n\}$ corresponds to the communication matrix $\mathbb{1}_n$ so that there exists an observable $\mathsf{M} \in \mathcal{O}([n], \mathcal{S})$ such that $\mathsf{M}_j(s_i) = \delta_{ij}$ for all $i, j \in [n]$. For more examples see Publication **V**.

**Remark 2** (*Behaviour*)**.** We note that in this prepare-and-measure scenario we only give Bob one choice of observable to measure so that the entire scheme is described by the corresponding communication matrix. The resulting collection of communication matrices in the case where Bob has multiple observables to choose from is often called a *behaviour*, and works on them can be found in [74–79, 81]. However, we will not focus on behaviours in this work.

**Remark 3** (*Convexity*). Often in this type of prepare-and-measure scenario it is assumed that Alice and Bob have access to a common source of randomness which they can use in the implementation of their task. In the case where there is this type of *shared randomness* between Alice and Bob one can mix communication matrices (of same size) so that the set of communication matrices (of a given size) is convex. With shared randomness it can be shown that the set of communication matrices in $d$-dimensional quantum theory is the same as in $d$-dimensional classical theory [82]. As we will see later, this is a drastic difference to the case where shared randomness is not part of the prepare-and-measure scenario, and because of this difference one can view shared randomness as an additional resource. For an explicit example showing the non-convexity of the set of communication matrices (without the shared randomness), see Publication **V**.

## 2.2 Ultraweak matrix majorization

Next we will consider when a communication task is more difficult than some other task by introducing a preorder on the set of communication matrices. We will see that a given task can be used to implement all other tasks that correspond to communication matrices that are below the communication matrix of the given task with respect to this preorder.

In order to proceed with the formal definition of the preorder let us denote the set of $n \times m$ row-stochastic matrices by $\mathcal{M}^{row}_{n,m}$ and the set of all finite row-stochastic matrices by $\mathcal{M}^{row}$ so that $\mathcal{M}^{row} = \cup_{n,m \in \mathbb{N}} \mathcal{M}^{row}_{n,m}$.

**Definition 8.** A matrix $D \in \mathcal{M}^{row}_{n,m}$ is *ultraweakly majorized* by matrix $C \in \mathcal{M}^{row}_{j,k}$ if there exist two other matrices $L \in \mathcal{M}^{row}_{n,j}$ and $R \in \mathcal{M}^{row}_{k,m}$ such that $LCR = D$. In this case we denote $D \preceq C$, and if also $C \preceq D$, then we say that $C$ and $D$ are *ultraweakly equivalent* and denote it $C \simeq D$.

The ultraweak majorization relation was first introduced in [84] as *I/O degradation* but in the current physical context it was only considered (and renamed) recently in [85]. The reason behind the term ultraweak matrix majorization comes from the earlier investigation into *matrix majorization* [86] and *weak matrix majorization* [87]: the previous definition reduces to that of the matrix majorization in the case when $n = j$ and $L = \mathbb{1}_n$, and weak matrix majorization when $k = m$ and $R = \mathbb{1}_k$. Hence, if a matrix $D$ is majorized or weakly majorized by another matrix $C$, then $D$ is also ultraweakly majorized by $C$.

### 2.2.1 Operational interpretation

The operational interpretation of ultraweak matrix majorization is the following [85]. Let us consider a communication task that produces a communication matrix $C \in \mathcal{C}_{k,l}(\mathcal{S})$ with states $\{s_1, \ldots, s_k\}$ and an $l$-outcome observable $\mathsf{M}$ such that $C_{ab} = \mathsf{M}_b(s_a)$. Given row-stochastic matrices $L \in \mathcal{M}_{n,k}^{row}$ and $R \in \mathcal{M}_{l,m}^{row}$ we see that

$$(LCR)_{ab} = \sum_{i,j} L_{ai} \mathsf{M}_j(s_i) R_{jb} = \left( \sum_j R_{jb} \mathsf{M}_j \right) \left( \sum_i L_{ai} s_i \right) = \mathsf{M}'_b(s'_a),$$

where $\{s'_a := \sum_i L_{ai} s_i\}_a$ are convex combinations of the states $\{s_1, \ldots, s_k\}$ due to the stochasticity of $L$, and $\mathsf{M}'$, defined by $\mathsf{M}'_b := \sum_j R_{jb} \mathsf{M}_j$ for all $b \in [m]$, is a new $m$-outcome observable given as a *post-processing* of $\mathsf{M}$ by the matrix $R$ whose element $R_{jb}$ is interpreted as a transition probability for an outcome $j$ to be transformed into outcome $b$. While we take a closer look at the post-processing of measurements in Chapter 3, in this context we only need it as a classical manipulation of measurement outcomes to produce new observables out of known ones.

Hence, since $C$ is a communication matrix and can be realized by a prepare-and-measure scenario, $LCR$ can also be realized via a prepare-and-measure scenario by using the mixtures of the states $\{s_1, \ldots, s_j\}$ given by the matrix $L$ and the post-processed measurement $\mathsf{M}'$, where the post-processing is given by the matrix $R$. This makes also $LCR$ a communication matrix. If there exists a communication task with a communication matrix $D$ such that $D = LCR$, by using the previous interpretation of $L$ and $R$ we see that the communication task and matrix $D$ can be reproduced by the prepare-and-measure scenario of $C$, and thus by implementing $C$ we also get $D$ as well. This means that $D$ itself must be at least as easy to implement as $C$. A demonstration of the ultraweak majorization is depicted in Fig. 2.2.

Based on the operational interpretation of ultraweak matrix majorization of communication matrices it is straightforward to see why two row-stochastic matrices related to two prepare-and-measure scenarios are ultraweakly equivalent in the cases where one of them is obtained from the other one by relabeling bijectively states and measurement outcomes, by adding the zero effect to the measurement device, by adding a state that is a mixture of the existing states or by splitting an outcome of the measurement device into several outcomes possibly with different probabilistic weights. We note that since we are considering ultraweakly equivalent matrices the reverse operations of the ones listed above also lead to the equivalence. We note that with the operations listed above every row-stochastic matrix can be reduced into a form where there are no zero columns, no row is a mixture of any other rows and where no two columns are proportional.

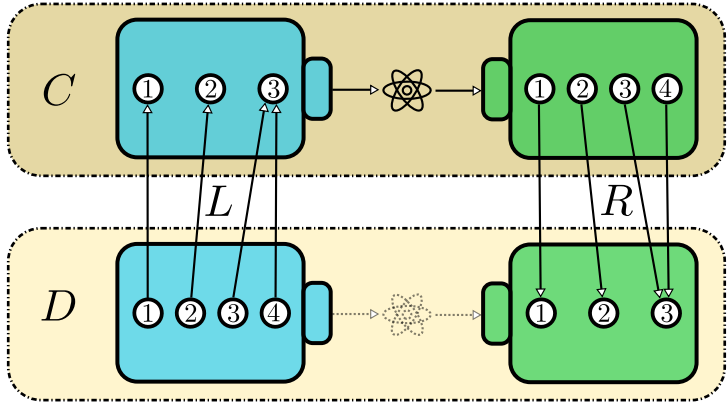

Figure 2.2: The prepare-and-measure scenario that implements $C$ is used to implement $D$ via ultraweak matrix majorization so that $D = LCR$. The matrix $L$ ($R$) holds the information of how the states (measurement) that are used to implement $C$ are manipulated in order to implement $D$.

### 2.2.2 The order structure

The relation of ultraweak majorization is reflexive and transitive and thus it is a preorder on the set of row-stochastic matrices $\mathcal{M}^{row}$. By considering the equivalence classes consisting of ultraweakly equivalent matrices the preorder can be extended to a partial order between the equivalence classes. A natural task is to consider the minimal and maximal elements with respect to this partial order. It was shown in [85] that for a row-stochastic matrix $M \in \mathcal{M}^{row}_{n,m}$ we have that

$$V_{\min(n,m)} \simeq V_{\max(n,m)} \preceq M \preceq \mathbb{1}_{\min(n,m)} \preceq \mathbb{1}_{\max(n,m)}, \qquad (2.1)$$

where $V_k$ denotes the $k \times k$ matrix with all entries $1/k$. The last inequality is strict if $n \neq m$. Thus, in general it means that the set $\mathcal{M}^{row}$ has no maximal elements but it does have a unique minimal element, i.e., the least element, namely $V_1 = 1$.

More interestingly, by considering a theory with a state space $\mathcal{S}$, we can examine the maximal and minimal elements on the set of communication matrices $\mathcal{C}(\mathcal{S}) \subset \mathcal{M}^{row}$. While the least element stays the same, maximal elements can also now exist. In particular, the communication matrices in a $d$-dimensional classical theory are all majorized by one maximal element.

**Proposition 2.** *For a row-stochastic matrix $C$ we have that $C \preceq \mathbb{1}_d$ if and only if $C \in \mathcal{C}(\mathcal{S}_d^{cl})$.*

The proof relies on the facts that there are $d$ pure states in $\mathcal{S}_d^{cl}$ and that every other state has a unique convex decomposition into those pure states, as well as that there exists a $d$-outcome observable $\mathsf{M}$ that distinguishes those pure states and that every other observable is a post-processing of $\mathsf{M}$ (see Chapters 3 and 4). We can now express $\mathcal{C}(\mathcal{S}_d^{cl})$ as

$$\mathcal{C}(\mathcal{S}_d^{cl}) = \{C \in \mathcal{M}^{row} \,|\, C \preceq \mathbb{1}_d\}.$$

What follows is that for other theories, such as quantum theory, there must be other maximal elements as well.

## 2.3　Ultraweak monotones and physical dimensions

To study the set of communication matrices and their order structure in different theories we consider the monotones related to the ultraweak order.

**Definition 9.** A function $f : \mathcal{M}^{row} \to \mathbb{R}$ is an *ultraweak monotone* if $D \preceq C$ implies $f(D) \leq f(C)$ for any $C, D \in \mathcal{M}^{row}$.

The ultraweak monotones can be used to detect when there is not an ultraweak majorization relation between two matrices and, in particular, when the matrices are not equivalent. Namely, if we have an ultraweak monotone $f$ and two row-stochastic matrices $C$ and $D$ and we have that $f(D) > f(C)$ (or $f(D) \neq f(C)$), then we know that $D \npreceq C$ (or $D \not\simeq C$).

### 2.3.1　Operationally motivated monotones

Next we introduce some ultraweak monotones and see how they link to some important mathematical and physical properites of different theories.

**Distinguishability monotone**

As was mentioned earlier, the task of errorless distinguishing of $n$ states corresponds to the communication matrix $\mathbb{1}_n$. By the operational interpretation of ultraweak matrix majorization, all communication matrices $C$ with $\mathbb{1}_n \preceq C$ can also be used for this task as well. Thus, if we set

$$\iota(C) := \max\{n \in \mathbb{N} \,|\, \mathbb{1}_n \preceq C\}$$

we see that it gives the maximum number of messages that can be sent and perfectly distinguished by implementing the matrix $C$. We call $\iota$ the *distinguishability monotone* and it is easy to see that it is a monotone.

The *operational dimension* $d_{op}(\mathcal{S})$ of a theory with state space $\mathcal{S}$ is determined as the maximum number of (pure) states that can be distinghuished (see, e.g., [88]). Thus, in the language of communication matrices, it is defined as the maximum $d \in \mathbb{N}$ such that $\mathbb{1}_d \in \mathcal{C}(\mathcal{S})$. Equivalently, we can express the operational dimension as the maximum value of the distinguishability monotone in the whole theory, i.e.,

$$d_{op}(\mathcal{S}) = \sup\{\iota(C) \,|\, C \in \mathcal{C}(\mathcal{S})\}.$$

For calculating the distinguishability monotone, we can prove the following result:

**Proposition 3.** *For a row-stochastic matrix $C$ we have that $\mathbb{1}_k \preceq C$ for some $k \in \mathbb{N}$ if and only if $C$ has $k$ orthogonal rows.*

Thus, $\iota(C)$ is then just the maximum number of orthogonal rows of $C$.

**Max monotone**

Instead of considering perfect distingusihability of states, we can allow for some error in the distinguishing process and try to minimize that error. This is known as the *minimum-error discrimination* of states, and we will see how our next monotone is related to it. We define the *max monotone* $\lambda_{\max}$ as $\lambda_{\max}(C) := \sum_j \max_i C_{ij}$ for any nonnegative matrix $C$. Once again it is straightforward to check that $\lambda_{\max}$ indeed is an ultraweak monotone.

We will show how $\lambda_{\max}$ is related to the task of minimum-error discrimination and decoding information. Suppose Alice is encoding $n$ equally likely messages into $n$ states $\{s_1, \ldots, s_n\} \subset \mathcal{S}$ and Bob decodes the message by using an $n$-outcome observable $\mathsf{M} \in \mathcal{O}([n], \mathcal{S})$ so that the standard minimum-error discrimination success probability reads as $\frac{1}{n} \sum_i \mathsf{M}_i(s_i)$. By fixing the observable $\mathsf{M}$ and optimizing over the states, we can consider the decoding (or discrimination) power of $\mathsf{M}$ by defining $\lambda_{\max}(\mathsf{M}) := \sum_j \max_{s \in \mathcal{S}} \mathsf{M}_j(s)$ so that the average success probability for decoding using $\mathsf{M}$ is given by $\lambda_{\max}(\mathsf{M})/n$ for the optimal set of states. Now if we have a communication matrix $C \in \mathcal{C}(\mathcal{S})$ so that $C_{ij} = \mathsf{M}_j(s_i)$ for some states $\{s_i\}_i \subset \mathcal{S}$ and an observable $\mathsf{M} \in \mathcal{O}(\Omega, \mathcal{S})$, we see that $\lambda_{\max}(C) = \sum_{j \in \Omega} \max_i \mathsf{M}_j(s_i)$. Furthermore, if $\{s_i\}_i$ are the maximizing states for the effects of $\mathsf{M}$, then $\lambda_{\max}(C) = \lambda_{\max}(\mathsf{M})$.

Let us generalize the definition of $\lambda_{\max}$ for the whole operational theory. If we maximize $\lambda_{\max}$ over the communication matrices (or equivalently over the observables) of the whole theory, i.e.,

$$\lambda_{\max}(\mathcal{S}) := \sup\{\lambda_{\max}(C) \,|\, C \in \mathcal{C}(\mathcal{S})\} = \sup\{\lambda_{\max}(\mathsf{M}) \,|\, \mathsf{M} \in \mathcal{O}(\mathcal{S})\},$$

we can generalize what is known as the *basic decoding theorem* in quantum theory [89] to any operational theory with state space $\mathcal{S}$: in the decoding task of $n$ states, Bob's probability of error $P_E^n$ is bounded by

$$P_E^n \geq 1 - \frac{\lambda_{\max,n}(\mathcal{S})}{n},$$

where $\lambda_{\max,n}(\mathcal{S}) := \sup\{\lambda_{\max}(\mathsf{M}) \,|\, \mathsf{M} \in \mathcal{O}(\Omega, \mathcal{S}) : |\Omega| \leq n\}$. We note that in [73] $\lambda_{\max}(\mathcal{S})$ was called the *information storability of $\mathcal{S}$* and they were able to relate it to the point-asymmetry of the state space so that in particular if $\mathcal{S}$ is point-symmetric, then we have $\lambda_{\max}(\mathcal{S}) = 2$.

**Min monotone**

On the other end of the previously defined max monotone we have the *min monotone* $\lambda_{\min}$ defined by $\lambda_{\min}(C) := -\sum_j \min_i C_{ij}$ for all $C \in \mathcal{M}^{row}$. By considering the possible values of $\lambda_{\min}$ we see that the minimum value it attains is $-1$ and it is attained on all matrices that have only identical columns, i.e., if they are considered as communication matrices, the measured observables are trivial on the measured states as they provide no distinguishing information between them. On the other hand, the maximum value is clearly zero and can be attained for example on an identity matrix of any size.

Although the maximal value of $\lambda_{\min}$ is not found to be so useful as it is always attained in every theory, we see that $\lambda_{\min}$ can still be linked to an important operational concept, namely the *noise content of an observable*. For a subset $\mathcal{N} \subseteq \mathcal{O}(\mathcal{S})$ and an observable $\mathsf{A} \in \mathcal{O}(\Omega, \mathcal{S})$ let us define

$$w(\mathsf{A}; \mathcal{N}) := \sup\{t \in [0,1] \,|\, \exists \mathsf{N} \in \mathcal{N}, \mathsf{B} \in \mathcal{O}(\mathcal{S}) : \ \mathsf{A} = t\mathsf{N} + (1-t)\mathsf{B}\} \quad (2.2)$$

which we call the *noise content of the observable* $\mathsf{A}$ *with respect to the noise set* $\mathcal{N}$. The noise content $w(\mathsf{A}; \mathcal{N})$ thus quantifies how much of $\mathsf{A}$ is in $\mathcal{N}$, and is taken to describe noise in the measurements. Contrary to external noise, i.e., noise that is added to the observables, the noise content gives us the amount of intrinsic noise that is already contained in the observable.

The typical choice for the set of noisy observables is $\mathcal{N} = \mathcal{T} := \mathcal{T}(\mathcal{S})$, the set of trivial observables. In this case we can show the following (proof can be found in Publication **I**):

**Proposition 4.** *Let* $\mathsf{A} \in \mathcal{O}(\Omega, \mathcal{S})$. *Then* $w(\mathsf{A}; \mathcal{T}) = \sum_{x \in \Omega} \inf_{s \in \mathcal{S}} \mathsf{A}_x(s)$.

Since $\mathcal{S}$ is compact, the infimum in the previous proposition is always attained. Now, let us consider $\lambda_{\min}$ similarly to $\lambda_{\max}$ as a property of the observables: if $C \in \mathcal{C}(\mathcal{S})$ so that $C_{ij} = \mathsf{M}_j(s_i)$ for some observable $\mathsf{M} \in \mathcal{O}(\Omega, \mathcal{S})$ and some states $\{s_i\}_i \subset \mathcal{S}$, we have $\lambda_{\min}(C) = -\sum_j \min_i \mathsf{M}_j(s_i)$ so that for observables we can define $\lambda_{\min}(\mathsf{M}) := -\sum_{j \in \Omega} \min_{s \in \mathcal{S}} \mathsf{M}_j(s)$. By the previous proposition we then have that $\lambda_{\min}(\mathsf{M}) = -w(\mathsf{M}; \mathcal{T})$ and the interpretation of $\lambda_{\min}$ as characterizing the intrinsic noise follows.

**Rank**

Our next ultraweak monotone is the *rank* of a matrix. Denoted by rank$(C)$, the rank of an $n \times m$ matrix $C$ is defined as the smallest integer $k \in \mathbb{N}$ such that $C$ can be factored as $C = LR$, where $L$ is an $n \times k$ matrix and $R$ is a $k \times m$ matrix. Since rank of a matrix cannot be increased in matrix multiplication, rank is indeed an ultraweak monotone. Rank is an easily computable and important mathematical quantity but without a clear physical meaning. However, if we consider the maximum rank that can be achieved in the set of communication matrices of a given theory with a state space $\mathcal{S} \subset \mathcal{V}$, we see that it can be related to the linear dimension $d_{lin}(\mathcal{S})$ of the theory defined as $d_{lin}(\mathcal{S}) := \dim(\operatorname{aff}(\mathcal{S})) + 1 = \dim(\mathcal{V})$ in the following way:

**Proposition 5.** $d_{lin}(\mathcal{S}) = \sup\{\text{rank}(C) \,|\, C \in \mathcal{C}(\mathcal{S})\}$.

The proof relies on the fact that we can find at most $d_{lin}(\mathcal{S})$ affinely independent states and an observable with maximally $d_{lin}(\mathcal{S})$ linearly independent effects such that the $d_{lin}(\mathcal{S}) \times d_{lin}(\mathcal{S})$ communication matrix formed using those states and that observable is full rank. This also shows that the supremum is always attained.

**Nonnegative rank**

Another rank-like quantity is the *nonnegative rank* of a nonnegative matrix, i.e., a matrix with nonnegative elements. Denoted by $\text{rank}_+(C)$, the nonnegative rank of a nonnegative $n \times m$ matrix $C$ is defined as the smallest $k \in \mathbb{N}$ such that $C$ can be decomposed as $C = LR$, where $L$ and $R$ are $n \times k$ and $k \times m$ nonnegative matrices, respectively. If $C \in \mathcal{M}^{row}_{n,m}$, then $L$ and $R$ can also be chosen such that $L \in \mathcal{M}^{row}_{n,k}$ and $R \in \mathcal{M}^{row}_{k,m}$ [90].

The definition of nonnegative rank for a row-stochastic matrix $C \in \mathcal{M}^{row}$ can be expressed as $\text{rank}_+(C) = \min\{k \in \mathbb{N} \,|\, C \preceq \mathbb{1}_k\}$, from which it becomes clear that $\text{rank}_+$ is an ultraweak monotone. If we define the *classical dimension* of the state space $\mathcal{S}$ as

$$d_{cl}(\mathcal{S}) := \inf\{d \in \mathbb{N} \,|\, \mathcal{C}(\mathcal{S}) \subseteq \mathcal{C}(\mathcal{S}^{cl}_d)\},$$

then we can conclude from Prop. 2 the following result:

**Proposition 6.** $d_{cl}(\mathcal{S}) = \sup\{\text{rank}_+(C) \,|\, C \in \mathcal{C}(\mathcal{S})\}$

It is clear from the definition of the nonnegative rank that

$$\text{rank}(C) \leq \text{rank}_+(C) \leq \min(n, m) \tag{2.3}$$

for any nonnegative $n \times m$ matrix $C$. However, calculating the exact nonnegative rank of a matrix turns out to be a computationally hard problem [91].

**Positive semidefinite rank**

Our last rank-like quantity is the *positive semidefinite rank* of a nonnegative matrix. Denoted by $\text{rank}_{\text{psd}}(C)$, the positive semidefinite rank of a nonnegative $n \times m$ matrix $C$ is defined as the smallest $k \in \mathbb{N}$ such that $C$ can be decomposed as $C_{ij} = \operatorname{tr}[A_i B_j]$, where $A_1, \ldots, A_n$ and $B_1, \ldots, B_m$ are $k \times k$ positive semidefinite matrices [92]. Based on the definition of $\operatorname{rank}_{\operatorname{psd}}$, it is straigtforward to show that it is an ultraweak monotone.

Let us denote by $\mathcal{S}_d^q$ the state space of a quantum system with a $d$-dimensional Hilbert space. The physical meaning for the positive semidefinite rank was shown in [93]: *a row-stochastic matrix $C \in \mathcal{M}^{row}$ has $\operatorname{rank}_{\operatorname{psd}}(C) \leq d$ if and only if $C \in \mathcal{C}(\mathcal{S}_d^q)$*. Thus, if we define the *quantum dimension* $d_q(\mathcal{S})$ of a state space $\mathcal{S}$ as the dimension of the minimal quantum system that can produce all the communication matrices of that theory, or more precisely as

$$d_q(\mathcal{S}) := \inf\{d \in \mathbb{N} \,|\, \mathcal{C}(\mathcal{S}) \subseteq \mathcal{C}(\mathcal{S}_d^q)\},$$

we arrive at the following result

**Proposition 7.** $d_q(\mathcal{S}) = \sup\{\operatorname{rank}_{\operatorname{psd}}(C) \,|\, C \in \mathcal{C}(\mathcal{S})\}$

Compared to the other ranks, we have the following inequalities [92]:

$$\sqrt{\operatorname{rank}(C)} \leq \operatorname{rank}_{\operatorname{psd}}(C) \leq \operatorname{rank}_+(C) \tag{2.4}$$

for all nonnegative matrices $C$. Similar to the nonnegative rank, calculating the positive semidefinite rank of a matrix is also a computationally hard problem [94].

### 2.3.2 Physical dimensions

Next we will demonstrate how one can use the introduced monotones to characterize different theories via the physical dimensions they induce. Let us start by collecting and comparing the values of the introduced monotones.

**Proposition 8.** *For an operation theory with a state space $\mathcal{S}$ we have*

$$\lambda_{\min}(C) < \iota(C) \leq \lambda_{\max}(C) \leq \operatorname{rank}_{\operatorname{psd}}(C) \leq \operatorname{rank}_+(C) \tag{2.5}$$

*and*

$$\operatorname{rank}(C) \leq \operatorname{rank}_{\operatorname{psd}}(C)^2, \quad \operatorname{rank}(C) \leq \operatorname{rank}_+(C) \tag{2.6}$$

*for all $C \in \mathcal{S}$.*

| $\mathcal{S}$ | $d_{op}(\mathcal{S})$ | $\lambda_{\max}(\mathcal{S})$ | $d_q(\mathcal{S})$ | $d_{cl}(\mathcal{S})$ | $d_{lin}(\mathcal{S})$ |
|---|---|---|---|---|---|
| $\mathcal{S}_d^{cl}$ | $d$ | $d$ | $d$ | $d$ | $d$ |
| $\mathcal{S}_d^q$ | $d$ | $d$ | $d$ | $\geq d^2$ $(*)$ | $d^2$ |
| $\mathcal{S}_n$, even $n$ | 2 | 2 $(\star)$ | $\geq 2$ | $\geq 3$ | 3 |
| $\mathcal{S}_n$, odd $n$ | 2 | $1 + \sec\left(\frac{\pi}{n}\right) \in (2,3)$ $(\dagger)$ | $\geq 3$ | $\geq 3$ | 3 |

Table 2.1: The values (or bounds) of the physical dimensions for $d$-dimensional quantum and classical theories and polygon theories for $n \geq 4$. Our conjecture is that the inequality $(*)$ is an equality. The value in $(\star)$ was shown in [73] to hold for all point-symmetric state spaces, while the value in $(\dagger)$ is an unpublished result and part of a future manuscript that is currently in development.

The first inequality from Eq. (2.5) is clear since $\lambda_{\min}(C) \in [-1, 0]$ and $\iota(C) \geq 1$. The second inequality follows from the fact that $C \succeq \mathbb{1}_{\iota(C)}$ so that $\lambda_{\max}(C) \geq \lambda_{\max}(\mathbb{1}_{\iota(C)}) = \iota(C)$. The third inequality was shown in [93] and the fourth is the same as in Eq. (2.4). The inequalities in Eq. (2.6) are just from Eq. (2.4) and (2.3). Taking the supremum over all communication matrices we arrive at the following Corollary:

**Corollary 1.** *For an operational theory with a state space $\mathcal{S}$ we have that*

$$d_{op}(\mathcal{S}) \leq \lambda_{\max}(\mathcal{S}) \leq d_q(\mathcal{S}) \leq d_{cl}(\mathcal{S}),$$
$$d_{lin}(\mathcal{S}) \leq d_q(\mathcal{S})^2, \quad d_{lin}(\mathcal{S}) \leq d_{cl}(\mathcal{S}).$$

As was discussed earlier, each monotone has an important physical (or mathematical) interpretation so the maximal values of the monotones are helpful when characterizing a given theory. One can use the introduced physical 'dimensions' induced by the monotones to form classes of theories where certain communication tasks are possible to implement. Using this type of characterization one can compare theories to each other based on the communication they can be used to perform. We demonstrate this concept by presenting the values for the dimensions in the known cases (and bounds given by the above Corollary in the unknown cases) for quantum and classical theory as well as the polygon theories in Table 2.1.

# Chapter 3

# Post-processing of measurements

As a step towards simulating a measurement with some collection of (different) measurements, we first look at the case of simulating a measurement with a single measurement. This is what is usually called the *post-processing of measurements*. Post-processing captures the idea that after a measurement process, having obtained the output(s) of a measurement device, one may want to process the classical information and other outputs that were obtained. One can, for instance, see if it is possible to reveal some other property of the system by manipulating the measurement outcome data (or other outputs) and obtain the outcome statistics of some other measurement (or alter the possible post-measurement state). We start this Chapter by considering the post-processing of measurement devices with only classical outputs, i.e., observables, and then continue to generalize the post-processing relation to instruments as well in accordance with Publication **VI**.

## 3.1 Post-processing of observables

In the measurement of an observable we are only interested in the classical measurement outcomes. Hence, for post-processing this means that we must consider a way to process these outcomes. As was mentioned in Chapter 1, the transformations between classical states are performed with classical channels that correspond to row-stochastic matrices. Thus, for the post-processing of observables we make the following formal definition:

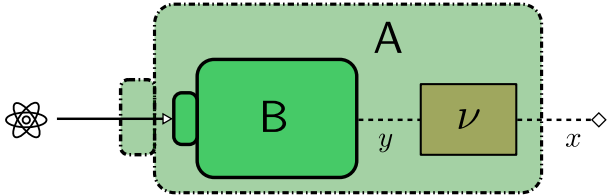

Figure 3.1: The system is measured with an observable B and after obtaining an outcome $y$ it is mapped to an outcome $x$ of an observable $\mathsf{A} = \nu \circ \mathsf{B}$ with the transition probability $\nu_{yx}$.

**Definition 10.** An observable $\mathsf{A} \in \mathcal{O}(\Omega, \mathcal{H})$ is a *post-processing* of an observable $\mathsf{B} \in \mathcal{O}(\Lambda, \mathcal{H})$, denoted by $\mathsf{B} \to \mathsf{A}$, if there exists a row-stochastic post-processing matrix $\nu = (\nu_{yx})_{y \in \Lambda, x \in \Omega} \in \mathcal{M}^{row}$ such that $\mathsf{A}_x = \sum_{y \in \Lambda} \nu_{yx} \mathsf{B}_y$.

The post-processing of an observable is demonstrated in Fig. 3.1. We may also denote $\mathsf{A} = \nu \circ \mathsf{B}$ if $\mathsf{A}$ is a post-processing of $\mathsf{B}$ via a post-processing matrix $\nu$. The matrix element $\nu_{yx}$ is interpreted as the transition probability that an outcome $y$ is mapped to outcome $x$. If both $\mathsf{B} \to \mathsf{A}$ and $\mathsf{A} \to \mathsf{B}$ hold we say that $\mathsf{A}$ and $\mathsf{B}$ are *post-processing equivalent* and denote it by $\mathsf{A} \leftrightarrow \mathsf{B}$. In quantum theory the post-processing relation of POVMs in the current context was first considered in [95]. For the case of POVMs with more general outcome sets, as well as a historical note, see [96] and the references therein.

**Example 7.** As an example of a post-processing we consider the case of *relabeling* where all of the elements of a post-processing matrix are either 0 or 1. Formally, following [97], we say that an observable $\mathsf{A} \in \mathcal{O}(\Omega, \mathcal{S})$ is a *refinement* of an observable $\mathsf{B} \in \mathcal{O}(\Lambda, \mathcal{S})$ if there exists a function $f : \Omega \to \Lambda$ such that $\mathsf{B}_y = \sum_{x \in f^{-1}(y)} \mathsf{A}_x$ for all $y \in \Lambda$. Relabeling thus consists of not only literal bijective relabeling of outcomes but also merging of different outcomes.

### 3.1.1 Post-processing partial order and structure

Post-processing is a reflexive and transitive relation, and so it is a preorder on the set $\mathcal{O}(\mathcal{S})$ of all observables on $\mathcal{S}$. Furthermore, it can be extended to a partial order on the equivalence classes of post-processing-equivalent observables. Let us consider this partial order more carefully. For POVMs in quantum theory this partial order has been studied in [95, 98].

First of all, it is easy to see that the trivial observables are the minimal elements with respect to this preorder: if $\mathsf{A} \in \mathcal{O}(\Omega, \mathcal{S})$ is an observable and $\mathsf{T} \in \mathcal{O}(\Lambda, \mathcal{S})$ is any trivial observable defined as $\mathsf{T}_y = p_y u$ for all $y \in \Lambda$ for some probability distribution $(p_y)_{y \in \Lambda}$, then we have $\mathsf{T} = \nu \circ \mathsf{A}$ by a post-processing matrix $\nu$ defined as $\nu_{xy} = p_y$ for all $x \in \Omega$ and $y \in \Lambda$ so that $\mathsf{A} \to \mathsf{T}$. Furthermore, the trivial observables form a single class of post-processing observables because if $\mathsf{B} \in \mathcal{O}(\Gamma, \mathcal{S})$ is an observable such that $\mathsf{B} = \mu \circ \mathsf{T}$ for some trivial observable $\mathsf{T} \in \mathcal{O}(\Lambda, \mathcal{S})$ by a post-processing matrix $\mu$, then $\mathsf{B}_z = \left( \sum_{y \in \Lambda} \mu_{yz} p_y \right) u$ for all $z \in \Gamma$, i.e., $\mathsf{B}$ must be trivial. Thus, any trivial observable can be post-processed from any observable and the set of trivial observables is closed with respect to post-processing. This shows that the only minimal element, i.e., the least element, in the set of equivalence classes with respect to the post-processing partial order is the equivalence class of trivial observables.

For the maximal elements, we make the following formal definition which was first considered in [95] and later renamed in [98]:

**Definition 11.** An observable $\mathsf{A}$ is *post-processing clean* if, whenever we have $\mathsf{B} \to \mathsf{A}$ for some observable $\mathsf{B}$, then $\mathsf{A} \to \mathsf{B}$ also holds.

We saw that the minimal elements formed a single equivalence class such that the least element is the equivalence class of trivial observables. To see that this is not the case with maximal elements, i.e., the post-processing clean observables, we give the following characterization (see Publication **II** for the proof):

**Proposition 9.** *An observable is post-processing clean if and only if it is indecomposable.*

We recall that in quantum theory a POVM is indecomposable if and only if it is rank-1. The previous result was first proved for POVMs in [95].

**Example 8.** Let us consider two indecomposable, i.e., rank-1, 4-outcome qubit POVMs $A \in \mathcal{O}([4], \mathbb{C}^2)$ and $B \in \mathcal{O}([4], \mathbb{C}^2)$ defined by

$$A(1) = \frac{1}{2} |\varphi_1\rangle\langle\varphi_1|, \ A(2) = \frac{1}{2} |\varphi_2\rangle\langle\varphi_2|, \ A(3) = \frac{1}{2} |\psi_1\rangle\langle\psi_1|, \ A(4) = \frac{1}{2} |\psi_2\rangle\langle\psi_2|,$$

$$B(1) = \frac{1}{3} |\varphi_1\rangle\langle\varphi_1|, \ B(2) = \frac{1}{3} |\varphi_2\rangle\langle\varphi_2|, \ B(3) = \frac{2}{3} |\psi_1\rangle\langle\psi_1|, \ B(4) = \frac{2}{3} |\psi_2\rangle\langle\psi_2|,$$

where $\{\varphi_1, \varphi_2\}$ and $\{\psi_1, \psi_2\}$ are some two orthonormal bases of $\mathbb{C}^2$ such that none of the vectors $\varphi_1, \varphi_2, \psi_1, \psi_2 \in \mathbb{C}^2$ are proportional to each other. With this rather trivial example we want to raise two points: First, although both POVMs are indecomposable, it is straightforward to see that they are not post-processing equivalent in that there is no single maximal element, i.e., a greatest element, with respect to the post-processing partial order in the set of equivalence classes of observables. Second, although the effects of $A$ and $B$ are proportional to each other, it is not a sufficient condition for $A$ and $B$ to be post-processing equivalent. We will discuss this point later.

As we saw in the previous example, there is no greatest element in the post-processing partial order but rather a class of maximal elements, namely the post-processing clean, i.e., the indecomposable, observables. As was mentioned in Chapter 1, indecomposability is an important concept for the structure of the theory because indecomposable effects characterize the extreme rays of the positive dual cone within which all the effects lie in. As is evident by now, another important structural concept is extremality since all the basic structures (states, observables, channels) we consider are convex and characterized by their extreme points. For indecomposable observables we can show the following necessary and sufficient condition for extremality (see Publication **II** for the proof or [99] in the quantum case):

**Proposition 10.** *A post-processing clean observable is extreme if and only if its nonzero effects are linearly independent.*

Having characterized the minimal and maximal elements of the post-processing partial order and the structures they have, next we will take a closer look inside the equivalence classes of observables.

### 3.1.2 Minimally sufficient observables

In addition to considering post-processing as a way to construct new observables out of known ones, one can look from the resource theoretic point of view, according to which the relation $A \to B$ can be interpreted as the observable A is more useful or more informative than the observable B. As we consider post-processing as a partial order, the comparison is between the equivalence classes of observables, so post-processing equivalent observables A and B can be seen as equally informative. However, even though A and B can be seen having essentially the same information content, the notion of *minimal sufficiency* captures the idea that inside the equivalence class there are observables with minimum informational redundancy.

**Definition 12.** An observable A is minimally sufficient if, whenever A is post-processing equivalent with some observable B, then B is a refinement of A.

The concept of minimal sufficiency originates from minimally sufficient statistics [100, 101] and was generalized to POVMs in [102] and observables in GPTs in [34]. It has been established that in every equivalence class there exists a minimally sufficient observable, called the *minimally sufficient representative*, and that it is unique up to isomorphic relabeling of outcomes. Furthermore, an observable $A \in \mathcal{O}(\Omega, \mathcal{S})$ is minimally sufficient if and only if it is *non-vanishing*, i.e., $A_x \neq o$ for all $x \in \Omega$, and it is *pairwise linearly independent*, i.e., $A_x \neq cA_y$ for any $c > 0$ for all $x \neq y$, $x, y \in \Omega$ [95, 102]. We can construct such observable as follows: Let $A \in \mathcal{O}(\Omega, \mathcal{S})$ and let us define an equivalence relation $\sim$ in $\Omega$ by denoting $x \sim y$ if and only if $A_x = cA_y$ for some $c > 0$. We denote the set of equivalence classes $\Omega/\sim$ by $\tilde{\Omega}$ and define a minimally sufficient observable $\tilde{A} \in \mathcal{O}(\tilde{\Omega}, \mathcal{S})$ by setting $\tilde{A}_{\tilde{x}} := \sum_{y \in \tilde{x}} A_y$ for all $\tilde{x} \in \tilde{\Omega}$. It can be easily checked that $\tilde{A}$ is pairwise linearly independent and that $A \leftrightarrow \tilde{A}$.

In particular, the uniqueness of the minimally sufficient representative in each equivalence class implies the following result (the proof can be extracted from [102] and is explicitly shown in Publication **VI** for POVMs):

**Proposition 11.** *If $A \in \mathcal{O}(\Omega, \mathcal{S})$ and $B \in \mathcal{O}(\Lambda, \mathcal{S})$ are two post-processing equivalent observables, then for all $x \in \Omega$ such that $A_x \neq o$ there is $y_x \in \Lambda$ and $c_{xy_x} > 0$ such that $A_x = c_{xy_x} B_{y_x}$.*

Furthermore, one can actually construct the post-processings $\nu$ and $\mu$ for $A \to B$ and $B \to A$ respectively such that the post-processing elements $\nu_{xy}$ and $\mu_{yx}$ are nonzero only if $A_x$ and $B_y$ are proportional. However, as we saw in Example 8 not all observables whose effects are proportional are post-processing equivalent.

## 3.2 Post-processing of instruments

Let us generalize the concept of post-processing to instruments. The essence of post-processing is that by manipulating the output(s) of a measurement device one wishes to learn some new information about the measured system. In this way one can think of the post-processed device as a new measurement device altogether and the act of post-processing as a recipe to obtain new devices out of a known one. For observables this meant stochastic processing of the classical outcomes but for instruments we have to also consider the post-measurement output state and post-process that as well.

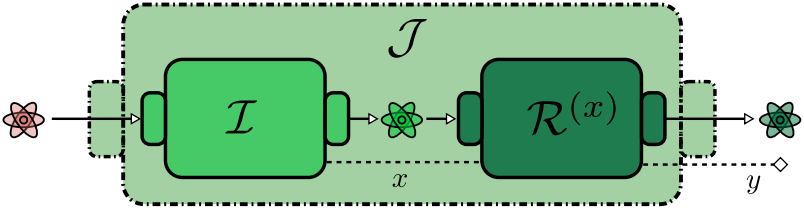

Figure 3.2: The system in state $s$ is measured with an instrument $\mathcal{I}$ and an outcome $x$ is obtained so that the (unnormalized) conditional output state is $\mathcal{I}_x(s)$. Based on the classical outcome $x$, the output state of $\mathcal{I}$ serves as an input for the instrument $\mathcal{R}^{(x)}$, which measures it and obtains an outcome $y$ so that the (unnormalized) conditional output state is then $\mathcal{R}^{(x)}_y(\mathcal{I}_x(s))$. The total resulting instrument is denoted by $\mathcal{J}$.

In Publication **VI** we consider the post-processing of a *quantum* instrument to be a sequential measurement, described by a set of other quantum instruments, conditioned on the measurement outcome of the original instrument. This way the classical outcome of the original instrument determines the resulting post-processing instrument and the output state is transformed accordingly. The post-processed classical outcome is obtained as the classical output of the post-processing instrument.

In this work we consider the same post-processing relation in the GPT framework and generalize some of the results in Publication **VI** while explicitly denoting the results which hold (at least so far) only in quantum theory. To formalize the above discussion, we make the following definition:

**Definition 13.** An instrument $\mathcal{J} \in \mathrm{Ins}(\Lambda, \mathcal{S}, \mathcal{S}'')$ is a post-processing of an instrument $\mathcal{I} \in \mathrm{Ins}(\Omega, \mathcal{S}, \mathcal{S}')$ if there exists a collection of instruments $\{\mathcal{R}^{(x)}\}_{x \in \Omega} \subset \mathrm{Ins}(\Lambda, \mathcal{S}', \mathcal{S}'')$ such that $\mathcal{J}_y = \sum_{x \in \Omega} \mathcal{R}^{(x)}_y \circ \mathcal{I}_x$ for all $y \in \Lambda$, where $\circ$ denotes the composition of maps.

Similarly to the case of observables, if $\mathcal{J}$ is a post-processing of $\mathcal{I}$, we denote it by $\mathcal{I} \to \mathcal{J}$, and if also $\mathcal{J} \to \mathcal{I}$, then $\mathcal{I}$ and $\mathcal{J}$ are post-processing equivalent which we denote by $\mathcal{I} \leftrightarrow \mathcal{J}$. The post-processing of instruments is demonstrated in Fig. 3.2.

In the case of channels our definition reduces to the relation considered in [103, 104] for quantum channels and [34, 47] for channels in GPTs: a channel $\mathcal{D} \in \mathrm{Ch}(\mathcal{S}, \mathcal{S}'')$ is a post-processing of a channel $\mathcal{C} \in \mathrm{Ch}(\mathcal{S}, \mathcal{S}')$ if there exists another channel $\mathcal{E} \in \mathrm{Ch}(\mathcal{S}', \mathcal{S}'')$ such that $\mathcal{D} = \mathcal{E} \circ \mathcal{C}$. Next we will see how the defined post-processing relation for instruments relates to the post-processing of observables.

### 3.2.1 Observables as measure-and-prepare instruments

Let $\mathsf{A} \in \mathcal{O}(\Omega, \mathcal{S})$ be an observable on a state space $\mathcal{S}$. We can identify the points of $\Omega$ with the extreme points $\{\delta_x\}_{x \in \Omega}$ of a simplex $\mathcal{S}^{cl}_{|\Omega|}$, which allows us to consider the observable $\mathsf{A}$ as an instrument $\mathcal{A} \in \mathrm{Ins}(\Omega, \mathcal{S}, \mathcal{S}^{cl}_{|\Omega|})$ with operations $\mathcal{A}_x : \mathcal{S} \to \mathcal{S}^{cl}_{|\Omega|}$ defined as $\mathcal{A}_x(s) := \mathsf{A}_x(s)\delta_x$ for all $x \in \Omega$ and $s \in \mathcal{S}$. Clearly the induced observable $\mathsf{A}^{\mathcal{A}}$ of $\mathcal{A}$ is $\mathsf{A}$. As we recall from Section 1.2.3, the instrument $\mathcal{A}$ defined above is a measure-and-prepare instrument, and thus *all observables can be identified with measure-and-prepare instruments whose outcome space is a simplex and whose prepared states are the pure states of the simplex.*

A direct generalization of the proof given in Publication **VI** shows that in the case of measure-and-prepare instruments, the post-processing relation is completely characterized by the post-processing relation of the induced observables.

**Proposition 12.** *Let $\mathcal{I}$ and $\mathcal{J}$ both be measure-and-prepare instruments. Then $\mathcal{I} \to \mathcal{J}$ if and only if $\mathsf{A}^{\mathcal{I}} \to \mathsf{A}^{\mathcal{J}}$.*

For the more specific class of measure-and-prepare instruments that are associated to the observables, we can show a stronger result which also characterizes the specific form of the instruments that one uses for post-processing.

**\*Proposition 13.** *Let $\mathsf{A} \in \mathcal{O}(\Omega, \mathcal{S})$ and $\mathsf{B} \in \mathcal{O}(\Lambda, \mathcal{S})$ be two observables and $\mathcal{A} \in \mathrm{Ins}(\Omega, \mathcal{S}, \mathcal{S}^{cl}_{|\Omega|})$ and $\mathcal{B} \in \mathrm{Ins}(\Lambda, \mathcal{S}, \mathcal{S}^{cl}_{|\Lambda|})$ their associated measure-and-prepare instruments. Then $\mathcal{A} \to \mathcal{B}$ if and only if there exist instruments*

$\{\mathcal{R}^{(x)}\}_{x \in \Omega} \subset \text{Ins}(\Lambda, \mathcal{S}_{|\Omega|}^{cl}, \mathcal{S}_{|\Lambda|}^{cl})$ *such that*

$$\mathcal{R}_y^{(x)}(\delta_x) = \nu_{xy}\delta_y \tag{3.1}$$

*for all $x \in \Omega$ for which $\mathsf{A}_x \neq o$ and all $y \in \Lambda$ for some row-stochastic matrix $\nu \in \mathcal{M}_{|\Omega|,|\Lambda|}^{row}$ such that $\mathsf{B} = \nu \circ \mathsf{A}$.*

*Proof.* First let $\mathsf{B} = \nu \circ \mathsf{A}$ hold for some $\nu \in \mathcal{M}_{|\Omega|,|\Lambda|}^{row}$ and let us define the post-processing instruments $\{\mathcal{R}^{(x)}\}_{x \in \Omega} \subset \text{Ins}(\Lambda, \mathcal{S}_{|\Omega|}^{cl}, \mathcal{S}_{|\Lambda|}^{cl})$ by setting $\mathcal{R}_y^{(x)}(\delta) = \nu_{xy}\delta_y$ for all $\delta \in \mathcal{S}_{|\Omega|}^{cl}$, $x \in \Omega$ and $y \in \Lambda$ so that in particular Eq. (3.1) is satisfied. Now we see that

$$\sum_{x \in \Omega} \mathcal{R}_y^{(x)}(\mathcal{A}_x(s)) = \sum_{x \in \Omega} \mathsf{A}_x(s)\mathcal{R}_y^{(x)}(\delta_x) = \sum_{x \in \Omega} \nu_{xy}\mathsf{A}_x(s)\delta_y = \mathsf{B}_y(s)\delta_y = \mathcal{B}_y(s)$$

for all $s \in \mathcal{S}$ and $y \in \Lambda$, and so $\mathcal{A} \to \mathcal{B}$.

Now let $\mathcal{A} \to \mathcal{B}$ so that there exists instruments $\{\mathcal{R}^{(x)}\}_{x \in \Omega} \subset \text{Ins}(\Lambda, \mathcal{S}_{|\Omega|}^{cl}, \mathcal{S}_{|\Lambda|}^{cl})$ such that for all $s \in \mathcal{S}$ and $y \in \Lambda$ we have that

$$\mathsf{B}_y(s)\delta_y = \mathcal{B}_y(s) = \sum_{x \in \Omega} \mathcal{R}_y^{(x)}(\mathcal{A}_x(s)) = \sum_{x \in \Omega} \mathsf{A}_x(s)\mathcal{R}_y^{(x)}(\delta_x) \tag{3.2}$$

$$= \sum_{y' \in \Lambda} \sum_{x \in \Omega} u^\Lambda(\mathcal{R}_y^{(x)}(\delta_x))\mathsf{A}_x(s) p_{y'}^{\mathcal{R}_y^{(x)}} \delta_{y'}, \tag{3.3}$$

where $\mathcal{R}_y^{(x)}(\delta_x) = u^\Lambda(\mathcal{R}_y^{(x)}(\delta_x))\left[\sum_{y' \in \Lambda} p_{y'}^{\mathcal{R}_y^{(x)}} \delta_{y'}\right]$ is the unique base and convex decomposition of the subnormalized state $\mathcal{R}_y^{(x)}(\delta_x)$. Because $\{\delta_y\}_{y \in \Lambda}$ is linearly independent in the $|\Lambda|$-dimensional vector space that $\mathcal{S}_{|\Lambda|}^{cl}$ is embedded in as a $(|\Lambda|-1)$-simplex, we must have for all $s \in \mathcal{S}$ and $y \in \Lambda$ that

$$\mathsf{B}_y(s)\delta_y = \left[\sum_{x \in \Omega} u^\Lambda(\mathcal{R}_y^{(x)}(\delta_x)) p_y^{\mathcal{R}_y^{(x)}} \mathsf{A}_x(s)\right]\delta_y. \tag{3.4}$$

Let us denote $\Omega_0 = \{x \in \Omega \,|\, \mathsf{A}_x = o\}$. For all $x \in \Omega \setminus \Omega_0$ we denote $\nu_{xy}^\mathcal{R} := u^\Lambda(\mathcal{R}_y^{(x)}(\delta_x)) p_y^{\mathcal{R}_y^{(x)}} \in [0,1]$ for all $y \in \Lambda$, and for all $x \in \Omega_0$ we denote $\nu_{xy}^\mathcal{R} := q_y$ for all $y \in \Lambda$ for some fixed probability distribution $(q_y)_{y \in \Lambda}$. By taking the unit functional $u^\Lambda$ on both sides of Eq. (3.4) we get that

$$\mathsf{B}_y(s) = \sum_{x \in \Omega} \nu^{\mathcal{R}}_{xy} \mathsf{A}_x(s) \tag{3.5}$$

for all $s \in \mathcal{S}$ and $y \in \Lambda$. We still have to check the row-stochasticity of $\nu^{\mathcal{R}}$. Clearly $\sum_{y \in \Lambda} \nu^{\mathcal{R}}_{xy} = 1$ for all $x \in \Omega_0$. To see this also for all $x \in \Omega \setminus \Omega_0$, let us take the sum over $\Lambda$ in Eq. (3.5) so that

$$\sum_{x \in \Omega \setminus \Omega_0} \left( \sum_{y \in \Lambda} \nu^{\mathcal{R}}_{xy} \right) \mathsf{A}_x(s) = 1 \tag{3.6}$$

for all $s \in \mathcal{S}$. We can fix a state $s \in \mathcal{S}$ such that $\mathsf{A}_x(s) > 0$ for all $x \in \Omega \setminus \Omega_0$ so that from the normalization of $\mathsf{A}$ and Eq. (3.6) it also follows that $\sum_{y \in \Lambda} \nu^{\mathcal{R}}_{xy} = 1$ for all $x \in \Omega \setminus \Omega_0$.

To see Eq. (3.1) let us denote $\Lambda_0^x = \{y \in \Lambda \,|\, u^\Lambda(\mathcal{R}_y^{(x)}(\delta_x)) = 0\}$ for all $x \in \Omega \setminus \Omega_0$. If $y \in \Lambda \setminus \Lambda_0^x$, then $u^\Lambda(\mathcal{R}_y^{(x)}(\delta_x)) > 0$, and it follows from the row-stochasticity of $\nu^{\mathcal{R}}$ that

$$\sum_{y \in \Lambda \setminus \Lambda_0^x} u^\Lambda(\mathcal{R}_y^{(x)}(\delta_x)) p_y^{\mathcal{R}_y^{(x)}} = 1. \tag{3.7}$$

Because $\sum_{y \in \Lambda} \mathcal{R}_y^{(x)}$ is a channel for all $x \in \Omega$ so that $\sum_{y \in \Lambda} u^\Lambda(\mathcal{R}_y^{(x)}(\delta_x)) = \sum_{y \in \Lambda \setminus \Lambda_0^x} u^\Lambda(\mathcal{R}_y^{(x)}(\delta_x)) = 1$, it follows from Eq. (3.7) that $p_y^{\mathcal{R}_y^{(x)}} = 1$ for all $y \in \Lambda \setminus \Lambda_0^x$ for all $x \in \Omega \setminus \Omega_0$. Hence, $\mathcal{R}_y^{(x)}(\delta_x) = \nu^{\mathcal{R}}_{xy} \delta_y = 0$ for all $y \in \Lambda_0^x$ and $\mathcal{R}_y^{(x)}(\delta_x) = \nu^{\mathcal{R}}_{xy} \delta_y \neq 0$ for all $y \in \Lambda \setminus \Lambda_0^x$ for all $x \in \Omega \setminus \Omega_0$, hence Eq. (3.1) is satisfied. $\square$

We see from the previous proposition that not only is the post-processing relation of the associated instruments of observables completely characterized by the post-processing relation of the observables themselves, but also the post-processing instruments $\mathcal{R}^{(x)}$ are limited to just classical post-processing of outcomes given by a row-stochastic matrix. A similar result for the post-processing of channels is given in [58]. Hence, we have recovered the traditional notion of post-processing of observables as a special case of the post-processing of instruments.

### 3.2.2 Order structure of instruments

Just as in the case of observables, the defined post-processing relation induces a partial order on the equivalence classes of instruments. This means that once again the first things to consider are the order structure and the maximal and minimal elements with respect to the post-processing partial order.

For observables the least element was seen to be the equivalence class of all trivial observables. By considering trivial observables as instruments in the manner that was presented in the previous section, one sees that they correspond to the trash-and-prepare instruments (with simplex output spaces), i.e., measure-and-prepare instruments that measure a trivial observable. We can also show that in the general case of post-processing of instruments the situation is the same (the proof is a direct generalization of the one presented in Publication **VI** for quantum instruments):

**Proposition 14.** *Any instrument can be post-processed into any trash-and-prepare instrument, and the set of trash-and-prepare instruments is closed with respect to post-processing.*

For the maximal elements we make the same definition of post-processing cleanness that we made for observables: *An instrument $\mathcal{I}$ is post-processing clean if whenever $\mathcal{J} \to \mathcal{I}$ for some other instrument $\mathcal{J}$, then $\mathcal{I} \to \mathcal{J}$.* Unlike in the case of observables, for the general case of instruments we find that there exists a unique maximal element. Again the proof is a direct analogue of the one given in Publication **VI** for quantum instruments. This coincides with the result known for (quantum) channels [47, 103, 104].

**Proposition 15.** *An instrument $\mathcal{I} \in \mathrm{Ins}(\Omega, \mathcal{S}, \mathcal{S}')$ is post-processing clean if and only if it is post-processing equivalent with the identity channel $id \in \mathrm{Ch}(\mathcal{S})$.*

Obviously the difference in the maximal elements of observables and instruments comes from the structural difference of observables (considered as instruments) as the post-processing is only limited to the specific type of instruments given by *Prop. 13, whilst in general we have no limitations.

Furthermore, in quantum theory we can give the following characterization for the equivalence class of the identity channel.

**Proposition 16** (Quantum theory). *A quantum instrument $\mathcal{I} \in \text{Ins}(\Omega, \mathcal{H}, \mathcal{K})$ is post-processing equivalent with the identity channel $\text{id} \in \text{Ch}(\mathcal{H})$ if and only if for all $x \in \Omega$ and $\varrho \in \mathcal{S}(\mathcal{H})$ it is of the form*

$$\mathcal{I}_x(\varrho) = \sum_{i=1}^{n_x} p_{xi} V_{xi} \varrho V_{xi}^* \tag{3.8}$$

*for some $n_x \in \mathbb{N}$, some probability distribution $(p_{xi})_{x \in \Omega, i \in [n_x]}$ and some isometries $V_{xi} : \mathcal{H} \to \mathcal{K}$ such that $V_{xj}^* V_{xi} = 0$ for all $i \neq j$, $i, j \in [n_x]$.*

We note that from the previous result it is clear that *the induced POVM of a post-processing clean quantum instrument is trivial*: if $\mathcal{I}$ is a post-processing clean instrument of the form Eq. (3.8), then $A^{\mathcal{I}}(x) = \sum_i p_{xi} V_{xi}^* V_{xi} = \sum_i p_{xi} \mathbb{1}_{\mathcal{H}}$ for all $x \in \Omega$ because $V_{xi}$ is an isometry, i.e., $V_{xi}^* V_{xi} = \mathbb{1}_{\mathcal{H}}$ for all $i \in [n_x]$ and $x \in \Omega$.

## 3.3 Case study: indecomposable instruments

For observables the indecomposable elements were seen to be the maximal ones with respect to post-processing, so we find it meaningful to consider the concept of indecomposablility in the case of instruments as well. We start by generalizing indecomposability of effects and observables to operations and instruments.

Similarly to the case of effects, we call a (nonzero) operation $\mathcal{M}$ *indecomposable* if $\mathcal{M} = \mathcal{N} + \mathcal{N}'$ for some other (nonzero) operations only when there exists $\alpha, \alpha' > 0$ such that $\mathcal{M} = \alpha \mathcal{N} = \alpha' \mathcal{N}'$. An instrument is said to be indecomposable if all of its nonzero operations are indecomposable.

### 3.3.1 Existence of indecomposable instruments

The result proved in [37] showing the decomposability of any effect into a finite sums of indecomposable effects (thereby showing the existence of indecomposable effects) can be directly generalized to operations as well so that, in particular, every operation can be represented as a finite sum of indecomposable operations.

**\*Proposition 17.** *Every nonzero operation can be decomposed into a finite sum of indecomposable operations.*

*Proof.* The set of all positive linear maps (which operations are a subset of) from a state space $\mathcal{S} \subset \mathcal{V}_+ \subset \mathcal{V}$ to a state space $\mathcal{S}' \subset \mathcal{V}'_+ \subset \mathcal{V}'$ forms a closed proper cone, denoted by $\mathcal{P}(\mathcal{V}_+, \mathcal{V}'_+)$, so that by [105, Thm. 3.3.15] it has a convex compact base $\mathcal{B}$ that can be expressed as $\mathcal{B} = \{\mathcal{N} \in \mathcal{P}(\mathcal{V}_+, \mathcal{V}'_+) \,|\, g(\mathcal{N}) = 1\}$ for some positive linear functional $g \in \mathcal{P}(\mathcal{V}_+, \mathcal{V}'_+)^*$. If we now take a nonzero operation $\mathcal{N}$ from $\mathcal{P}(\mathcal{V}_+, \mathcal{V}'_+)$, it has a base/convex decomposition (due to convexity and compactness of $\mathcal{B}$) of the form $\mathcal{N} = \lambda \sum_x p_x \tilde{\mathcal{N}}_x$, where $\lambda > 0$, $(p_x)_x$ is a (non-vanishing) probability distribution and $\{\tilde{\mathcal{N}}_x\}_x$ is some finite set of extreme points of $\mathcal{B}$. Let us denote $\mathcal{N}_x = \lambda p_x \tilde{\mathcal{N}}_x$ for all $x$ so that $\mathcal{N} = \sum_x \mathcal{N}_x$. From the subnormalization criteria $u'(\mathcal{N}(v)) \leq u(v)$ for all $v \in \mathcal{V}_+$ for the operation $\mathcal{N}$ it follows that also $u'(\mathcal{N}_x(v)) \leq u(v)$ for all $x$ and $v \in \mathcal{V}_+$ so that also $\mathcal{N}_x$ is an operation for all $x$. What remains to show is that $\mathcal{N}_x$ is indecomposable for all $x$. Suppose that for some $y$ we have that $\mathcal{N}_y = \mathcal{M}_y + \mathcal{Q}_y$ for some nonzero operations $\mathcal{M}_y$ and $\mathcal{Q}_y$. Since $\mathcal{M}_y, \mathcal{Q}_y \in \mathcal{P}(\mathcal{V}_+, \mathcal{V}'_+)$, they have base decompositions $\mathcal{M}_y = m_y \hat{\mathcal{M}}_y$ and $\mathcal{Q}_y = q_y \hat{\mathcal{Q}}_y$ for some $m_y, q_y > 0$ and $\hat{\mathcal{M}}_y, \hat{\mathcal{Q}}_y \in \mathcal{B}$ so that we can now write $\tilde{\mathcal{N}}_y = m_y/(\lambda p_y) \hat{\mathcal{M}}_y + q_y/(\lambda p_y) \hat{\mathcal{Q}}_y$. Since $\tilde{\mathcal{N}}_y, \hat{\mathcal{M}}_y, \hat{\mathcal{Q}}_y \in \mathcal{B}$ so that $g(\tilde{\mathcal{N}}_y) = g(\hat{\mathcal{M}}_y) = g(\hat{\mathcal{Q}}_y) = 1$, it follows that $m_y/(\lambda p_y) + q_y/(\lambda p_y) = 1$ so that the former expression is a convex decomposition for $\tilde{\mathcal{N}}_y$. Because $\tilde{\mathcal{N}}_y$ is an extreme point of $\mathcal{B}$, we must have $\hat{\mathcal{M}}_y = \hat{\mathcal{Q}}_y$ so that $\mathcal{N}_y$, $\mathcal{M}_y$ and $\mathcal{Q}_y$ are all proportional to each other, and the claim follows. □

Furthermore, in quantum theory the indecomposable POVMs are exactly those that are rank-1, and, indeed, in the case of quantum instruments we can show the following:

**Proposition 18** (Quantum theory)**.** *A quantum instrument is indecomposable if and only if it is Kraus rank-1.*

### 3.3.2 Indecomposability ≠ post-processing cleanness

From the previous result (for quantum instruments) it is immediate that unlike in the case of observables, for instruments the set of indecomposable elements does not coincide with the post-processing clean elements in general. In order to emphasize this point and make it explicit, we present the following Corollary for quantum instruments:

**\*Corollary 2** (Quantum theory)**.** *An indecomposable quantum instrument $\mathcal{I}$ is post-processing clean if and only if the induced POVM $A^{\mathcal{I}}$ is trivial.*

The result can be easily seen as a consequence of Prop. 16 and 18. What we also see is that *a quantum instrument $\mathcal{I} \in \mathrm{Ins}(\Omega, \mathcal{H}, \mathcal{K})$ is post-processing clean and indecomposable if and only if it is of the form $\mathcal{I}_x(\varrho) = p_x V_x \varrho V_x^*$ for all $x \in \Omega$ and $\varrho \in \mathcal{S}(\mathcal{H})$ for some probability distribution $(p_x)_{x \in \Omega}$ and some isometries $V_x : \mathcal{H} \to \mathcal{K}$.*

By using the above results for quantum instruments, we immediately see that not all indecomposable instruments are equivalent with the identity channel and so the notion of indecomposability does not characterize the notion of post-processing cleanness like it did in the case of observables. As a concrete example we give the following class of quantum instruments.

**Example 9** (Quantum theory)**.** Let $A \in \mathcal{O}(\Omega, \mathcal{H})$. We define the *Lüders instrument* $\mathcal{I}^A \in \mathrm{Ins}(\Omega, \mathcal{H})$ of the POVM $A$ by setting $\mathcal{I}_x^A(\varrho) := \sqrt{A(x)} \varrho \sqrt{A(x)}$ for all $x \in \Omega$ and $\varrho \in \mathcal{S}(\mathcal{H})$. Since $\sqrt{A(x)}$ is the only Kraus operator of $\mathcal{I}_x^A$, the instrument $\mathcal{I}^A$ is indecomposable. However, it is not post-processing clean unless $A$ is a trivial POVM.

**Remark 4.** Although in general indecomposability of instruments does not coincide with maximality with respect to the post-processing partial order, we can still obtain every instrument as a post-processing of some indecomposable instrument: Let $\mathcal{I} \in \mathrm{Ins}(\Omega, \mathcal{S}, \mathcal{S}')$ be an instrument so that by *Prop. 17 every operation $\mathcal{I}_x$ has a decomposition into a sum of some indecomposable operations $\mathcal{I}_{xi}$ such that $\mathcal{I}_x = \sum_{i \in [n_x]} \mathcal{I}_{xi}$ for some $n_x \in \mathbb{N}$ for all $x \in \Omega$. If we now define the *detailed instrument $\hat{\mathcal{I}}$ of $\mathcal{I}$ related to the decomposition* $\{\mathcal{I}_{xi}\}_{i,x}$ by setting $\hat{\mathcal{I}}_{(x,i)} = \mathcal{I}_{xi}$ for all $i \in [n_x]$ and $x \in \Omega$, we see that clearly $\hat{\mathcal{I}}$ is indecomposable and furthermore $\hat{\mathcal{I}} \to \mathcal{I}$, where the post-processing is given by just a classical merging of outcomes. Hence, every instrument can be obtained as a post-processing of its detailed instrument.

We note that in quantum theory it is known that every instrument can be expressed as a composition of the Lüders instrument related to its induced POVM with some channel (see, e.g., [106]). This is another way of post-processing all quantum instruments from the indecomposable ones.

### 3.3.3 Measure-and-prepare instruments

As was explained in Chapter 1, in addition to being measurement devices, instruments can also be considered as conditional state preparators, and instruments whose conditional output state is only dependent on the classical measurement outcome form the class of the measure-and-prepare instruments. As a particular case of indecomposability we take a closer look at indecomposable measure-and-prepare instruments and the post-processing equivalence of indecomposable and measure-and-prepare instruments.

As we know, if $\mathcal{I} \in \text{Ins}(\Omega, \mathcal{S}, \mathcal{S}')$ is a measure-and-prepare instrument, then $\mathcal{I}$ is of the form $\mathcal{I}_x(s) = \mathsf{A}_x(s)s'_x$ for all $x \in \Omega$ and $s \in \mathcal{S} \subset \mathcal{V}_+ \subset \mathcal{V}$ for some observable $\mathsf{A} \in \mathcal{O}(\Omega, \mathcal{S})$ and some set of states $\{s'_x\}_{x \in \Omega} \subset \mathcal{S}' \subset \mathcal{V}'_+ \subset \mathcal{V}'$. In what follows, if we consider an instrument $\mathcal{I} \in \text{Ins}(\Omega, \mathcal{S}, \mathcal{S}')$ to be measure-and-prepare, we use the previous expression to represent it.

**\*Proposition 19.** *A measure-and-prepare instrument $\mathcal{I} \in \text{Ins}(\Omega, \mathcal{S}, \mathcal{S}')$ is indecomposable if and only if the induced observable $\mathsf{A}^{\mathcal{I}}$ is indecomposable and the prepared states $s'_x \in \mathcal{S}'$ are all pure for all $x \in \Omega$ for which $\mathsf{A}^{\mathcal{I}}_x \neq o$.*

*Proof.* Let first $\mathsf{A}^{\mathcal{I}} = \mathsf{A}$ be indecomposable and the states $\{s'_x\}_{x \in \Omega}$ be pure. For any $x \in \Omega$ such that $\mathcal{I}_x \neq 0$ (which holds if and only if $\mathsf{A}_x \neq o$) let $\mathcal{I}_x = \mathcal{N}_x + \mathcal{M}_x$ for some nonzero operations $\mathcal{N}_x$ and $\mathcal{M}_x$ from $\mathcal{S}$ to $\mathcal{S}'$. Thus, $\mathsf{A}_x(s)s'_x = \mathcal{N}_x(s) + \mathcal{M}_x(s)$ holds for all $s \in \mathcal{S}$. Let $s \in \mathcal{S}$ be such that $\mathcal{N}_x(s) \neq 0$ and $\mathcal{M}_x(s) \neq 0$ and hence $\mathsf{A}_x(s) \neq 0$. Since $s'_x$ is a pure state, $\mathsf{A}_x(s)s'_x$ lies on the extreme ray of $\mathcal{V}'_+$ that is generated by $s'_x$. Thus, it follows that also $\mathcal{N}_x(s)$ and $\mathcal{M}_x(s)$ lie on the same extreme ray so that in particular they can be expressed as

$$\mathcal{N}_x(s) = u'(\mathcal{N}_x(s))s'_x, \quad \mathcal{M}_x(s) = u'(\mathcal{M}_x(s))s'_x. \tag{3.9}$$

Also, if we have a state $s \in \mathcal{S}$ such that $\mathcal{N}_x(s) = 0$ and/or $\mathcal{M}_x(s) = 0$, then Eq. (3.9) also holds. It follows that functionals $N_x$ and $M_x$ defined by $N_x(s) = u'(\mathcal{N}_x(s))$ and $M_x(s) = u'(\mathcal{M}_x(s))$ for all $s \in \mathcal{S}$ are effects on $\mathcal{S}$ so that $N_x, M_x \in \mathcal{E}(\mathcal{S})$. Furthermore, since $\mathcal{I}_x = \mathcal{N}_x + \mathcal{M}_x$, we have that $\mathsf{A}_x = N_x + M_x$. Because $\mathsf{A}_x \neq o$ is indecomposable there exist $n_x, m_x > 0$ such that $\mathsf{A}_x = n_x N_x = m_x M_x$. Then we see that also $\mathcal{I}_x = n_x \mathcal{N}_x = m_x \mathcal{M}_x$ so that $\mathcal{I}_x$ is indecomposable.

On the other hand, if $\mathsf{A}$ is not indecomposable and/or the states $\{s'_x\}_{x \in \Omega}$ are not pure, it is straightforward to use some decomposition of some decomposable effect $\mathsf{A}_x$ into a sum of indecomposable effects and/or some nontrivial convex decomposition of some mixed state $s'_x$ into pure states to construct operations $\mathcal{N}_x$ and $\mathcal{M}_x$ such that $\mathcal{I}_x = \mathcal{N}_x + \mathcal{M}_x$, where $\mathcal{N}_x$ and $\mathcal{M}_x$ are not proportional so that $\mathcal{I}_x$ cannot be indecomposable. □

Having characterized the measure-and-prepare instruments that them-selves are indecomposable we are next interested in the measure-and-prepare instruments that may not be indecomposable but that are post-processing equivalent with an indecomposable instrument. We saw earlier that even though the indecomposable instruments are not maximal with respect to the post-processing partial order in general, every instrument can still be post-processed from the indecomposable ones. In the case of measure-and-prepare instruments we want to consider which cases the equivalence with an indecomposable instrument also holds.

Because $\mathcal{J} \to \mathcal{I}$ always holds for some instruments $\mathcal{I}$ and $\mathcal{J}$ where $\mathcal{J}$ is indecomposable, let us continue by considering the relation $\mathcal{I} \to \mathcal{J}$ when $\mathcal{J}$ is indecomposable. We can show the following.

**\*Proposition 20.** *Let $\mathcal{I} \in \mathrm{Ins}(\Omega, \mathcal{S}, \mathcal{S}')$ and $\mathcal{J} \in \mathrm{Ins}(\Lambda, \mathcal{S}, \mathcal{S}'')$ be two instruments such that $\mathcal{J}$ is indecomposable. If $\mathcal{I} \to \mathcal{J}$, then $\mathsf{A}^{\mathcal{J}} \to \mathsf{A}^{\mathcal{I}}$. Furthermore, if $\mathcal{I}$ is a measure-and-prepare instrument then so is $\mathcal{J}$.*

*Proof.* Since $\mathcal{I} \to \mathcal{J}$, there exist post-processing instruments $\{\mathcal{R}^{(x)}\}_{x\in\Omega} \in \mathrm{Ins}(\Lambda, \mathcal{S}', \mathcal{S}'')$ such that $\mathcal{J}_y = \sum_{x\in\Omega} \mathcal{R}_y^{(x)} \circ \mathcal{I}_x$ for all $y \in \Lambda$. Let us denote $\Lambda_0 = \{y \in \Lambda \,|\, \mathcal{J}_y = 0\}$ and $\Omega_0^y = \{x \in \Omega \,|\, \mathcal{R}_y^{(x)} \circ \mathcal{I}_x = 0\}$. Since $\mathcal{J}$ is indecomposable, we must have that $\mathcal{R}_y^{(x)} \circ \mathcal{I}_x = \nu_{yx} \mathcal{J}_y$ for some numbers $\nu_{yx} \geq 0$ for all $x \in \Omega$ and $y \in \Lambda \setminus \Lambda_0$, where $\nu_{yx} = 0$ if and only if $x \in \Omega_0^y$ for some $y \in \Lambda \setminus \Lambda_0$. We note that for all $y \in \Lambda \setminus \Lambda_0$ we must have $\Omega \setminus \Omega_0^y \neq \emptyset$. For $y \in \Lambda_0$ it is clear that $\mathcal{R}_y^{(x)} \circ \mathcal{I}_x = 0 = \mathcal{J}_y$ for all $x \in \Omega$ so that we can set $\nu_{yx} = p_x$ for all $y \in \Lambda_0$ for some fixed (non-vanishing) probability distribution $(p_x)_{x\in\Omega}$, and we see that $\mathcal{R}_y^{(x)} \circ \mathcal{I}_x = \nu_{yx} \mathcal{J}_y$ still holds. Clearly also $\sum_{x\in\Omega} \nu_{yx} = 1$ for all $y \in \Lambda$ so that $\nu := (\nu_{xy})_{x,y} \in \mathcal{M}^{row}$. We now see that

$$\mathsf{A}_x^{\mathcal{I}}(s) = u'(\mathcal{I}_x(s)) = u''\left(\left(\sum_{y\in\Lambda} \mathcal{R}_y^{(x)}\right)(\mathcal{I}_x(s))\right) = \sum_{y\in\Lambda} u''\left(\mathcal{R}_y^{(x)}(\mathcal{I}_x(s))\right)$$
$$= \sum_{y\in\Lambda} u''\left(\nu_{yx}\mathcal{J}_y(s)\right) = \sum_{y\in\Lambda} \nu_{yx} \mathsf{A}_y^{\mathcal{J}}(s)$$

for all $x \in \Omega$ and $s \in \mathcal{S}$ so that $\mathsf{A}^{\mathcal{J}} \to \mathsf{A}^{\mathcal{I}}$.

Now let $\mathcal{I}$ be a measure-and-prepare instrument. For each $y \in \Lambda \setminus \Lambda_0$ let us fix some $x_y \in \Omega \setminus \Omega_0^y$. By following the previous part of the proof, we see that if we denote $\mu_{yx_y} := 1/\nu_{yx_y} > 0$, then $\mathcal{J}_y = \mu_{yx_y} \mathcal{R}_y^{(x_y)} \circ \mathcal{I}_{x_y}$ for all $y \in \Lambda \setminus \Lambda_0$. Let us set $s''_{yx_y} := \mathcal{R}_y^{(x_y)}(s'_{x_y})/u''(\mathcal{R}_y^{(x_y)}(s'_{x_y})) \in \mathcal{S}''$ for all $y \in \Lambda \setminus \Lambda_0$, where now $u''(\mathcal{R}_y^{(x_y)}(s'_{x_y})) \neq 0$. It follows from the uniqueness of the base decomposition of $\mathcal{J}_y(s)$ that we can express $\mathcal{J}_y$ as $\mathcal{J}_y(s) = \mathsf{A}_y^{\mathcal{J}}(s)s''_y$ for all $s \in \mathcal{S}$, where $\mathsf{A}_y^{\mathcal{J}}(s) = \mu_{yx_y}\mathsf{A}_{x_y}(s)u''(\mathcal{R}_y^{(x_y)}(s'_{x_y}))$ and $s''_y := s''_{yx_y} \in \mathcal{S}''$ are independent from the choice of $x_y \in \Omega \setminus \Omega_0^y$. On the other hand, if we take $y \in \Lambda$ such that $\mathcal{J}_y = 0$, then also $\mathsf{A}_y^{\mathcal{J}} = o$ so that we can set $s''_y = s''$ for some fixed state $s'' \in \mathcal{S}''$ and $\mathcal{J}_y(s) = \mathsf{A}_y^{\mathcal{J}}(s)s''_y$ still holds for all $s \in \mathcal{S}$. Hence, $\mathcal{J}$ is a measure-and-prepare instrument. □

To conclude this chapter, as a Corollary of our results we are able to characterize those measure-and-prepare instruments that are post-processing equivalent with an indecomposable instrument and even the form of that indecomposable instrument.

**\*Corollary 3.** *A measure-and-prepare instrument $\mathcal{I}$ is post-processing equivalent with an indecomposable instrument $\mathcal{J}$ if and only if the induced observable $\mathsf{A}^{\mathcal{I}}$ is indecomposable. In this case also $\mathcal{J}$ is a measure-and-prepare instrument.*

*Proof.* First let $\mathsf{A} = \mathsf{A}^{\mathcal{I}}$ be indecomposable. Let us take some set of pure states $\{t'_x\}_{x \in \Omega} \subset \mathcal{S}'$ and let us define instruments $\{\mathcal{R}^{(x)}\}_{x \in \Omega} \subset \mathrm{Ins}(\Omega, \mathcal{S}')$ by setting $\mathcal{R}_y^{(x)}(s') = \delta_{xy}t'_x$ for all $s' \in \mathcal{S}'$ and $x, y \in \Omega$, where $\delta$ is the Kronecker symbol. Now we see that $\mathcal{J}_y(s) := \sum_{x \in \Omega} \mathcal{R}_y^{(x)}(\mathcal{I}_x(s)) = \mathsf{A}_y(s)t'_y$ for all $s \in \mathcal{S}$ and $y \in \Omega$ defines a measure-and-prepare instrument $\mathcal{J} \in \mathrm{Ins}(\Omega, \mathcal{S}, \mathcal{S}')$ such that $\mathsf{A}^{\mathcal{J}} = \mathsf{A}$ is indecomposable and the prepared states are pure so that by Prop. 19 we have that $\mathcal{J}$ is indecomposable. Furthermore, if we define instruments $\{\mathcal{Q}^{(y)}\}_{y \in \Omega} \subset \mathrm{Ins}(\Omega, \mathcal{S}')$ by setting $\mathcal{Q}_x^{(y)}(s') = \delta_{yx}s'_y$ for all $s' \in \mathcal{S}'$ and $x, y \in \Omega$, we see that $\sum_{y \in \Omega} \mathcal{Q}_x^{(y)}(\mathcal{J}_y(s)) = \mathsf{A}_x(s)s'_x = \mathcal{I}_x(s)$ for all $s \in \mathcal{S}$ and $x \in \Omega$. Thus, $\mathcal{I} \leftrightarrow \mathcal{J}$, where $\mathcal{J}$ is indecomposable.

Now let $\mathcal{I}$ be post-processing equivalent with an indecomposable instrument $\mathcal{J} \in \mathrm{Ins}(\Lambda, \mathcal{S}, \mathcal{S}'')$. From *Prop. 20 it follows that also $\mathcal{J}$ is a measure-and-prepare instrument so that by *Prop. 19 the induced observable $\mathsf{A}^{\mathcal{J}}$ is indecomposable. Since $\mathcal{I} \leftrightarrow \mathcal{J}$, by Prop. 12 we then have that $\mathsf{A} = \mathsf{A}^{\mathcal{I}} \leftrightarrow \mathsf{A}^{\mathcal{J}}$ so that Prop. 11 implies that also $\mathsf{A}$ is indecomposable. □

# Chapter 4

# Simulation of observables

In the previous Chapter we considered how to obtain a new measurement device from a known one by the means of post-processing the output(s) of the known device. In the case of observables this meant that we are relabeling the classical measurement outcomes with conditional probabilities given by a stochastic matrix. Unlike in the case of instruments, the post-processing of observables is purely a classical process and thus does not require any additional resources to be implemented (apart from some source of randomness). Similarly to classical post-processing, the mixing of observables is also a purely classical process in this sense. The scheme of combining these two classical operations on a set of observables has been coined the *simulation of observables* [30, 107]. Simulation thus captures the idea of classically manipulating a set of observables in order to obtain some new observables.

In the beginning of this Chapter we present the simulation scheme and the structure it imposes, as was studied in Publication **II**. As an application of simulability we consider the topics of Publication **IV**, where along with providing examples of the simulation process, theories beyond the no-restriction hypothesis are considered. We argue that all operational restrictions on observables must be closed with respect to the simulation process. Other works on the simulation of observables include [108–110].

## 4.1 Simulation scheme and structure

Following Publication **II**, we start by formalizing the simulation process, and then proceed to show that the simulation of all observables can be reduced to a particular class of observables and study their structure.

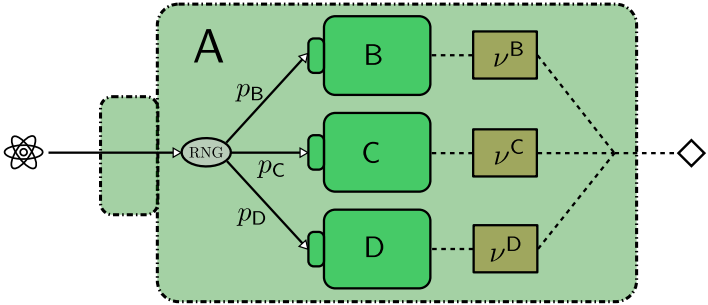

Figure 4.1: Simulation of observable A with observables B, C and D by using post-processings $\nu^B$, $\nu^C$ and $\nu^D$ and mixing weights $p_B, p_C, p_D \in [0,1]$ such that $p_B + p_C + p_D = 1$.

Let us consider a set of $n$ observables $\{B^{(i)}\}_{i=1}^n$ such that $B^{(i)} \in \mathcal{O}(\Omega_i, \mathcal{S})$ for all $i \in [n]$. The chronological process of simulation then goes as follows: after preparing a system i) we pick an observable $B^{(i)}$ with some probability $p_i$, ii) we measure the system with the observable $B^{(i)}$, iii) we obtain an outcome $x \in \Omega_i$, and iv) we perform a post-processing $\nu^{(i)} : \Omega_i \to \Omega$ for some other outcome set $\Omega$ so that the obtained outcome $x \in \Omega_i$ is mapped to outcome $y \in \Omega$ with probability $\nu_{xy}^{(i)}$. We can formalize this process with the following definition:

**Definition 14.** A set of observables $\mathcal{A} \subseteq \mathcal{O}(\mathcal{S})$ can be *simulated* by another set of observables $\mathcal{B} \subseteq \mathcal{O}(\mathcal{S})$ if for all $A \in \mathcal{A}$ with an outcome set $\Omega_A$ there exist a finite set of observables $\{B^{A,(i)}\}_{i \in [n_A]} \subset \mathcal{B}$ with outcome sets $\Omega_i$ for some $n_A \in \mathbb{N}$, a probability distribution $(p_i^A)_{i \in [n_A]}$ and row-stochastic post-processing matrices $\{\nu^{A,(i)}\}_{i \in [n_A]} \subset \mathcal{M}_{|\Omega_A|,|\Omega_i|}^{row}$ such that

$$A = \sum_{i=1}^{n_A} p_i^A \left( \nu^{A,(i)} \circ B^{A,(i)} \right). \tag{4.1}$$

The simulation scheme is depicted in Fig. 4.1. We denote the set of all observables simulable by a subset $\mathcal{B} \subseteq \mathcal{O}(\mathcal{S})$ by $\mathfrak{sim}(\mathcal{B})$, and call the observables in $\mathcal{B}$ *simulators*. We note that in the case when $\mathcal{B} = \{B\}$ for some single observable $B \in \mathcal{O}(\mathcal{S})$, the mixing part of the simulation scheme becomes trivial so that $\mathfrak{sim}(B)$ consists just of post-processings of $B$.

Considered as a map on the power set $2^{\mathcal{O}(\mathcal{S})}$ of all subsets of observables, $\mathfrak{sim}(\cdot)$ can be shown to be a *closure operator* on $\mathcal{O}(\mathcal{S})$ such that it satisfies the following properties for all subsets $\mathcal{B}, \mathcal{C} \subseteq \mathcal{O}(\mathcal{S})$:

(sim1) $\mathcal{B} \subseteq \mathfrak{sim}(\mathcal{B})$,

(sim2) $\mathfrak{sim}(\mathfrak{sim}(\mathcal{B})) = \mathfrak{sim}(\mathcal{B})$,

(sim3) $\mathcal{B} \subseteq \mathcal{C} \implies \mathfrak{sim}(\mathcal{B}) \subseteq \mathfrak{sim}(\mathcal{C})$.

If equality holds in (sim1), i.e., $\mathcal{B} = \mathfrak{sim}(\mathcal{B})$, we say that $\mathcal{B}$ is *simulation closed*. As we see from (sim2), $\mathfrak{sim}(\mathcal{B})$ is always simulation closed for any $\mathcal{B} \subseteq \mathcal{O}(\mathcal{S})$. In particular this means that $\mathfrak{sim}(\mathcal{B})$ is convex and closed with respect to post-processing.

## 4.2 Simulation irreducible observables

We have seen separately in the case of mixing and post-processing of observables that all other observables can be reduced into convex decompositions of extreme observables and post-processings of indecomposable observables. Let us now consider the task of reducing all observables into simulations of a specific class of observables, namely the *simulation irreducible observables*.

As a first (trivial) remark, we note that every observable $\mathsf{A}$ can be simulated by a set of simulators $\mathcal{B}$ whenever $\mathsf{A}$ belongs to $\mathcal{B}$. The property of those observables for which this is (essentially) the only way that they can be simulated is captured by the following definition:

**Definition 15.** An observable $\mathsf{A}$ is simulation irreducible if whenever $\mathsf{A} \in \mathfrak{sim}(\mathcal{B})$ for some subset $\mathcal{B} \subseteq \mathcal{O}(\mathcal{S})$ then there exists a single observable $\mathsf{B} \in \mathcal{B}$ such that $\mathsf{A} \leftrightarrow \mathsf{B}$.

Thus, a simulation irreducible observable can only be simulated by itself (up to post-processing equivalence). The set of simulation irreducible observables on $\mathcal{S}$ is denoted by $\mathcal{O}_{irr}(\mathcal{S})$. As was mentioned, the tasks of mixing and post-processing can be reduced to only mixing and post-processing extreme and indecomposable observables, respectively. We can show that in the task of simulating observables, all other observables can be reduced to simulations of simulation irreducible observables.

**Proposition 21.** *For every observable $\mathsf{A} \in \mathcal{O}(\mathcal{S})$ there exists a finite set of simulation irreducible observables $\mathcal{B}^{\mathsf{A}} \subset \mathcal{O}_{irr}(\mathcal{S})$ such that $\mathsf{A} \in \mathfrak{sim}(\mathcal{B}^{\mathsf{A}})$.*

The proof, presented in Publication **II**, is an adaptation of the one given in [111], and is based on the following characterization of the simulation irreducible observables:

**Proposition 22.** *An observable is simulation irreducible if and only if it is indecomposable and post-processing equivalent with an extreme observable.*

Thus, perhaps unsurprisingly, the task of simulating all observables can be reduced to that of simulating just those observables that are 'maximal' in both of its main components, namely mixing and post-processing, so that simulation irreducible observables are post-processing clean and extreme (up to post-processing equivalence).

**Example 10** (Quantum theory)**.** In quantum theory the indecomposable observables correspond to POVMs with rank-1 effects. To check if such a POVM $A$ is simulation irreducible, we can construct the minimally sufficient representative $\tilde{A}$ and check if it has linearly independent effects, i.e., if it is extreme. For the $d$-dimensional quantum theory $\mathcal{S}_d^q = \mathcal{S}(\mathcal{H})$ for some $d$-dimensional Hilbert space $\mathcal{H}$ we have that $d_{lin}(\mathcal{S}_d^q) = d^2$ so the maximal number of outcomes for an extreme simulation irreducible POVM is $d^2$. In fact, it can be shown that for any $n \in \{d, \ldots, d^2\}$ one can construct an extreme simulation irreducible POVM with $n$ outcomes [111]. For the case $n = d$, we can take any orthonormal basis $\{\varphi_i\}_{i=1}^d$ of $\mathcal{H}$ and define the (projective) extreme rank-1 POVM $A \in \mathcal{O}([d], \mathcal{H})$ as $A(i) := |\varphi_i\rangle\langle\varphi_i|$ for all $i \in [d]$. Since there is a continuum of bases of $\mathcal{H}$ and since the previously defined extreme rank-1 POVMs related to two different bases of $\mathcal{H}$ are never post-processing equivalent (see Prop. 11), we see that there exists a continuum of simulation irreducible POVMs on $\mathcal{S}_d^q$.

By characterizing the extreme rays of the positive dual cone $\mathcal{V}_+^* \subset \mathcal{V}^*$ of the positive cone $\mathcal{V}_+ \subset \mathcal{V}$ of which a state space $\mathcal{S} \subset \mathcal{V}_+$ is a base of, one is able to characterize the cone $\mathcal{V}_+^*$ and the effect space $\mathcal{E}(\mathcal{S})$ (under the no-restriction hypothesis) given by Eq. (1.2) as $\mathcal{E}(\mathcal{S}) = \mathcal{V}_+^* \cap (u - \mathcal{V}_+^*)$. Because indecomposable effects correspond to the extreme rays of $\mathcal{V}_+^*$, the characterization of the indecomposable and in particular the simulation irreducible observables provides valuable information about the geometric structure of the effect and state spaces. On the other hand, in some simple state spaces the geometric structure of the effect space is simple as well and the characterization of the simulation irreducible observables becomes easy.

**Example 11** (Polygon state spaces)**.** In the polygon state spaces $\mathcal{S}_n$ we can show the following: If $n = 2m$ for some $m \in \mathbb{N}$, i.e., $n$ is even, then there are exactly $m$ dichotomic and $m(m-1)(m-2)/3$ trichotomic extreme simulation irreducible observables on $\mathcal{S}_n$. Thus, in particular, every observable on an even polygon can be simulated with observables with three outcomes, and in the case of the square state space $\mathcal{S}_4$ there are exactly two dichotomic simulation irreducible observables that can simulate every other observable on $\mathcal{S}_4$. On the other hand, if $n = 2m + 1$ for some $m \in \mathbb{N}$, i.e., $n$ is odd, then there are exactly $m(m+1)(2m+1)/6$ trichotomic extreme simulation irreducible observables on $\mathcal{S}_n$. In odd polygon state spaces there are no dichotomic simulation irreducible observables. Details about the simulation irreducible observables on polygons can be found in Publication **II**.

As a demonstration of the usefulness of finding the simulation irreducible observables in order to characterize the structure of the theory, we can show the following important result.

**Proposition 23.** *A state space is non-classical if and only if there exists at least two post-processing inequivalent simulation irreducible observables.*

Geometrically we can reformulate the previous result as follows: a state space $\mathcal{S}$ is a simplex if and only if there exists only one observable on $\mathcal{S}$ with linearly independent indecomposable effects. Given Prop. 21, we see that in classical state spaces every observable can be simulated, i.e., in the case of one observable, post-processed, from a single observable. This property, commonly called the *joint measurability* (or *compatibility*) of all observables, is a unique feature of classical theories [28]. In all other theories there exist (at least) two observables which are not jointly measurable (i.e., they are incompatible,) so that they cannot be post-processed from any single observable. We consider the connections between compatibility and simulability more closely in the last Chapter of this thesis.

## 4.3 Simulation closed restrictions of observables

When we introduced the set of effects in Section 1.2.2 we assumed the no-restriction hypothesis according to which all mathematically valid effects (affine functionals $e : \mathcal{S} \to [0, 1]$ that give probabilities on states) are also physical effects in every theory. Usually the no-restriction hypothesis is assumed for mathematical convenience and is loosely justified by the fact that it holds in classical and quantum theory. However, there are no universal operational principles that would justify the validity of the no-restriction hypothesis in every theory.

By placing restrictions on the effects one is able to define new interesting models and show that it can affect the way composite systems may be formed [112]. Furthermore, recently it has been shown that the no-restriction hypothesis may play a crucial role in singling out the quantum correlations from other non-signalling theories [113]. Other works beyond the no-restriction hypothesis include [18, 114, 115].

However, the common theme in previous works is that the restrictions have been focused solely on effects. In our work in Publication **IV** we extend the restrictions to the set of observables and see that there are restrictions that are not induced solely by some restricted set of effects. Furthermore, we argue that every such restriction on observables, represented by some subset of observables, must be simulation closed in order to be operationally valid. Namely, if we determine our set of physical observables, by the process of simulation we are able to produce other observables that by the operational nature of simulation must also be physically feasible. We classify the operational restrictions in three disjoint classes based on whether the restriction takes place in the level of effects, observables or both.

### 4.3.1 Three types of operational restrictions

An *operational restriction* on observables is represented by a simulation closed subset $\tilde{\mathcal{O}} \subset \mathcal{O}(\mathcal{S})$ of observables so that $\tilde{\mathcal{O}} = \mathfrak{sim}(\tilde{\mathcal{O}})$. We note that every such (nonempty) restriction $\tilde{\mathcal{O}}$ contains the set of trivial observables $\mathcal{T}(\mathcal{S})$ since any trivial observable can be post-processed from any other observable.

By considering an operational restriction given by a simulation closed subset $\tilde{\mathcal{O}}$, we can consider which are the physically feasible effects given by the restriction. We use the following notation: for a subset $\tilde{\mathcal{O}} \subset \mathcal{O}(\mathcal{S})$ we denote by $\mathcal{E}_{\tilde{\mathcal{O}}} \subseteq \mathcal{E}(\mathcal{S})$ the set of all effects $e \in \mathcal{E}(\mathcal{S})$ such that $e \in \operatorname{ran}(\mathsf{A})$[1] for some $\mathsf{A} \in \tilde{\mathcal{O}}$. Physically this means that $\mathcal{E}_{\tilde{\mathcal{O}}}$ consists of those effects that can be obtained as relabelings from the effects of the observables.

On the other hand, we can consider this as we are given a set of physically feasible effects $\tilde{\mathcal{E}}$ and the operational restriction is given by the set

---

[1] We recall that for an observable $\mathsf{A} \in \mathcal{O}(\Omega, \mathcal{S})$ we have $\operatorname{ran}(\mathsf{A}) = \left\{ \sum_{x \in \tilde{\Omega}} \mathsf{A}_x \mid \tilde{\Omega} \subseteq \Omega \right\}$.

of observables that can be constructed using those effects. Formally, for a subset $\tilde{\mathcal{E}} \subset \mathcal{E}(\mathcal{S})$ we denote by $\mathcal{O}_{\tilde{\mathcal{E}}} \subset \mathcal{O}(\mathcal{S})$ the set of all observables $\mathsf{A} \in \mathcal{O}(\mathcal{S})$ such that $\operatorname{ran}(\mathsf{A}) \subset \tilde{\mathcal{E}}$. Naturally, if we are to consider $\mathcal{O}_{\tilde{\mathcal{E}}}$ as an operational restriction, we require it to be simulation closed. Furthermore, we require the following consistency conditions to hold:

(E1) $u \in \tilde{\mathcal{E}}$ as the unit effect is an integral part of the definition of an observable, and

(E2) for every $e \in \tilde{\mathcal{E}}$ there exists $\mathsf{A} \in \mathcal{O}_{\tilde{\mathcal{E}}}$ such that $e \in \operatorname{ran}(\mathsf{A})$, i.e., for every physical effect there exists a physical observable that we can implement it as a part of.

We are ready to classify the operational restrictions described by a simulation closed set $\tilde{\mathcal{O}} \subset \mathcal{O}(\mathcal{S})$ into three disjoint cases. First, we can have that

(R1) $\tilde{\mathcal{O}} = \mathcal{O}_{\tilde{\mathcal{E}}}$ for some $\tilde{\mathcal{E}} \subset \mathcal{E}(\mathcal{S})$.

The restrictions in class (R1) thus consist of those that are completely induced by restrictions on effects. We note that it is not guaranteed that any effect restriction $\tilde{\mathcal{E}}$ induces a simulation closed restriction $\mathcal{O}_{\tilde{\mathcal{E}}}$ on observables. We characterize all suitable effect restrictions (that satisfy the consistency conditions (E1) and (E2)) in the next section. However, it is easy to see that the effect restriction inducing a restriction of type (R1) is always unique as is shown by the following result:

**Proposition 24.** *Let $\tilde{\mathcal{E}} \subset \mathcal{E}(\mathcal{S})$ be an effect restriction satisfying the consistency conditions (E1) and (E2). Then $\mathcal{E}_{\mathcal{O}_{\tilde{\mathcal{E}}}} = \tilde{\mathcal{E}}$.*

Second, in addition to effect restrictions we can have

(R2) $\mathcal{E}_{\tilde{\mathcal{O}}} = \mathcal{E}(\mathcal{S})$ but $\tilde{\mathcal{O}} \neq \mathcal{O}(\mathcal{S})$.

Restrictions of type (R2) thus do not restrict the set of effects in any way but rather how those effects can be formed into observables. We note that if $\tilde{\mathcal{O}}$ is a restriction of type (R2) it cannot be of type (R1): namely, otherwise by Prop. 24 we would have that the effect restriction $\tilde{\mathcal{E}}$ inducing $\tilde{\mathcal{O}} = \mathcal{O}_{\tilde{\mathcal{E}}}$ would have to be $\tilde{\mathcal{E}} = \mathcal{E}(\mathcal{S})$ so that $\tilde{\mathcal{O}} = \mathcal{O}(\mathcal{S})$.

Third, we can finally have that the restriction is not of type (R1) or (R2) so that (R3) $\mathcal{E}_{\tilde{\mathcal{O}}} \subset \mathcal{E}(\mathcal{S})$ and $\tilde{\mathcal{O}} \neq \mathcal{O}_{\tilde{\mathcal{E}}}$ for any $\tilde{\mathcal{E}} \subset \mathcal{E}(\mathcal{S})$.

Hence, for restrictions in class (R3) there are restrictions already on the level of effects but also some restrictions that only manifest in the level of observables. We continue to consider all three distinct types of restrictions separately.

### 4.3.2   (R1) Restrictions purely on effects

The restrictions in class (R1) are purely induced by restrictions on effects. However, in order for an effect restriction $\tilde{\mathcal{E}}$ to induce a simulation closed restriction $\tilde{\mathcal{O}}$ on observables, we must choose $\tilde{\mathcal{E}}$ suitably. This is the content of the following Proposition:

**Proposition 25.** *Let $\tilde{\mathcal{E}} \subset \mathcal{E}(\mathcal{S})$ be an effect restriction satisfying the consistency conditions (E1) and (E2). Then $\mathcal{O}_{\tilde{\mathcal{E}}}$ is simulation closed if and only if $\tilde{\mathcal{E}}$ is convex.*

The intuition behind this result is the following: from the definition of $\mathcal{O}_{\tilde{\mathcal{E}}}$ it follows that if $\mathsf{A} \in \mathcal{O}_{\tilde{\mathcal{E}}}$, then $\mathsf{A}_x + \mathsf{A}_y \in \tilde{\mathcal{E}}$ for all outcomes $x \neq y$ so that in particular $\mathcal{O}_{\tilde{\mathcal{E}}}$ is already closed with respect to relabelings, which are the extreme elements in the set of post-processing matrices. Since simulation is just a combination of mixing and post-processing, it can be shown that the convexity of $\tilde{\mathcal{E}}$ is both a necessary and sufficient condition to make $\mathcal{O}_{\tilde{\mathcal{E}}}$ closed with respect to mixing and general post-processings.

**Convex effect subalgebras**

When we consider the effect space $\mathcal{E}(\mathcal{S})$ of a state space $\mathcal{S}$ in the GPT framework, the effects are defined through their action on the states. If we want to abstractly define and consider effects as an elementary concept without relying on the concept of a state, the *effect algebra* [116] (see also [117–120]) is a natural notion for this. In this section we will consider abstract effect algebras only conceptually without going into the mathematical details. The formal mathematical definitions and treatment can be found in Publication **IV**.

Conceptually, an effect algebra $\mathcal{E}$ is a non-empty set consisting of a collection of events $e \in \mathcal{E}$ and it includes a partial operation $\oplus$ that describes the joining of such events so that $e \oplus f$ is a new event corresponding to the joining of events $e, f \in \mathcal{E}$ whenever it is defined. Furthermore, we require that there are two special events of which one, o, corresponds to the event that never happens and the other, 1, to the event that always happens.

In the case of effect algebras we are still considering the operational notion of events and the joining of events. Thus, one of the most important operational notions behind the present GPT framework, namely mixing, is a valid and reasonable concept to study in the case of effect algebras as well. For an effect algebra $\mathcal{E}$, and for each $\alpha \in [0,1]$ and $e \in \mathcal{E}$ we define a process of forming a new element $\alpha e$ which is interpreted as a splitting of the event $e$ into two events, $\alpha e$ and $(1-\alpha)e$. An effect algebra with this property (along with some natural consistency conditions) is called a *convex effect algebra* [117]. In particular, if we have the splittings $\alpha e$ and $(1-\alpha)f$ for some $\alpha \in [0,1]$ and $e, f \in \mathcal{E}$ in a convex effect algebra $\mathcal{E}$, then $\alpha e \oplus (1-\alpha)f \in \mathcal{E}$ also holds such that we can indeed implement mixtures of events.

In order to consider restrictions on an effect algebra $\mathcal{E}$ we need to consider subsets of $\mathcal{E}$ that would correspond to physically feasible effects. As with any algebraic structure, the natural subsets to consider are the ones which inherit the structure of the original set. Thus, for an (convex) effect algebra $\mathcal{E}$ we focus on the subsets of $\mathcal{E}$ that are themselves (convex) effect algebras, namely the (*convex) effect subalgebras* of $\mathcal{E}$ [120].

Coming back to the GPT framework, it can be confirmed that the effect space $\mathcal{E}(\mathcal{S}) = \mathcal{V}_+^* \cap (u - \mathcal{V}_+^*)$ of a state space $\mathcal{S} \subset \mathcal{V}_+ \subset \mathcal{V}$ always forms a convex effect algebra. On the other hand it can be shown [117] that *for every convex effect algebra $\mathcal{E}$ there exists a real vector space $\mathcal{W}$, a proper generating cone $\mathcal{W}_+$ and a nonzero element $u \in \mathcal{W}_+$ such that $\mathcal{E}$ is affinely isomorphic to $\mathcal{W}_+ \cap (u - \mathcal{W}_+)$.*

In this linear representation we can show the following characterization:

**Proposition 26.** *Let $\mathcal{E}(\mathcal{S})$ be an effect space of a state space $\mathcal{S} \subset \mathcal{V}_+ \subset \mathcal{V}$. A subset $\tilde{\mathcal{E}} \subseteq \mathcal{E}(\mathcal{S})$ is a convex effect subalgebra of $\mathcal{E}(\mathcal{S})$ if and only if $\tilde{\mathcal{E}} = U \cap \mathcal{E}(\mathcal{S})$ for some linear subspace $U \subseteq \mathcal{V}^*$ such that $u \in U$.*

Using this characterization, it is easy to confirm that every convex effect subalgebra $\tilde{\mathcal{E}}$ of an effect space $\mathcal{E}(\mathcal{S})$ induces a simulation-closed restriction $\mathcal{O}_{\tilde{\mathcal{E}}} \subseteq \mathcal{O}(\mathcal{S})$. Namely, it is straightforward to see that $\tilde{\mathcal{E}}$ is a convex subset of $\mathcal{E}(\mathcal{S})$ satisfying the consistency conditions (E1) and (E2) so that, from Prop. 25, it follows that $\mathcal{O}_{\tilde{\mathcal{E}}}$ is simulation closed.

However, we point out that not every convex subset $\tilde{\mathcal{E}} \subset \mathcal{E}(\mathcal{S})$ is a convex effect subalgebra of $\mathcal{E}(\mathcal{S})$. This is demonstrated in the following

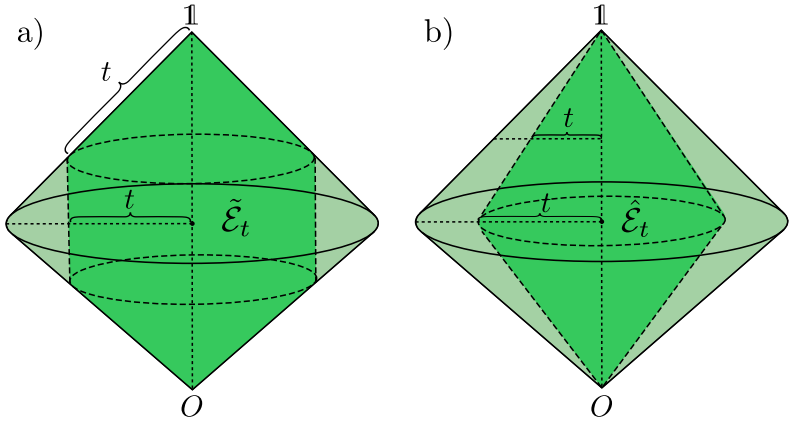

Figure 4.2: Noise restriction a) with random noise and b) with depolarizing noise on a three-dimensional cross section of the qubit effect space (also the same as the rebit effect space).

section along with providing concrete examples of restrictions of type (R1).

**Noise restrictions on effects**

Let $\mathcal{S} \subset \mathcal{V}_+ \subset \mathcal{V}$ be a state space. For each $t \in [0, 1]$ let us define

$$\tilde{\mathcal{E}}_t = \{te + (1-t)pu \,|\, e \in \mathcal{E}(\mathcal{S}),\ p \in [0,1]\}. \tag{4.2}$$

The interpretation of $\tilde{\mathcal{E}}_t$ is that it describes the set of noisy effects, where the amount of noise mixed with an effect is described by the noise parameter $t$. We note that in this case we are mixing effects with random noise given by the parameter $p \in [0, 1]$. For $t = 1$ we clearly have $\tilde{\mathcal{E}}_1 = \mathcal{E}(\mathcal{S})$, and for $t = 0$ we have that $\tilde{\mathcal{E}}_0 = \{pu \,|\, p \in [0,1]\}$ which consists of completely noisy trivial effects. We note that $\tilde{\mathcal{E}}_1$ and $\tilde{\mathcal{E}}_0$ are both trivial effect subalgebras of $\mathcal{E}(\mathcal{S})$. On the other hand, clearly $\tilde{\mathcal{E}}_t$ is convex for all $t \in [0, 1]$ and it satisfies the consistency conditions (E1) and (E2) so that by Prop. 25 the induced restriction $\mathcal{O}_{\tilde{\mathcal{E}}_t}$ is simulation closed. The noise restriction is depicted in Fig. 4.2 a).

Let $t \neq 0$. Clearly $\tilde{\mathcal{E}}_t$ includes all elements of the form $te$ for all $e \in \mathcal{E}(\mathcal{S})$. If we choose $e$ to be an extreme ray of the dual cone $\mathcal{V}_+^*$ so that $te$ spans the same ray, it follows that because $\mathcal{E}(\mathcal{S})$ spans $\mathcal{V}_+^*$ and because $\mathcal{V}_+^*$ is a generating cone we must have that also $\tilde{\mathcal{E}}_t$ spans the whole dual space $\mathcal{V}^*$. Suppose that $\tilde{\mathcal{E}}_t$ is a convex subalgebra of $\mathcal{E}(\mathcal{S})$. Then by Prop. 26 we must have that $\tilde{\mathcal{E}}_t = U_t \cap \mathcal{E}(\mathcal{S})$ for some linear subspace $U_t \subseteq \mathcal{V}^*$ such that $u \in U_t$. It then follows that also $U_t$ must span $\mathcal{V}^*$ so that in fact $U_t = \mathcal{V}^*$ and $\tilde{\mathcal{E}}_t = \mathcal{E}(\mathcal{S})$. Hence, $\tilde{\mathcal{E}}_t$ is a convex subalgebra of $\mathcal{E}(\mathcal{S})$ if and only if $t = 0$ or $t = 1$.

**Example 12** (Point-symmetric state spaces). Next we will construct another noise restriction on the set of effects on point-symmetric state spaces. The restriction is analogous to the action of depolarization in quantum theory (see, e.g., [63]) in the sense that it constricts the effect space, and it can thus be interpreted as adding specific noise to the effects and measurements. This restriction also generalizes the noisy boxworld restriction that was studied in [112].

Let us take $t \in [0, 1]$ and consider the restricted set of effects

$$\hat{\mathcal{E}}_t = \{te + (1-t)e(s_0)u \,|\, e \in \mathcal{E}(\mathcal{S})\} \quad (4.3)$$

on a $d$-dimensional point-symmetric state space $\mathcal{S} \subset \mathbb{R}^{d+1}$ with an inversion point $s_0$. Similar to before, $\hat{\mathcal{E}}_t \subseteq \mathcal{E}(\mathcal{S})$ for all $t \in (0, 1]$ and the equality holds only if $t = 1$, and $\hat{\mathcal{E}}_0 = \{pu \,|\, p \in [0, 1]\}$. We note that, contrary to $\tilde{\mathcal{E}}_t$ as defined in Eq. (4.2), here we are mixing noise that is dependent on each effect. It follows that $\hat{\mathcal{E}}_t \subseteq \tilde{\mathcal{E}}_t$ for all $t \in [0, 1]$. Again we see that $\hat{\mathcal{E}}_t$ is convex for all $t \in [0, 1]$ and it satisfies the consistency conditions (E1) and (E2) so that by Prop. 25 the induced restriction $\mathcal{O}_{\hat{\mathcal{E}}_t}$ is simulation closed.

If we define the *depolarising map* $\Phi_t : \mathbb{R}^{d+1} \to \mathbb{R}^{d+1}$ as $\Phi_t(x) = tx + (1-t)x(s_0)u$, we see that $\hat{\mathcal{E}}_t = \Phi_t(\mathcal{E}(\mathcal{S}))$ for each $t \in [0, 1]$. For $t \in (0, 1]$, it is straigthforward to verify that $\Phi_t$ is a positive-linear isomorphism between $\mathcal{E}(\mathcal{S})$ and $\hat{\mathcal{E}}_t$ that preserves the unit effect $u$. From this isomorphism it follows that $\hat{\mathcal{E}}_t$ spans the same space as $\mathcal{E}(\mathcal{S})$ which is the whole $\mathbb{R}^{d+1}$. Suppose that $t \neq 0$ and $\hat{\mathcal{E}}_t$ is a convex subalgebra of $\mathcal{E}(\mathcal{S})$. Since $\hat{\mathcal{E}}_t$ spans the whole $\mathbb{R}^{d+1}$, the only subspace $U$ from Prop. 26 such that $\hat{\mathcal{E}}_t = U \cap \mathcal{E}(\mathcal{S})$ is $U = \mathbb{R}^{d+1}$. Thus, again $\hat{\mathcal{E}}_t$ is a convex subalgebra of $\mathcal{E}(\mathcal{S})$ if and only if $t = 0$ or $t = 1$. We depict the depolarizing noise restriction in Fig. 4.2 b).

### 4.3.3 (R2) Restrictions purely on observables

Let us now consider restrictions of type (R2), i.e., simulation closed restrictions $\tilde{\mathcal{O}} \subset \mathcal{O}(\mathcal{S})$ such that $\mathcal{E}_{\tilde{\mathcal{O}}} = \mathcal{E}(\mathcal{S})$. Hence, in this class the effects are not restricted in any way but the process of obtaining physical observables from effects is.

**Effectively dichotomic observables**

As an example of the simulation process, and as a restriction of type (R2), we consider a scenario where we restrict the number of outcomes of our simulators. This type of restriction may happen for example in experimental set-ups where one may only be able to directly implement measurement devices with a specific number of outcomes (for example measurements on superconducting qubits and polarized photons). However, given these devices, one can still consider simulating other devices with possibly more outcomes.

If an observable $\mathsf{A}$ can be simulated with observables that have at most $n$, we call $\mathsf{A}$ *effectively n-tomic*. Formally we define the set of $n$-tomic observables $\mathcal{O}_{n-eff}(\mathcal{S})$ by setting $\mathcal{O}_{n-eff}(\mathcal{S}) := \mathfrak{sim}(\mathcal{O}([n], \mathcal{S}))$, where we consider the inclusion $\mathcal{O}([n], \mathcal{S}) \subset \mathcal{O}([n+1], \mathcal{S})$ by adding zero effects to the observables in $\mathcal{O}([n], \mathcal{S})$. It is then also clear that $\mathcal{O}_{n-eff}(\mathcal{S}) \subset \mathcal{O}_{(n+1)-eff}(\mathcal{S})$. In the simplest case we are restricting ourselves to using only observables with two outcomes which results in the set of effectively dichotomic observables $\mathcal{O}_{2-eff}(\mathcal{S})$. Foundational and experimental motivation for studying $n$-tomicity can be found in [108, 121, 122].

If we now have any restriction $\tilde{\mathcal{O}} \subset \mathcal{O}(\mathcal{S})$ of type (R2), then by definition $\mathcal{E}_{\tilde{\mathcal{O}}} = \mathcal{E}(\mathcal{S})$. The definition of $\mathcal{E}_{\tilde{\mathcal{O}}}$ therefore implies that for each effect $e \in \mathcal{E}(\mathcal{S}) = \mathcal{E}_{\tilde{\mathcal{O}}}$ there exists an observable $\mathsf{A} \in \tilde{\mathcal{O}}$ such that $e \in \mathrm{ran}(\mathsf{A})$. In particular this means that for each effect $e \in \mathcal{E}(\mathcal{S})$, the dichotomic observable with effects $e$ and $u - e$ must be in $\tilde{\mathcal{O}}$. Thus, $\mathcal{O}([2], \mathcal{S}) \subset \tilde{\mathcal{O}}$ such that from the simulation closedness of $\tilde{\mathcal{O}}$ it follows that also $\mathcal{O}_{2-eff}(\mathcal{S}) \subseteq \tilde{\mathcal{O}}$. We have shown the following:

**Proposition 27.** *If $\tilde{\mathcal{O}} \subset \mathcal{O}(\mathcal{S})$ is of type (R2), then $\mathcal{O}_{2-eff}(\mathcal{S}) \subseteq \tilde{\mathcal{O}}$.*

In some theories it can happen that $\mathcal{O}_{2-eff}(\mathcal{S}) = \mathcal{O}(\mathcal{S})$ (for example the square state space). Hence, we cannot characterize the restriction class (R2) any further in general, and the specific nature of these types of restrictions is different in different theories.

However, what we will do is explore a few necessary and sufficient conditions for an observable to be effectively dichotomic. For an effect $e \in \mathcal{E}(\mathcal{S})$ let us denote $\lambda_{\max}(e) = \sup_{s \in \mathcal{S}} e(s)$, where the supremum is always attained due to the compactness of $\mathcal{S}$. We also remember from Section 2.3 that $\lambda_{\max}(\mathsf{A})$ for an observable $\mathsf{A} \in \mathcal{O}(\Omega, \mathcal{S})$ is then $\lambda_{\max}(\mathsf{A}) = \sum_{x \in \Omega} \lambda_{\max}(\mathsf{A}_x)$. Now we can show the following.

**Proposition 28.** *Let $\mathsf{A} \in \mathcal{O}(\Omega, \mathcal{S})$.*

   a) *If there exists $y \in \Omega$ such that $\lambda_{\max}(\mathsf{A}) - \lambda_{\max}(\mathsf{A}_y) \leq 1$, then $\mathsf{A}$ is effectively dichotomic.*

   b) *If $\lambda_{\max}(\mathsf{A}) > 2$, then $\mathsf{A}$ is not effectively dichotomic.*

   c) *If $\{\mathsf{A}_y\}_{y \in \tilde{\Omega}}$ consists of indecomposable effects for some $\tilde{\Omega} \subseteq \Omega$ such that $t\mathsf{A}_y \neq \mathsf{A}_z$ and $t\mathsf{A}_y + r\mathsf{A}_z \neq u$ for all $y, z \in \tilde{\Omega}$ and $t, r > 0$, and furthermore $\sum_{x \in \tilde{\Omega}} \lambda_{\max}(\mathsf{A}_x) > 1$, then $\mathsf{A}$ is not effectively dichotomic.*

The first result *a)* is a direct generalization of a result in [107]. As an application of effectively dichotomic observables, let us apply them to an operational task similar to those considered in Chapter 2.

**Unambiguous discrimination of two pure qubit states**

Let us consider the task of the *unambiguous discrimination of states* (see, e.g., [123] for details in the quantum case). The task consists of identifying an unkown state $s$ out of some set of possible states $\{s_i\}_{i=1}^n \in \mathcal{S}$ without an error. This means that we are looking for an observable $\mathsf{A} \in \mathcal{O}(\mathcal{S})$ such that $\mathsf{A}_j(s_j) > 0$ and $\mathsf{A}_j(s_i) = 0$ for all $i \neq j$, $i, j \in [n]$. In this way if we measure $s$ with observable $\mathsf{A}$ and detect the outcome $j$, we know that $s = s_j$. Furthermore, if $\mathsf{A}_j(s_j) = 1$ so that $\mathsf{A} \in \mathcal{O}([n], \mathcal{S})$, the states $\{s_i\}_{i=1}^n$ can be *perfectly discriminated*. On the other hand, if the states $\{s_i\}_{i=1}^n$ can only be unambiguously dicriminated but not perfectly, then we can have an inconclusive outcome ? such that $\mathsf{A} \in \mathcal{O}([n] \cup \{?\}, \mathcal{S})$ and $\mathsf{A}_? = u - \sum_{j=1}^n \mathsf{A}_j$. Given that the states $\{s_i\}_{i=1}^n$ are assigned with prior probabilities $(p_i)_{i=1}^n$, the average success probability for the unambiguous discrimination of the states $\{s_i\}_{i=1}^n$ using an observable $\mathsf{A} \in \mathcal{O}([n] \cup \{?\}, \mathcal{S})$ is then $p_{succ}(\mathsf{A}) = \sum_{j=1}^n p_j \mathsf{A}_j(s_j)$. The optimal success probability for the states is thus obtained by optimizing over all the observables, $p_{succ} = \sup_{\mathsf{A} \in \mathcal{O}(\mathcal{S})} p_{succ}(\mathsf{A})$.

In quantum theory $\mathcal{S}(\mathcal{H})$ it is known that two pure states $\varrho_1 = |\varphi_1\rangle\langle\varphi_1|$ and $\varrho_2 = |\varphi_2\rangle\langle\varphi_2|$ for some unit vectors $\varphi_1, \varphi_2 \in \mathcal{H}$, $\varphi_1 \neq \varphi_2$, can be discriminated perfectly if and only if the states are orthogonal, i.e., $\langle\varphi_1|\varphi_2\rangle = 0$ (see, e.g., [63]). Let us now focus on the case when $\varrho_1, \varrho_2 \in \mathcal{S}(\mathbb{C}^2)$ are two

non-orthogonal pure qubit states. Thus, the task is to a find 3-outcome POVM $A \in \mathcal{O}(\{1,2,?\},\mathbb{C}^2)$ such that $\text{tr}\,[A(1)\varrho_2] = \text{tr}\,[A(2)\varrho_1] = 0$. These conditions imply that

$$A(1) = q_1(\mathbb{1}_2 - |\varphi_2\rangle\langle\varphi_2|), \quad A(2) = q_2(\mathbb{1}_2 - |\varphi_1\rangle\langle\varphi_1|) \qquad (4.4)$$

for some $q_1, q_2 > 0$ such that $A(?) = \mathbb{1}_2 - A(1) - A(2) \geq O$.

Let us suppose now that the discriminating POVM A is effectively dichotomic. We note that both $A(1)$ and $A(2)$ are rank-1, i.e., indecomposable, and that $q_1 = \lambda_{\max}(A(1))$ and $q_2 = \lambda_{\max}(A(2))$. Clearly, since $\varphi_1 \neq \varphi_2$, we have that $A(1) \neq tA(2)$ for any $t > 0$. Suppose that $t_1 A(1) + t_2 A(2) = \mathbb{1}_2$ for some $t_1, t_2 > 0$. Since $A(1)$ and $A(2)$ are rank-1, the previous expression holds if and only if $\varrho_1$ and $\varrho_2$ are orthogonal and $t_1 = 1/q_1$ and $t_2 = 1/q_2$. Since this is not the case, we conclude that the conditions of Prop. 28 *c)* are satisfied so that by the effective dichotomicity of A we have that

$$q_1 + q_2 = \lambda_{\max}(A(1)) + \lambda_{\max}(A(2)) \leq 1. \qquad (4.5)$$

If we use equal a priori probabilities $p_1 = p_2 = 1/2$ for the states $\varrho_1$ and $\varrho_2$, we see that the average success probability then reads

$$p_{succ}(A) = \frac{1}{2}\text{tr}\,[A(1)\varrho_1] + \frac{1}{2}\text{tr}\,[A(2)\varrho_2] = \frac{q_1+q_2}{2}\left(1 - |\langle\varphi_1|\varphi_2\rangle|^2\right) \qquad (4.6)$$

$$\leq \frac{1}{2}\left(1 - |\langle\varphi_1|\varphi_2\rangle|^2\right) \qquad (4.7)$$

for any effectively dichotomic POVM $A \in \mathcal{O}_{2-eff}(\{1,2,?\},\mathbb{C}^2)$.

However, it is well-known that when optimized over all POVMs in $\mathcal{O}(\{1,2,?\},\mathbb{C}^2)$, the optimal success probability is found to be $p_{succ} = p_{succ}(A) = 1 - |\langle\varphi_1|\varphi_2\rangle|$ with the optimal POVM $A$ defined as in Eq. (4.4) with $q_1 = q_2 = 1 + |\langle\varphi_1|\varphi_2\rangle|$ (see, e.g., [63]). Since $\frac{1}{2}\left(1 - |\langle\varphi_1|\varphi_2\rangle|^2\right) \leq 1 - |\langle\varphi_1|\varphi_2\rangle|$ where the equality holds if and only if $\varrho_1 = \varrho_2$, we conclude that *the restriction to use only effectively dichotomic POVMs results in a decrease in the success probability of the unambiguous discrimination of two non-orthogonal pure qubit states.*

### 4.3.4 (R3) Restrictions on both effects and observables

Lastly we focus on the restrictions in the class (R3) for which the restriction is not induced by any restriction on effects but which still restricts the set of physical effects. Because (R3) is defined as restrictions that are not (R1) or (R2), and since the restrictions in (R2) are dependent of the theory, we cannot give a convenient characterization for (R3) either in all different theories. Thus, we will focus only on providing an example of a restriction of type (R3). We see that we can provide such an example by generalizing one of the noisy effect restrictions that were considered earlier.

**Noise restrictions on observables**

In Section 4.3.2 we introduced noise restriction on effects resulting in restrictions of type (R1). Let us generalize these restrictions to observables. Let $\mathcal{T} = \mathcal{T}(\mathcal{S})$ be the set of trivial observables on a state space $\mathcal{S}$. Following Eq. (4.2), let us define

$$\tilde{\mathcal{O}}_t = \{t\mathsf{A} + (1-t)\mathsf{T} \,|\, \mathsf{A} \in \mathcal{O}(\mathcal{S}),\ \mathsf{T} \in \mathcal{T}(\mathcal{S})\}. \tag{4.8}$$

The physical interpretation of an observable $t\mathsf{A}+(1-t)\mathsf{T} \in \tilde{\mathcal{O}}_t$ is that it is a noisy version of the observable $\mathsf{A} \in \mathcal{O}(\Omega, \mathcal{S})$, where the magnitude and the distribution of the noise are given by the noise parameter $t$ and probability distribution $(p_x)_{x \in \Omega}$ respectively, where $p_x = \mathsf{T}_x(s)$ for all $s \in \mathcal{S}$ and $x \in \Omega$. Clearly, for $t = 0$ and $t = 1$ we have $\tilde{\mathcal{O}}_0 = \mathcal{T}(\mathcal{S})$ and $\tilde{\mathcal{O}}_1 = \mathcal{O}(\mathcal{S})$. It is straightforward to see that $\tilde{\mathcal{O}}_t$ is simulation closed for all $t \in [0,1]$ so that it is a valid operational restriction. We will show that $\tilde{\mathcal{O}}_t$ belongs to the class (R3) for all $t \in (0,1)$.

If we consider the set of physically feasible effects $\mathcal{E}_{\tilde{\mathcal{O}}_t}$ given by the restriction $\tilde{\mathcal{O}}_t$, we see that $\mathcal{E}_{\tilde{\mathcal{O}}_t} = \tilde{\mathcal{E}}_t$, where $\tilde{\mathcal{E}}_t$ is given by Eq. (4.2). Thus, if $t \in (0,1)$, $\mathcal{E}_{\tilde{\mathcal{O}}_t} \neq \mathcal{E}(\mathcal{S})$ and so $\tilde{\mathcal{O}}_t$ cannot be a restriction of type (R2). On the other hand, if we assume that it is of type (R1), then from Prop. 24 it follows that the effect restriction $\tilde{\mathcal{E}}$ inducing $\tilde{\mathcal{O}}_t$ must be $\tilde{\mathcal{E}} = \mathcal{E}_{\tilde{\mathcal{O}}_t} = \tilde{\mathcal{E}}_t$. Thus, if we can show that there exists an observable $\mathsf{A} \in \mathcal{O}_{\tilde{\mathcal{E}}_t}$ such that $\mathsf{A} \notin \tilde{\mathcal{O}}_t$, then we see that the restriction $\tilde{\mathcal{O}}_t$ cannot be of type (R1), which would imply that it must be (R3).

By using the noise content $w(\,\cdot\,;\mathcal{T})$, we see that $\tilde{\mathcal{O}}_t$ consists exactly of those observables $\mathsf{B} \in \mathcal{O}(\mathcal{S})$ with $w(\mathsf{B};\mathcal{T}) \geq 1-t$. On the other hand $\mathcal{O}_{\tilde{\mathcal{E}}_t}$ consists of observables $\mathsf{A} \in \mathcal{O}(\Omega, \mathcal{S})$ of the form $\mathsf{A}_x = ta_x + (1-t)r_x u$ for some $a_x \in \mathcal{E}(\mathcal{S})$ and $r_x \in [0,1]$ for all $x \in \Omega$. The construction of an observable $\mathsf{A} \in \mathcal{O}_{\tilde{\mathcal{E}}_t}$ such that $\mathsf{A} \notin \tilde{\mathcal{O}}_t$ presented in Publication **IV** is rather cumbersome, but the main point is that we do not have to necessarily have that $\sum_x a_x = u$ and $\sum_x r_x = 1$ as we would have if $\mathsf{A} \in \tilde{\mathcal{O}}_t$. It turns out that we can actually choose the effects $a_x$ and the positive numbers $r_x$ such that $\sum_x \inf_{s \in \mathcal{S}} a_x(s) = 0$ and $\sum_x a_x = \alpha u$ for some $\alpha > 1$ as well as $r := \sum_x r_x < 1$. From the properties of the noise content it follows that $w(\mathsf{A}; \mathcal{T}) = (1-t)r < 1-t$ so that $\mathsf{A} \notin \tilde{\mathcal{O}}_t$. This shows the following:

**Proposition 29.** *Let $t \in (0,1)$. Then $\tilde{\mathcal{O}}_t \neq \mathcal{O}_{\tilde{\mathcal{E}}}$ for any $\tilde{\mathcal{E}} \subset \mathcal{E}(\mathcal{S})$.*

Thus, we have demonstrated that the noise restriction given by Eq. (4.8) is not induced by any effect restriction but on the other hand it does restrict the set of physically feasible effects. In other words, there are restrictions on the level of effects as well as on the level of observables. One can easily confirm that if one is to have a similar generalization, denoted by $\hat{\mathcal{O}}_t$, of the noise condition $\hat{\mathcal{E}}_t$ given by Eq. (4.3) to observables on point-symmetric state spaces, then $\hat{\mathcal{O}}_t$ does not have this feature so that it is in fact induced by the effect restriction $\hat{\mathcal{E}}_t$ and is thus of type (R1).

# Chapter 5

# Simulability and compatibility

In an operational theory a set of devices are called *incompatible* if they cannot be implemented with a joint device that would have all of them as its components. The study of incompatible observables in quantum theory has a long history and is based on Heisenberg's uncertainty principle [49] and Born's notion of complementarity [124], which in particular show that there are quantum observables that cannot be measured jointly. Although incompatibility seems like a restrictive property of quantum measurements, at the same time it has been found at the core of many of the theory's distinguishing features, such as Bell inequality violations [125, 126] and the no-broadcasting theorem [127]. For incompatibility (mostly) in quantum theory we refer the reader to [48]. In GPTs incompatibility has been considered, e.g., in [27–35].

As we will see below and as was originally proven in [28], not only in quantum theory but in all non-classical theories one can find incompatible observables. In this Chapter we study the incompatibility of observables, show how it is linked to the simulation of observables and see how the previously introduced simulation structure can be used in topics related to incompatibilty. The results of this Chapter are largely based on Publications **I** and **III** but we also present some new results about incompatible observables on point-symmetric state spaces.

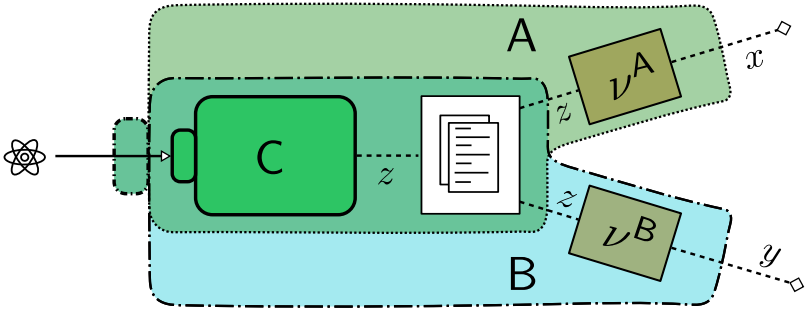

Figure 5.1: A joint measurement of observables A and B means that there exists an observable C from which both A and B can be post-processed. The outcomes $x$ and $y$ for observables A and B can be obtained simultaneously by copying the outcome $z$ of the observable C and by then applying the respective post-processings $\nu^A$ and $\nu^B$.

## 5.1 Compatibility of observables

Let us start with a formal definition of compatibility, i.e., joint measurability, of observables, and proceed to consider some of the resulting basic properties.

### 5.1.1 Definition and properties

**Definition 16.** A collection of observables $\mathcal{A} \subseteq \mathcal{O}(\mathcal{S})$ is *compatible* (or *jointly measurable*) if there exists a single observable $C \in \mathcal{O}(\mathcal{S})$ such that every observable $A \in \mathcal{A}$ is a post-processing of C, i.e., $\mathcal{A} \subset \mathfrak{sim}(C)$. If a collection of observables is not compatible, then it is called *incompatible*.

We emphasize that the above definition captures the idea that an outcome is produced for *each* of the observables in $\mathcal{A}$ when we perform exactly *one* measurement of the observable C on a *single* system. This is in contrast to just measuring an *informationally complete observable* whose whole measurement statistics (collected by performing the same measurement many times) specifies the measured state uniquely from which one can calculate the measurement statistics of any other observable [48]. We illustrate the compatibility of two observables with Fig. 5.1.

Another equivalent condition for compatibility of $A \in \mathcal{O}(\Omega, \mathcal{S})$ and $B \in$

$\mathcal{O}(\Lambda, \mathcal{S})$ is the existence of a *joint observable* $\mathsf{G} \in \mathcal{O}(\Omega \times \Lambda, \mathcal{S})$ such that $\mathsf{A}$ and $\mathsf{B}$ can be obtained as marginals of $\mathsf{G}$ so that $\sum_{y \in \Lambda} \mathsf{G}_{xy} = \mathsf{A}_x$ for all $x \in \Omega$ and $\sum_{x \in \Omega} \mathsf{G}_{xy} = \mathsf{B}_y$ for all $y \in \Lambda$. To see this, let us first note that clearly taking marginals is a post-processing so that both $\mathsf{A}$ and $\mathsf{B}$ can be obtained from the joint observable $\mathsf{G}$ by post-processing. On the other hand if $\mathsf{A}$ and $\mathsf{B}$ can be obtained from an observable $\mathsf{C} \in \mathcal{O}(\Gamma, \mathcal{S})$ by post-processing matrices $\nu^\mathsf{A}$ and $\nu^\mathsf{B}$ respectively, we see that an observable $\mathsf{G}$ defined by $\mathsf{G}_{xy} := \sum_{z \in \Gamma} \nu^\mathsf{A}_{zx} \nu^\mathsf{B}_{zy} \mathsf{C}_z$ for all $x \in \Omega$ and $y \in \Lambda$ is a joint observable for $\mathsf{A}$ and $\mathsf{B}$. The equivalence between compatibility and the existence of a joint observable can be readily generalized to any (finite) number of observables.

By considering classical theories, from Prop. 23 and 21 we see that all the observables on a classical state space $\mathcal{S}_d^{cl}$ are compatible as they can all be simulated from the single extreme simulation irreducible observable on $\mathcal{S}_d^{cl}$. On the other hand, if $\mathcal{S}$ is not a simplex, i.e., it is non-classical, then there exists (at least) two inequivalent simulation irreducible observables $\mathsf{A}, \mathsf{B} \in \mathcal{O}_{irr}(\mathcal{S})$. Suppose that $\mathsf{A}$ and $\mathsf{B}$ are compatible so that there exists $\mathsf{C} \in \mathcal{O}(\mathcal{S})$ such that $\mathsf{A}, \mathsf{B} \in \mathfrak{sim}(\mathsf{C})$. By the definition of simulation irreducibility it follows that then $\mathsf{A} \leftrightarrow \mathsf{C} \leftrightarrow \mathsf{B}$ which contradicts the inequivalence of $\mathsf{A}$ and $\mathsf{B}$. Thus, $\mathsf{A}$ and $\mathsf{B}$ must be incompatible so that *in every non-classical state space there exists (at least) two incompatible observables.* This was originally proven in [28] whilst also showing that the two incompatible observables can actually be chosen to be dichotomic (although in this case they might not be simulation irreducible).

Already from the definition of compatibility we see the connection to simulability. For a compatible set, such as the set of observables in a classical theory, the minimal number of observables needed to simulate is just one so we call it 1-*simulable*. By generalization, following [30], considering sets that are 2-, 3-, 4-, ..., and $n$-simulable, we can create a hierarchy that generalizes the notion of compatibility. For example, in classical state space $\mathcal{S}_d^{cl}$ the set of all observables $\mathcal{O}(\mathcal{S}_d^{cl})$ is 1-simulable, while in the square state space $\mathcal{S}_4$ we have that $\mathcal{O}(\mathcal{S}_4)$ is 2-simulable. As only one step away from being compatible in this generalized compatibility hierarchy, in this sense the square state space can thus be considered as being closest to a classical state space that a non-classical state space can be. Before making more connections to simulability, let us demonstrate compatibility in the case of two dichotomic obsevables.

### 5.1.2 Two dichotomic observables

For two dichotomic observables $\mathsf{A}, \mathsf{B} \in \mathcal{O}(\{+,-\}, \mathcal{S})$ we can characterize compatibility in the following way [25, 28]: First let $\mathsf{A}$ and $\mathsf{B}$ be compatible so that there exists a joint observable $\mathsf{G} \in \mathcal{O}(\{+,-\} \times \{+,-\}, \mathcal{S})$ such that $\mathsf{A}_+ = \mathsf{G}_{++} + \mathsf{G}_{+-}$, $\mathsf{A}_- = \mathsf{G}_{-+} + \mathsf{G}_{--}$, $\mathsf{B}_+ = \mathsf{G}_{++} + \mathsf{G}_{-+}$ and $\mathsf{B}_- = \mathsf{G}_{+-} + \mathsf{G}_{--}$. Let us denote $a := \mathsf{A}_+$, $b := \mathsf{B}_+$ and $g := \mathsf{G}_{++}$. Then from the previous equations it follows that

$$g \geq o, \quad a \geq g, \quad b \geq g, \quad u \geq a + b - g. \tag{5.1}$$

On the other hand, if there exists a function $g \in \mathcal{V}^*$ such that the conditions in Eq. (5.1) are satisfied for two dichotomic observables $\mathsf{A}$ and $\mathsf{B}$ with $a = \mathsf{A}_+$ and $b = \mathsf{B}_+$, then we can define a joint observable $\mathsf{G} \in \mathcal{O}(\{+,-\} \times \{+,-\}, \mathcal{S})$ for $\mathsf{A}$ and $\mathsf{B}$ by setting $\mathsf{G}_{++} = g$, $\mathsf{G}_{+-} = a - g$, $\mathsf{G}_{-+} = b - g$ and $\mathsf{G}_{--} = u - a - b + g$ so that the positivity of the effects is guaranteed.

Thus, we can conclude the following: *on a state space $\mathcal{S} \subset \mathcal{V}$, two dichotomic observables $\mathsf{A}, \mathsf{B} \in \mathcal{O}(\{+,-\}, \mathcal{S})$ with $a = \mathsf{A}_+$ and $b = \mathsf{B}_+$ are compatible if and only if there exists a linear functional $g \in \mathcal{V}^*$ such that the conditions in Eq. (5.1) are satisfied.*

**Example 13** (Qubit)**.** In the case of $\mathcal{S} = \mathbb{C}^2$, it is known (see [128, 129]) that if two dichotomic qubit POVMs $A, B \in \mathcal{O}(\{+,-\}, \mathbb{C}^2)$ with $a := A(+) = \frac{1}{2}(\alpha \mathbb{1}_2 + \vec{a} \cdot \vec{\sigma})$ and $b := B(+) = \frac{1}{2}(\beta \mathbb{1}_2 + \vec{b} \cdot \vec{\sigma})$ are compatible, then

$$\left\| \vec{a} + \vec{b} \right\|_2 + \left\| \vec{a} - \vec{b} \right\|_2 \leq 2. \tag{5.2}$$

Furthermore, if $\alpha = \beta = 1$, i.e., the POVMs are unbiased, then they are compatible if and only if Eq. (5.2) is satisfied.

#### Incompatiblity in point-symmetric state spaces

Let us generalise the well-known condition for compatibility of dichtomic qubit POVMs that was considered above. Recently in [35] the authors proved an even more general result from which this can be obtained as a special case. On the other hand, our result generalizes the result in [29] for even polygon state spaces.

**\*Proposition 30.** *Let $\mathsf{A}, \mathsf{B} \in \mathcal{O}(\{+,-\}, \mathcal{S})$ be two dichotomic observables with effects $\mathsf{A}_+ = a = \frac{1}{2}(\vec{a}, \alpha)$ and $\mathsf{B}_+ = b = \frac{1}{2}(\vec{b}, \beta)$ on a point-symmetric*

state space $\mathcal{S} \subset \mathbb{R}^{d+1}$. *If the observables* A *and* B *are compatible, then*

$$\left\|\vec{a}+\vec{b}\right\|_{\mathcal{E}} + \left\|\vec{a}-\vec{b}\right\|_{\mathcal{E}} \leq 2. \tag{5.3}$$

*Proof.* As it was discussed above, A and B are compatible if and only if there exists a linear functional $g \in \mathbb{R}^{d+1}$, parametrized by $g = \frac{1}{2}(\vec{g}, \gamma)$, such that the conditions in Eq. (5.1) are satisfied. From the *Prop. 1 one can see that this is the case if and only if

$$\|\vec{g}\|_{\mathcal{E}} \leq \gamma, \qquad \|\vec{a}-\vec{g}\|_{\mathcal{E}} \leq \alpha - \gamma,$$
$$\left\|\vec{b}-\vec{g}\right\|_{\mathcal{E}} \leq \beta - \gamma, \quad \left\|\vec{a}+\vec{b}-\vec{g}\right\|_{\mathcal{E}} \leq 2 - \alpha - \beta + \gamma.$$

Now

$$\left\|\vec{a}+\vec{b}\right\|_{\mathcal{E}} = \left\|\vec{a}+\vec{b}-\vec{g}+\vec{g}\right\|_{\mathcal{E}} \leq \left\|\vec{a}+\vec{b}-\vec{g}\right\|_{\mathcal{E}} + \|\vec{g}\|_{\mathcal{E}} \leq 2 - \alpha - \beta + 2\gamma$$

and

$$\left\|\vec{a}-\vec{b}\right\|_{\mathcal{E}} = \left\|\vec{a}-\vec{g}-(\vec{b}-\vec{g})\right\|_{\mathcal{E}} \leq \|\vec{a}-\vec{g}\|_{\mathcal{E}} + \left\|\vec{b}-\vec{g}\right\|_{\mathcal{E}} \leq \alpha + \beta - 2\gamma.$$

Thus, the existence of a joint observable for A and B implies the existence of $\gamma \in [0, 2]$ such that

$$\left\|\vec{a}+\vec{b}\right\|_{\mathcal{E}} + \alpha + \beta - 2 \leq 2\gamma \leq \alpha + \beta - \left\|\vec{a}-\vec{b}\right\|_{\mathcal{E}},$$

which leads to Eq. (5.3). $\square$

We recall that an effect $a = \frac{1}{2}(\vec{a}, \alpha) \in \mathcal{E}(\mathcal{S})$ on a point-symmetric state space $\mathcal{S}$ is unbiased if and only if $\alpha = 1$, and that an observable is unbiased if all of its nonzero effects are unbiased. For two unbiased observables we can show (as in the qubit case) that the inequality (5.3) is actually also a sufficient condition for compatibility.

***Proposition 31.*** *Two unbiased dichotomic observables* A *and* B *with effects* $A_+ = a = \frac{1}{2}(\vec{a}, 1)$ *and* $B_+ = b = \frac{1}{2}(\vec{b}, 1)$ *on a point-symmetric state space are compatible if and only if*

$$\left\|\vec{a}+\vec{b}\right\|_{\mathcal{E}} + \left\|\vec{a}-\vec{b}\right\|_{\mathcal{E}} \leq 2. \tag{5.4}$$

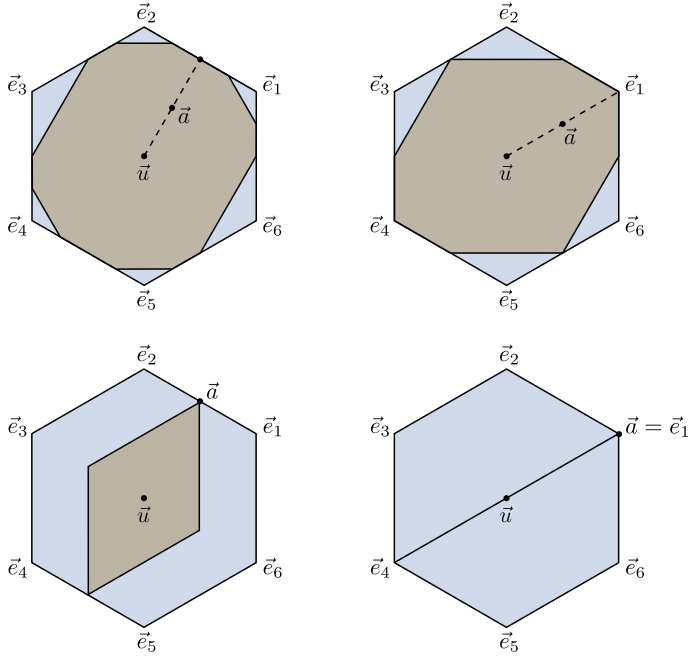

Figure 5.2: The compatibility regions (brown), i.e., the possible regions for the effects $b = \frac{1}{2}(\vec{b}, 1)$ in Eq. (5.4), for the unbiased effects on a regular hexagon state space $\mathcal{S}_6$ for some fixed effects $a = \frac{1}{2}(\vec{a}, 1)$. The set of unbiased effects (purple) is determined by the non-trivial extreme effects $e_k = \frac{1}{2}(\vec{e}_k, 1)$ for all $k \in \{1, 2, 3, 4, 5, 6\}$ with the point $\frac{1}{2}u = \frac{1}{2}(\vec{u}, 1)$ in the middle. We note that our result agrees with [29].

*Proof.* If A and B are compatible, then the inequality follows from *Prop. 30. Now let Eq. (5.4) hold. The following is a joint observable for A and B:

$$\mathsf{G}_{++} = \frac{1}{2}(t_+\vec{c}_+, t_+ + \frac{1}{2}t), \qquad \mathsf{G}_{+-} = \frac{1}{2}(t_-\vec{c}_-, t_- + \frac{1}{2}t),$$
$$\mathsf{G}_{--} = \frac{1}{2}(-t_+\vec{c}_+, t_+ + \frac{1}{2}t), \qquad \mathsf{G}_{-+} = \frac{1}{2}(-t_-\vec{c}_-, t_- + \frac{1}{2}t),$$

where $\vec{c}_\pm = (\vec{a} \pm \vec{b})/\left\|\vec{a} \pm \vec{b}\right\|_\mathcal{E}$, $t_+ = \left\|\vec{a} + \vec{b}\right\|_\mathcal{E}/2$, $t_- = \left\|\vec{a} - \vec{b}\right\|_\mathcal{E}/2$ and $t = 1 - t_+ - t_-$. □

A demonstration of the previous result is depicted in Fig. 5.2.

## 5.2 Simulation based condition for (in)compatibility

As we have seen, compatibility captures the idea that the measurement of multiple observables can be reduced to measuring and simulating just one observable. Let us consider a situation where one seemingly needs more than one observable to simulate a given set of observables. Thus, let $\mathcal{A} \subset \mathcal{O}(\mathcal{S})$ be a subset of observables such that $\mathcal{A} \subseteq \mathfrak{sim}(\mathcal{B})$ for some set of simulators $\mathcal{B} \subset \mathcal{O}(\mathcal{S})$. By definition $\mathcal{A} \subseteq \mathfrak{sim}(\mathcal{B})$ if and only if $\mathsf{A} = \sum_{i \in [n_\mathsf{A}]} p_i^\mathsf{A} \left( \nu^{\mathsf{A},(i)} \circ \mathsf{B}^{\mathsf{A},(i)} \right)$ for some observables $\{\mathsf{B}^{\mathsf{A},(i)}\}_{i=1}^{n_\mathsf{A}} \subset \mathcal{B}$ with outcome sets $\Omega_i$, some probability distribution $(p_i^\mathsf{A})_{i=1}^{n_\mathsf{A}}$ and post-processing matrices $\{\nu^{\mathsf{A},(i)}\}_{i=1}^{n_\mathsf{A}} \subset \mathcal{M}^{row}$ for some $n_\mathsf{A} \in \mathbb{N}$ for all $\mathsf{A} \in \mathcal{A}$ with outcome set $\Omega_\mathsf{A}$.

Let us consider a case where $n := n_\mathsf{A}$ and $(p_i)_{i=1}^n := (p_i^\mathsf{A})_{i=1}^{n_\mathsf{A}}$ are independent of the observable $\mathsf{A} \in \mathcal{A}$. We define a new post-processing $\nu^\mathsf{A}$ by setting $\nu^\mathsf{A}_{(i,x)y} = \nu^{\mathsf{A},(i)}_{xy}$ for all $x \in \Omega_i$, $y \in \Omega_\mathsf{A}$ and $i \in [n]$, and a new observable $\hat{\mathsf{B}}$ by setting $\hat{\mathsf{B}}_{(i,x)} = p_i \mathsf{B}_x^{(i)}$ for all $x \in \Omega_{\mathsf{B}^{(i)}}$ and $i \in [n]$. What follows is that then one can check that $\mathsf{A} = \nu^\mathsf{A} \circ \hat{\mathsf{B}}$ for all $\mathsf{A} \in \mathcal{A}$. Thus, in this case $\mathcal{A}$ is actually compatible as it can be simulated with only one observable. On the other hand, we see that if $\mathcal{A}$ is compatible, then clearly there exists a set $\mathcal{B} = \{\mathsf{B}\}$ such that $\mathcal{A} \subseteq \mathfrak{sim}(\mathcal{B})$, where the implemented (trivial) probability distribution is the same for all $\mathsf{A} \in \mathcal{A}$. Thus, we have shown the following result which can be extracted from Publication **I** and which was also explicitly shown in [30].

**Proposition 32.** *A set of observables $\mathcal{A} \subseteq \mathcal{O}(\mathcal{S})$ is compatible if and only if there exists a set of simulators $\mathcal{B} \subseteq \mathcal{O}(\mathcal{S})$ such that $\mathcal{A} \subseteq \mathfrak{sim}(\mathcal{B})$, where the mixing part of the simulation is achieved with the same probability distribution for every observable in $\mathcal{A}$.*

The joint measurement scheme for two observables given by the above result is depicted in Fig. 5.3. As a further demonstration of this result, let us consider a collection of $m$ observables $\{\mathsf{A}^{(i)}\}_{i=1}^m$, where $\mathsf{A}^{(i)} \in \mathcal{O}(\Omega_i, \mathcal{S})$, and suppose that

$$\sum_{i=1}^m w(\mathsf{A}^{(i)}; \mathcal{T}) \geq m - 1, \tag{5.5}$$

where $\mathcal{T} = \mathcal{T}(\mathcal{S})$ is the set of trivial observables on $\mathcal{S}$. We will show that the above noise condition is sufficient to satisfy the premise of Prop. 32. Let us start by defining a probability distribution $(p_i)_{i=1}^m$ by setting $p_i = 1 - w(\mathsf{A}^{(i)}; \mathcal{T})$ for all $i \in [m-1]$ and $p_m = 1 - \sum_{j=1}^{m-1} p_j$. Indeed, clearly $\sum_{i=1}^m p_j = 1$, and it can be checked that $p_i \geq 1 - w(\mathsf{A}^{(i)}; \mathcal{T})$ for all $i \in [m]$, where the inequality is an equality for $i \in [m-1]$ and the inequality for $i = m$ follows from Eq. (5.5).

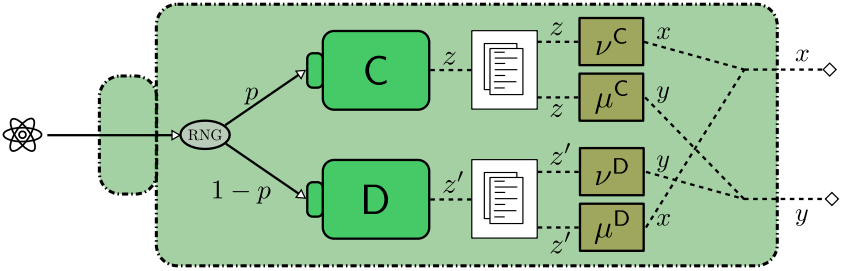

Figure 5.3: Simulation-based joint measurement scheme for two observables $\mathsf{A} = p(\nu^\mathsf{C} \circ \mathsf{C}) + (1-p)(\mu^\mathsf{D} \circ \mathsf{D})$ and $\mathsf{B} = (1-p)(\nu^\mathsf{D} \circ \mathsf{D}) + p(\mu^\mathsf{C} \circ \mathsf{C})$.

From the defining Eq. (2.2) of the noise content of $\mathsf{A}^{(i)}$ with respect to the set of trivial observables $\mathcal{T}$ (and since the supremum in that said definition is always attained for the noise set $\mathcal{T}$ as shown in Prop. 4) it follows that for each $i \in [m]$ there exists $\mathsf{B}^{(i)} \in \mathcal{O}(\Omega_i, \mathcal{S})$ and $\mathsf{T}^{(i)} \in \mathcal{T}(\Omega_i, \mathcal{S})$ such that

$$\mathsf{A}^{(i)} = w(\mathsf{A}^{(i)}; \mathcal{T})\mathsf{T}^{(i)} + (1 - w(\mathsf{A}^{(i)}; \mathcal{T}))\mathsf{B}^{(i)}.$$

For $i \in [m-1]$ we see that then $\mathsf{A}^{(i)} = p_i \mathsf{B}^{(i)} + (1-p_i)\mathsf{T}^{(i)}$. For $i = m$ it is straightforward to see from $p_m \geq 1 - w(\mathsf{A}^{(m)}; \mathcal{T})$ that there exists $\tilde{\mathsf{B}}^{(m)} \in \mathcal{O}(\Omega_m, \mathcal{S})$ such that $\mathsf{A}^{(m)} = p_m \tilde{\mathsf{B}}^{(m)} + (1-p_m)\mathsf{T}^{(m)}$. Let us redefine $\mathsf{B}^{(m)} := \tilde{\mathsf{B}}^{(m)}$ so that now $\mathsf{A}^{(i)} = p_i \mathsf{B}^{(i)} + (1-p_i)\mathsf{T}^{(i)}$ for all $i \in [m]$.

Let us set $\mathcal{B} = \{\mathsf{B}^{(i)}\}_{i=1}^m$. For each $i, j \in [m]$ let us define a post-processing $\nu^{(i,j)} \in \mathcal{M}^{row}$ by setting $\nu_{xy}^{(i,i)} = \delta_{xy}$ for all $x, y \in \Omega_i$, and $\nu_{xy}^{(i,j)} = q_y^{(i)}$ for all $x, y \in \Omega_i$ and $j \in [m] \setminus \{i\}$, where $(q_y^{(i)})_{y \in \Omega_i}$ is the probability distribution that defines $\mathsf{T}^{(i)}$, i.e., $\mathsf{T}_y^{(i)} = q_y^{(i)} u$ for all $y \in \Omega_i$. One can then easily confirm that

$$\sum_{j=1}^m p_j \left(\nu^{(i,j)} \circ \mathsf{B}^{(j)}\right) = p_i \mathsf{B}^{(i)} + (1-p_i)\mathsf{T}^{(i)} = \mathsf{A}^{(i)}$$

for all $i \in [m]$. Hence, $\{\mathsf{A}^{(i)}\}_{i=1}^m \subset \mathfrak{sim}(\mathcal{B})$, where the used probability distribution is the same for all $\mathsf{A}^{(i)}$ so that by Prop. 32 the set $\{\mathsf{A}^{(i)}\}_{i=1}^m$ is compatible. Hence, we have proven the following.

**Proposition 33.** *Let $\{\mathsf{A}^{(i)}\}_{i=1}^m \subset \mathcal{O}(\mathcal{S})$ be a collection of $m$ observables. If $\sum_{i=1}^m w(\mathsf{A}^{(i)}; \mathcal{T}) \geq m - 1$, then $\{\mathsf{A}^{(i)}\}_{i=1}^m$ is compatible.*

For example in the case of quantum theory, we can express the result as follows: *If the sum of the minimal eigenvalues of all of the effect operators of a collection of $m$ POVMs is greater than or equal to $m - 1$, then they are compatible.*

## 5.3 No-free-information principle

In quantum theory one can prove two powerful theorems: the *no-information-without-disturbance (NIWD)* theorem shows that those observables that do not cause any disturbance to the system when measured must be trivial [130], and the *no-free-information (NFI)* theorem says that those observables that can be measured jointly with any other observable must also be trivial (see, e.g., [63, Prop. 3.25]) . In other words, in quantum theory all nontrivial measurements cause disturbance, and there is no free information in the sense that a measurement of any nontrivial observable precludes the measurement of some other observable.

### 5.3.1 The principles

In Publication **III** we show that both of the claims described above do not hold in general in all GPTs. Thus, we can consider them as principles rather than theorems, and see in which theories the NIWD and NFI principles do and do not hold. To this end, we define the following three sets of observables on a state space $\mathcal{S}$:

$$\mathcal{T}(\mathcal{S}) = \{\mathsf{A} \in \mathcal{O}(\mathcal{S}) \,|\, \mathsf{A}_x(s) = \mathsf{A}_x(s') \,\forall x \in \Omega_\mathsf{A}, \,\forall s, s' \in \mathcal{S}\},$$
$$\mathcal{ND}(\mathcal{S}) = \{\mathsf{A} \in \mathcal{O}(\mathcal{S}) \,|\, \mathsf{A} \infty\, id\},$$
$$\mathcal{FC}(\mathcal{S}) = \{\mathsf{A} \in \mathcal{O}(\mathcal{S}) \,|\, \mathsf{A} \infty\, \mathsf{B} \,\forall \mathsf{B} \in \mathcal{O}(\mathcal{S})\},$$

where we have denoted the compatibility of two observables or an observable and a channel by the symbol $\infty$.

The first set $\mathcal{T}(\mathcal{S})$ is clearly just the set of trivial observables on $\mathcal{S}$, i.e., observables that do not provide any information about the measured state. The second set $\mathcal{ND}(\mathcal{S})$, the set of *non-disturbing observables*, consist of those observables that are compatible with the identity channel $id$ on $\mathcal{S}$. In this context the compatibility of a channel $\Phi$ and an observable $\mathsf{A}$ means that there exists an instrument $\mathcal{I}$ that can be used to implement both, i.e., $\sum_x \mathcal{I}_x = \Phi$ and $\mathsf{A}^{\mathcal{I}} = \mathsf{A}$ [48]. Thus, the set of observables that are compatible with the identity channel can be measured with some instrument in such way that the measured state is left unchanged and thus undisturbed. The last set $\mathcal{FC}(\mathcal{S})$, the set of *fully compatible observables*, is the set of all observables that can be measured jointly with any other observable on $\mathcal{S}$.

In general we can argue the following: First of all, any trivial observable $\mathsf{T} \in \mathcal{T}(\Omega, \mathcal{S})$, defined as $\mathsf{T}_x = p_x u$ for all $x \in \Omega$ for some probability distribution $(p_x)_{x \in \Omega}$, is non-disturbing as it can be implemented with the instrument $\mathcal{I} \in \mathrm{Ins}(\Omega, \mathcal{S})$ defined as $\mathcal{I}_x = p_x id$ for all $x \in \Omega$ so that $\sum_{x \in \Omega} \mathcal{I}_x = id$ and $\mathsf{A}^{\mathcal{I}} = \mathsf{T}$. Secondly, if an observable $\mathsf{A} \in \mathcal{O}(\Lambda, \mathcal{S})$ is non-disturbing so that it can be implemented with an instrument $\mathcal{J} \in \mathrm{Ins}(\Lambda, \mathcal{S})$ such that $\sum_{y \in \Lambda} \mathcal{J}_y = id$ and $\mathsf{A}^{\mathcal{J}} = \mathsf{A}$, we see that for any observable $\mathsf{B} \in \mathcal{O}(\Gamma, \mathcal{S})$ we can define an observable $\mathsf{G}^{\mathsf{B}} \in \mathcal{O}(\Lambda \times \Gamma, \mathcal{S})$ by setting $\mathsf{G}^{\mathsf{B}}_{yz} = \mathsf{B}_z \circ \mathcal{J}_y$ for all $y \in \Lambda$ and $z \in \Gamma$. Clearly then $\mathsf{G}^{\mathsf{B}}$ is a joint observable for $\mathsf{A}$ and $\mathsf{B}$ so that $\mathsf{A} \infty \mathsf{B}$ for all $\mathsf{B} \in \mathcal{O}(\mathcal{S})$. Thus, we can conclude the following

$$\mathcal{T}(\mathcal{S}) \subseteq \mathcal{ND}(\mathcal{S}) \subseteq \mathcal{FC}(\mathcal{S}). \tag{5.6}$$

In this way we get a concise formulation of the two principles: *the NIWD principle means that $\mathcal{ND}(\mathcal{S}) = \mathcal{T}(\mathcal{S})$ and the NFI principle means that $\mathcal{FC}(\mathcal{S}) = \mathcal{T}(\mathcal{S})$.* From Eq. (5.6) and the discussion above we also see that *the NFI principle implies the NIWD principle.*

We proceed by looking at the sets of non-disturbing and fully compatible observables separately, characterize them in a given theory and provide examples of theories where one, both or none of the principles hold. In previous works, the NIWD principle has been studied in GPTs in [131], but the NFI principle does not seem to have been studied before to the best of our knowledge.

### 5.3.2 Non-disturbing observables

Let us start by introducing a structure needed for the characterization of state spaces where the NIWD principle can be violated.

**Definition 17.** The *direct sum of state spaces* $\mathcal{S}_1 \subset \mathcal{V}_1, \ldots, \mathcal{S}_n \subset \mathcal{V}_n$, denoted by $\bigoplus_{i=1}^n \mathcal{S}_i$, is a state space in the vector space $\mathcal{V}_1 \times \cdots \times \mathcal{V}_n$ defined as the set of ordered and weighted pairs of states from $\mathcal{S}_1, \ldots, \mathcal{S}_n$, i.e.,

$$\bigoplus_{i=1}^n \mathcal{S}_i := \left\{ (\lambda_1 s_1, \ldots, \lambda_n s_n) \mid \forall i \in [n] : s_i \in \mathcal{S}_i, \lambda_i \in [0,1], \sum_{i=1}^n \lambda_i = 1 \right\}.$$

Some basic properties of direct sums of state spaces include that if $\mathcal{S} = \bigoplus_{i=1}^n \mathcal{S}_i \subset \mathcal{V}_1 \times \cdots \times \mathcal{V}_n$ for some state spaces $\mathcal{S}_1, \ldots, \mathcal{S}_n$, then $\mathcal{E}(\mathcal{S}) = \mathcal{E}(\mathcal{S}_1) \times \cdots \times \mathcal{E}(\mathcal{S}_n)$. Also, if $\mathsf{A} = (\mathsf{A}^{(1)}, \ldots, \mathsf{A}^{(n)}) \in \mathcal{O}(\mathcal{S})$ and $\mathsf{B} = (\mathsf{B}^{(1)}, \ldots, \mathsf{B}^{(n)}) \in \mathcal{O}(\mathcal{S})$ are two observables on $\mathcal{S} = \bigoplus_{i=1}^n \mathcal{S}_i$, then $\mathsf{A} \oo \mathsf{B}$ if and only if $\mathsf{A}^{(i)} \oo \mathsf{B}^{(i)}$ for all $i \in [n]$. Furthermore, the following is a useful condition for determining when a state space can be expressed as a direct sum of some other state spaces:

**Proposition 34.** *Let $\mathcal{S}$ be a state space. If $\mathcal{S}_1, \ldots, \mathcal{S}_n \subset \mathcal{S}$ are closed, convex subsets of $\mathcal{S}$ such that $\mathrm{conv}(\cup_{i=1}^n \mathcal{S}_i) = \mathcal{S}$ and for every $s \in \mathcal{S}$ there exists a unique convex decomposition $s = \sum_{i=1}^n \lambda_i s_i$ into elements of $\mathcal{S}_1, \ldots, \mathcal{S}_n$, then $\mathcal{S} = \bigoplus_{i=1}^n \mathcal{S}_i$.*

By using the direct sum structure (similar to [131]) we are able to show the following characterization of a state space $\mathcal{S}$ where the NIWD principle is violated that also contains a characterization for the set of non-disturbing observables $\mathcal{ND}(\mathcal{S})$:

**Proposition 35.** *An observable $\mathsf{A} \in \mathcal{O}(\Omega, \mathcal{S})$ is non-disturbing, i.e., $\mathsf{A} \in \mathcal{ND}(\mathcal{S})$, if and only if the state space $\mathcal{S}$ can be represented as a direct sum $\mathcal{S} = \bigoplus_{i=1}^n \mathcal{S}_i$ such that $\mathsf{A}_x$ is constant on each $\mathcal{S}_i$ for all $x \in \Omega$ and $i \in [n]$.*

The intuition behind the previous result is that the non-disturbing observables can only provide essentially classical information about in which summand state space $\mathcal{S}_i$ the state was prepared in, and naturally such observables only exist if the state space can be represented as a direct sum to begin with. Clearly every state space $\mathcal{S}$ can be represented as a direct sum with only one summand, $\mathcal{S}$, but in such a representation only the trivial observables are constant on $\mathcal{S}$ by definition. Thus, the violation of the NIWD principle indeed requires a non-trivial direct sum structure.

As a simple example, let $\mathcal{S} \subset \mathcal{V}$ be a 2-dimensional state space so that $\dim(\text{aff}(\mathcal{S})) = 2$ and $\dim(\mathcal{V}) = 3$. Suppose $\mathcal{S}$ is a direct sum of state spaces $\mathcal{S}_1$ and $\mathcal{S}_2$. Then $\mathcal{S} = \mathcal{S}_1 \oplus \mathcal{S}_2 \subset \mathcal{V}_1 \times \mathcal{V}_2$ so that $\dim(\mathcal{V}_1) + \dim(\mathcal{V}_2) = \dim(\mathcal{V}) = 3$. It follows that necessarily one of the state spaces $\mathcal{S}_1$ and $\mathcal{S}_2$ must be a point and the other one a line segment so that $\mathcal{S}$ is a triangle, $\mathcal{S} = \mathcal{S}_3^{cl}$. Thus, *the only 2-dimensional state space that admits non-disturbing observables is just the classical state space shaped as a triangle.* In particular this implies that for the polygon state spaces $\mathcal{S}_n$ we have that $\mathcal{ND}(\mathcal{S}_n) = \mathcal{T}(\mathcal{S}_n)$ for all $n > 3$.

Intuitively it is known that in the classical case one can perform every measurement without disturbing the system. To see this in the current context, let us consider a $d$-dimensional classical state space $\mathcal{S}_d^{cl}$. For $\mathcal{S}_d^{cl}$ we see that because $\mathcal{S}_d^{cl} = \text{conv}(\{\delta_1, \ldots, \delta_d\})$, where $\delta_1, \ldots, \delta_d$ are the extreme points of $\mathcal{S}_d^{cl}$, and because the convex decomposition of every state $\delta \in \mathcal{S}_d^{cl}$ into pure states $\delta_1, \ldots, \delta_d$ is unique, it follows from Prop. 34 that $\mathcal{S}_d^{cl} = \bigoplus_{i=1}^d \{\delta_i\}$. On the other hand, since every effect of every observable on $\mathcal{S}_d^{cl}$ is obviously constant on each pure state $\delta_i$, it follows that $\mathcal{ND}(\mathcal{S}_d^{cl}) = \mathcal{O}(\mathcal{S}_d^{cl})$ so that every observable on a classical state space is non-disturbing.

Using the direct sum of state spaces one can easily construct other state spaces where the NIWD principle is violated, i.e., $\mathcal{ND}(\mathcal{S}) \neq \mathcal{T}(\mathcal{S})$. For example, this is true in all 3-dimensional state spaces that are a direct sum of any 2-dimensional convex set and a point so that they have a pyramid type of shape with any convex base. Moreover, although in single system quantum theory we know that non-disturbing observables are always trivial, we can still consider direct sums of quantum state spaces which lead to different superselection sectors [132]. For a more detailed example of this see Publication **III**.

## 5.4  Fully compatible observables

For the set $\mathcal{FC}(\mathcal{S})$ of fully compatible observables it turns out that we can use the structure that we developed for studying the simulability of observables. As we have seen, simulation can be considered as a certain type of generalization to joint measurability, so this type of connection is only natural.

### 5.4.1 Characterization and properties

In particular, we can show that the simulation irreducible observables play a crucial role in the characterization of the set of fully compatible observables in a given theory.

**Proposition 36.** *An observable is compatible with every other observable if and only if it can be post-processed from every simulation irreducible observable, i.e.,*

$$\mathcal{FC}(\mathcal{S}) = \bigcap_{\mathsf{B} \in \mathcal{O}_{irr}(\mathcal{S})} \mathfrak{sim}(\mathsf{B}).$$

Since for each $\mathsf{B} \in \mathcal{O}_{irr}(\mathcal{S})$ there is an extreme simulation irreducible observable $\hat{\mathsf{B}}$ such that $\hat{\mathsf{B}} \leftrightarrow \mathsf{B}$ (namely the minimally sufficient representative of the equivalence class of $\mathsf{B}$), it follows that $\mathfrak{sim}(\mathsf{B}) = \mathfrak{sim}(\hat{\mathsf{B}})$ so that if we denote the set of extreme simulation irreducible observables by $\mathcal{O}_{irr}^{ext}(\mathcal{S})$, we must have that actually $\mathcal{FC}(\mathcal{S}) = \bigcap_{\mathsf{B} \in \mathcal{O}_{irr}^{ext}(\mathcal{S})} \mathfrak{sim}(\mathsf{B})$.

The previous result is analogous to the characterization of the trivial observables: $\mathcal{T}(\mathcal{S}) = \bigcap_{\mathsf{B} \in \mathcal{O}(\mathcal{S})} \mathfrak{sim}(\mathsf{B}) = \bigcap_{\mathsf{B} \in \mathcal{O}(\mathcal{S}) \setminus \mathcal{T}(\mathcal{S})} \mathfrak{sim}(\mathsf{B})$. We see that restricting the set $\mathcal{O}(\mathcal{S}) \setminus \mathcal{T}(\mathcal{S})$ to $\mathcal{O}_{irr}(\mathcal{S})$ is enough to make the difference, if such a difference exists in a given theory, between trivial and fully compatible observables. As was stated at the very beginning of this section, such a difference does not exist in quantum theory where the NFI principle holds. The following result provides another class of theories where the NFI principle does hold.

**Proposition 37.** *In every point-symmetric state space $\mathcal{S}$ we have that $\mathcal{FC}(\mathcal{S}) = \mathcal{ND}(\mathcal{S}) = \mathcal{T}(\mathcal{S})$.*

This result is based on the fact that in point-symmetric state spaces one is able to find two inequivalent dichotomic simulation irreducible observables $\mathsf{A}$ and $\mathsf{B}$ and by analysing the structure of the effects in $\mathfrak{sim}(\mathsf{A}) \cap \mathfrak{sim}(\mathsf{B})$ one can show that only trivial effects can exist in the intersection of the spans determined by the extreme rays corresponding to the two indecomposable effects of each observable.

To see that in general there is a difference between $\mathcal{FC}(\mathcal{S})$ and $\mathcal{T}(\mathcal{S})$, i.e., that there are theories where the NFI principle does not hold, we can show the following:

**Proposition 38.** *Let $\mathcal{S}$ be a d-dimensional state space. If $|\mathcal{O}^{ext}_{irr}(\mathcal{S})| < \infty$ and all of the extreme simulation irreducible observables have $d+1$ outcomes, then $\mathcal{FC}(\mathcal{S}) \neq \mathcal{T}(\mathcal{S})$.*

Although the previous result is quite limited in its effectiveness considering all of the different theories, we see that we can demonstrate its use in the case of classical state spaces: on $\mathcal{S}^{cl}_d$ there is only one extreme simulation irreducible observable with $d+1$ outcomes (the one that perfectly distinguishes all the $d+1$ pure states), and we see that $\mathcal{FC}(\mathcal{S}^{cl}_d) = \mathcal{O}(\mathcal{S}^{cl}_d)$. This confirms the result that we have already established before. As has also been established before, $\mathcal{FC}(\mathcal{S}) = \mathcal{O}(\mathcal{S})$ holds only in classical theories and in this case $\mathcal{ND}(\mathcal{S}) = \mathcal{O}(\mathcal{S})$ also holds.

So far we have seen that there are theories (such as quantum and point-symmetric theories) where $\mathcal{FC}(\mathcal{S}) = \mathcal{ND}(\mathcal{S}) = \mathcal{T}(\mathcal{S})$ so that the NFI, and thus also the NIWD, principles hold, and theories (such as the classical theory, superselected quantum theory or any theory with a direct sum state space) where $\mathcal{ND}(\mathcal{S}) \neq \mathcal{T}(\mathcal{S})$ so the NIWD principle is violated. Furthermore, if $\mathcal{S}$ is classical, then we have that $\mathcal{O}(\mathcal{S}) = \mathcal{FC}(\mathcal{S}) = \mathcal{ND}(\mathcal{S}) \neq \mathcal{T}(\mathcal{S})$. To conclude this Section and Chapter, we present the following example where we find that there are theories where $\mathcal{FC}(\mathcal{S}) \neq \mathcal{ND}(\mathcal{S}) = \mathcal{T}(\mathcal{S})$, and $\mathcal{FC}(\mathcal{S}) \neq \mathcal{ND}(\mathcal{S}) \neq \mathcal{T}(\mathcal{S})$.

### 5.4.2   (Odd) Polygon state spaces

Let us consider the polygon state spaces $\mathcal{S}_n$. As was discussed after Prop. 35, since $\mathcal{S}_n$ is 2-dimensional for all $n \in \mathbb{N}$, we have that $\mathcal{ND}(\mathcal{S}_n) = \mathcal{T}(\mathcal{S}_n)$ for all $n > 3$ and $\mathcal{ND}(\mathcal{S}_3) = \mathcal{O}(\mathcal{S}_3)$. Also, if $n = 2m$ for some $m \in \mathbb{N}$ so that $n$ is even, the state space is point-symmetric so that by Prop. 37 we have $\mathcal{FC}(\mathcal{S}_{2m}) = \mathcal{ND}(\mathcal{S}_{2m}) = \mathcal{T}(\mathcal{S}_{2m})$.

Let us focus on the case when $n$ is odd so that $n = 2m + 1$. As was explained in Ex. 11 in Sec. 4.2, the extreme simulation observables on $\mathcal{S}_{2m+1}$ are trichotomic and there are a finite number of them so that we can use Prop. 38 to establish that $\mathcal{FC}(\mathcal{S}_{2m+1}) \neq \mathcal{T}(\mathcal{S}_{2m+1})$. In fact, the extreme simulation irreducible observables are easy to find on $\mathcal{S}_{2m+1}$ and by looking at the structure of the intersections in characterization of $\mathcal{FC}(\mathcal{S}_{2m+1})$ from Prop. 36, the set of effects of the fully compatible observables, denoted by $\mathcal{E}_{\mathcal{FC}(\mathcal{S}_{2m+1})}$, can be fully characterized. While we leave the details of the full characterization to Publication **III**, we present this set of effects for the simplest non-classical odd polygon, the pentagon, in Fig. 5.4.

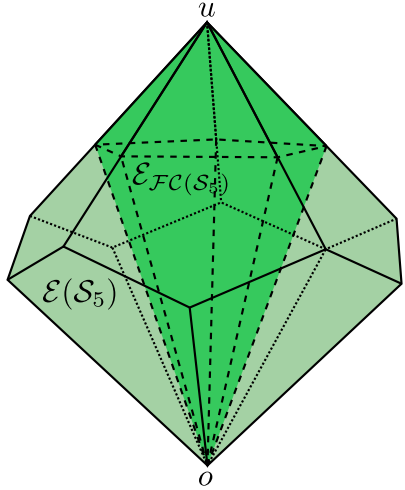

Figure 5.4: The set of effects $\mathcal{E}_{\mathcal{FC}(\mathcal{S}_5)}$ of the fully compatible observables $\mathcal{FC}(\mathcal{S}_5)$ in the effect space $\mathcal{E}(\mathcal{S}_5)$ of the pentagon state space $\mathcal{S}_5$.

From the Fig. 5.4 one can see that the set of effects $\mathcal{E}_{\mathcal{FC}(\mathcal{S}_{2m+1})}$ are heavily centred around the line segment $\mathcal{E}_{\mathcal{T}} = \{pu \,|\, p \in [0,1]\}$ of the trivial effects. Indeed, as we increase $m$ in $n = 2m+1$, we can show that $\mathcal{E}_{\mathcal{FC}(\mathcal{S}_{2m+1})}$ shrinks more around $\mathcal{E}_{\mathcal{T}}$ so that in the limit $m \to \infty$ when the polygon becomes a disc that is isomorphic to the Bloch disc of real qubits $\mathcal{S}(\mathbb{R}^2)$, we have that $\mathcal{E}_{\mathcal{FC}(\mathcal{S}(\mathbb{R}^2))} = \mathcal{E}_{\mathcal{T}}$ and $\mathcal{FC}(\mathcal{S}(\mathbb{R}^2)) = \mathcal{T}(\mathcal{S}(\mathbb{R}^2))$. Some of this process can be explained by the use of the noise content with which we can show the following:

**Proposition 39.** *Let* $\mathsf{A} \in \mathcal{O}(\Omega, \mathcal{S}_{2m+1})$ *be an observable on an odd polygon state space* $\mathcal{S}_{2m+1}$ *with effects* $\mathsf{A}_x = \alpha_x(\vec{a}_x, \sigma_{2m+1})$ *for all* $x \in \Omega$. *If* $\mathsf{A} \in \mathcal{FC}(\mathcal{S}_{2m+1})$, *then*

$$w(\mathsf{A}; \mathcal{T}) \geq 1 - \frac{\sin\left(\frac{\pi}{2(4l+3)}\right)}{\cos\left(\frac{\pi}{4l+3}\right)}, \tag{5.7}$$

*if* $m = 2l+1$ *for some* $l \in \mathbb{N} \cup \{0\}$, *and*

$$w(\mathsf{A}; \mathcal{T}) \geq 1 - \frac{\sin\left(\frac{2\pi}{2(4l+1)}\right)}{\cos^2\left(\frac{\pi}{4l+1}\right)}, \tag{5.8}$$

*if* $m = 2l$ *for some* $l \in \mathbb{N}$.

| $n$ | 3 | 5 | 7 | 9 | 11 | 13 | $\cdots$ | $\to \infty$ |
|---|---|---|---|---|---|---|---|---|
| R.H.S. of Eq. (5.7) | 0 | - | 0.753 | - | 0.852 | - | $\cdots$ | $\to 1$ |
| R.H.S. of Eq. (5.8) | - | 0.528 | - | 0.803 | - | 0.872 | $\cdots$ | $\to 1$ |

Table 5.1: The lower bounds for the noise contents of the fully compatible observables on odd polygon state spaces given by Prop. 39.

The values of the right-hand sides of Eq. (5.7) and (5.8) are listed in Table 5.1. We see that the fully compatible observables are indeed quite noisy as described above.

Lastly, let us consider a direct sum of two odd polygon state spaces, for example a pentagon $\mathcal{S}_5$ and a heptagon $\mathcal{S}_7$. Since $\mathcal{S}_5 \oplus \mathcal{S}_7$ is a direct sum one can construct a dichotomic observable $\mathsf{A} \in \mathcal{O}([2], \mathcal{S}_5 \oplus \mathcal{S}_7)$ defined by the effects $\mathsf{A}_1 = (u, o) \in \mathcal{E}(\mathcal{S}_5) \times \mathcal{E}(\mathcal{S}_7) = \mathcal{E}(\mathcal{S}_5 \oplus \mathcal{S}_7)$ and $\mathsf{A}_2 = (o, u) \in \mathcal{E}(\mathcal{S}_5 \oplus \mathcal{S}_7)$. Since the effects of $\mathsf{A}$ are not proportional to $u = (u, u) \in \mathcal{E}(\mathcal{S}_5 \oplus \mathcal{S}_7)$, the observable $\mathsf{A}$ is not trivial. However, the effects of $\mathsf{A}$ are constant on the summands $\mathcal{S}_5$ and $\mathcal{S}_7$ so that $\mathsf{A} \in \mathcal{ND}(\mathcal{S}_5 \oplus \mathcal{S}_7)$. Just as was explained before, the observable $\mathsf{A}$ only detects whether the state was prepared in $\mathcal{S}_5$ or $\mathcal{S}_7$ without disturbing the prepared state. On the other hand, we can define an observable $\mathsf{B} \in \mathcal{FC}(\mathcal{S}_5 \oplus \mathcal{S}_7)$ by setting $\mathsf{B} = (\mathsf{B}_5, \mathsf{B}_7)$, where $\mathsf{B}_5 \in \mathcal{FC}(\mathcal{S}_5) \setminus \mathcal{T}(\mathcal{S}_5)$ and $\mathsf{B}_7 \in \mathcal{FC}(\mathcal{S}_7) \setminus \mathcal{T}(\mathcal{S}_7)$. Since $\mathsf{B}_5$ and $\mathsf{B}_7$ are not trivial observables, neither is $\mathsf{B}$ and furthermore the effects of $\mathsf{B}_5$ and $\mathsf{B}_7$ are not constant on $\mathcal{S}_5$ and $\mathcal{S}_7$ respectively so that $\mathsf{B} \notin \mathcal{ND}(\mathcal{S}_5 \oplus \mathcal{S}_7)$. Thus, $\mathcal{FC}(\mathcal{S}_5 \oplus \mathcal{S}_7) \neq \mathcal{ND}(\mathcal{S}_5 \oplus \mathcal{S}_7) \neq \mathcal{T}(\mathcal{S}_5 \oplus \mathcal{S}_7)$.

# Conclusions

In this thesis we have considered measurements and their properties in general operational theories. As measurements are a defining part of any empirical physical theory, studying them gives us great detail about the properties of the theory. By considering various operational tasks involving measurements one is able to quantify how well they can be implemented in different theories, allowing one to compare theories to one another. This helps us to understand what might be the defining properties of quantum theory and what makes quantum theory so special. In this thesis the main focus was specifically measurement simulability and concepts related to it.

After having introduced the convex formulation of general probabilistic theories in Chapter 1, we started Chapter 2 by considering various communication tasks that are characterized by communication matrices which originated from the experimental prepare-and-measure scenarios. This allowed us to provide context to operational properties that can be considered in the framework of operational theories. Furthermore, we considered how the communication tasks could be simulated by other tasks by the means of ultraweak matrix majorization. This helped us determine that some tasks might be more difficult to implement than others. By considering various mathematically and physically motivated monotones on this induced order of difficulty we were able to derive physically meaningful characterizing dimensions for the theories. We identified (most) of the characterizing dimensions for quantum theory and compared it to other theories and saw differences in tasks such as minimum error state discrimination.

Chapter 3 was our first step towards simulation of measurements. As one of the two main components of simulability, we considered the classical manipulation of measurement outcomes known as the post-processing of observables. Furthermore, we generalized the concept of post-processing to instruments. We studied how post-processing can be used to obtain a new measurement from a known one, how this defines a way to compare measurements to each other and characterized the structure of this induced order. We saw that in the case of observables every observable can be post-processed from different indecomposable observables while for instruments only the identity channel can be used to post-process every other instrument from. We demonstrated the newly defined instrument post-processing by considering indecomposable instruments and measure-and-prepare instruments. For quantum theory we saw that indecomposable elements are exactly the rank-1 elements both for POVMs and quantum instruments and that random orthogonal isometric instruments can be used to post-process every other quantum instrument.

In Chapter 4 we reviewed the simulation scheme of observables, combining the classical operations of mixing and post-processing of observables. We found that every observable can be simulated by a class of simulation irreducible observables, and we characterized these observables in terms of the structural notions of indecomposability and extremality. We demonstrated the simulation of observables by considering theories beyond the no-restriction hypothesis in which we argued that the physical restriction imposed on the set of observables must be closed with respect to the simulation process. Furthermore, we characterize the restrictions based on which level of description of a measurement the restriction takes place and give examples of different types of restrictions. Although in quantum theory the general set of POVMs is unrestricted, experimental limitations might force one to use these types of restrictions on experimental set-ups even for quantum systems.

We close the thesis in Chapter 5 by considering the connections between compatibility, i.e., joint measurability, and simulability of observables. We see that simulability can be considered as a generalization of compatibility and derive a simulation– and noise–based sufficient condition for compatibility. We explore this connection further and see that the set of fully compatible observables, i.e., observables that can be measured jointly with any other observable, can be characterized in terms of the simulation irreducible observables. We see that two important principles that hold in quantum theory, namely the no-information-without-disturbance principle (according to which observables that do not cause any disturbance in the measured system must be trivial) and the no-free-information principle (according to which fully compatible observables must be trivial), do not hold in operational theories in general. This way we see that to some extent these principles hold some of the characteristics of quantum theory and may even be used as axioms to exclude unphysical theories.